\documentclass[11pt,a4paper]{article}
\usepackage{jcappub}

\title{Boltzmann hierarchies for self-interacting warm dark matter scenarios}
\author[a,b]{Rafael Yunis}
\author[c]{Carlos R. Arg\"uelles}
\author[d]{Diana L\'opez Nacir}

\affiliation[a]{ICRANet, Piazza della Repubblica 10, I--65122 Pescara, Italy}
\affiliation[b]{Physics Department, La Sapienza University of Rome, P.le Aldo Moro 5, I--00185 Rome, Italy}
\affiliation[c]{Facultad de Ciencias Astron\'omicas y Geof\'isicas, Universidad Nacional de La Plata, Paseo del Bosque, B1900FWA LaPlata, Argentina}
\affiliation[d]{Departamento de F\'isica  Juan Jos\'e Giambiagi, FCEyN UBA and IFIBA CONICET-UBA, Facultad de Ciencias Exactas y Naturales, Ciudad Universitaria, Pabell\'on I, 1428 Buenos Aires, Argentina}
\emailAdd{yunis121@gmail.com}

\usepackage[utf8]{inputenc}	
\usepackage[T1]{fontenc} 		
\usepackage{lmodern}				
\usepackage{microtype}			

\usepackage{cleveref}
\usepackage{siunitx}
\usepackage[all]{nowidow}
\usepackage{tabularx}
\usepackage{booktabs}

\usepackage{graphicx}
\usepackage{caption}
\usepackage{subcaption}

\usepackage[markup=nocolor,addedmarkup=bf,authormarkup=none]{changes}
\definechangesauthor[name={R.Y.}, color=red]{ry}
\definechangesauthor[name={C.A.}, color=blue]{ca}
\definechangesauthor[name={D.L.N.}, color=magenta]{dln}

\newcommand{\nocontentsline}[3]{}
\newcommand{\tocless}[2]{\bgroup\let\addcontentsline=\nocontentsline#1{#2}\egroup}

\abstract{{
We provide a general framework for self-interacting warm dark matter (WDM) in cosmological perturbations, by deriving from first principles a Boltzmann hierarchy which retains certain independence from a particular interaction Lagrangian. 
We consider elastic interactions among the massive particles, and obtain a hierarchy which is more general than the ones usually obtained for non-relativistic (as for cold DM) or for ultra-relativistic (as for neutrinos) approximations. The more general momentum-dependent kernel integrals in the Boltzmann collision terms, are explicitly calculated for different field-mediator models, including examples of a scalar field
or a massive vector field. 
As an application, we study the evolution of the interaction rate per particle under the relaxation time approximation, and assess when a given self-interaction is relevant in comparison with the Hubble expansion rate. Our framework aims to be a useful tool to evaluate DM self-interaction effects in the linear power spectrum, with the consequent imprints on non-linear scales of structure formation.
  
}}

\begin{document}
\thispdfpagelabel{Title}
\maketitle
\flushbottom


\sloppy
\clearpage
\section{Introduction}
\label{sec:Intro}


In the standard cosmological paradigm, $\Lambda$CDM, Dark Matter (DM) has long been a necessary ingredient in the Big Bang model of the universe and in understanding its evolution since the early stages
. 
While the evidence for its existence is implied by its gravitational effects in astrophysical, galactic and cosmological structures; understanding the nature and composition of this species is still an elusive subject~\cite{Bergstrom:2000pn, Munoz:2003gx, Bertone:2004pz, BertoneNature}
. 
Several attempts have been made to explain this phenomenon by macroscopic objects, yet a microscopic origin of the DM phenomenon by a new particle species remains as the most plausible hypothesis~\cite{McGaugh:2014nsa,Bull:2015stt,Bertone:2004pz,BertoneNature,Milgrom:2019cle}
. 
On the early stages of DM research, active neutrinos appeared as promising candidates for this particle species~\cite{Szalay:1976ef,Gershtein:1966gg}. However, neutrinos are ``hot'' Dark Matter (HDM) with a free-streaming length which erases structures up to large scales~\cite{Primack:2000iq}, while numerical simulations have shown that such ``top-down'' structure formation is incompatible with clustering constraints~\cite{Hut:1984qd}
.  
Cosmological data favored the adoption of the $\Lambda$CDM paradigm~\cite{Bahcall:1999xn}: in the standard scenario, DM is assumed to be produced in a thermal distribution and modeled as collisionless after it decouples from the other species. The effective decoupling is assumed to occur at a temperature smaller than the DM mass so that the distribution corresponds to non-relativistic particles. The traditional candidates are weakly interacting massive particles (WIMPS) which were in thermal equilibrium with the species in the cosmic plasma via weak interactions \cite{1999coph.book.....P}
.
In this scenario, galaxies form in a ``bottom-up'' fashion: small scales (favored by the small velocity dispersion of CDM particles) become non-linear and collapse first, and their merging and accretion leads to formation of structures on larger scales
. 
On these scales, data of the structure of the universe is consistent with CDM driving the formation of galaxies and clusters, however, no viable fundamental particle within the standard model (SM) fulfills these properties~\cite{Feng:2010gw,Bertone:2004pz} 
. 



The standing $\Lambda$CDM paradigm, is in remarkable agreement with large scale cosmological observations (see for instance \cite{Ivanov:2019hqk,Troster:2019ean,DAmico:2019fhj,2019arXiv190907951C}) and it is also compatible with an increasing amount of observed galaxy properties (e.g. \cite{Vogelsberger:2014dza} and \cite{2015MNRAS.450.1937C})%
.
However, it has been noted that in this paradigm it is challenging to describe some observables on smaller scales, such as the ``missing'' dark matter sub-halos or the so called core-cusp discrepancy~\cite{Bullock:2017xww}%
. 
High resolution cosmological simulations of average-sized halos in $\Lambda$CDM predicts~\cite{Griffen_2016} an overproduction of small-scale structures, significantly larger that the observed number of small satellite galaxies in the Local Group~\cite{Janesh_2019,Drlica-Wagner:2015ufc}
. 
Moreover, N-body simulations of CDM-only predict a singular density profile for virialized halos~\cite{Navarro:1995iw,Diemand:2005wv}, while observational evidence points to dwarf spheroidal galaxies (dSphs) having smooth cores in their central regions~\cite{Joung:2008zy,Battaglia:2008jz}
. 
Some other tensions have been raised between CDM-only predictions and observations (see for example a review in~\cite{Bullock:2017xww})%
. 

Among the earliest approaches to alleviate/resolve those conflicts is to consider two DM components, one ``cold'' and one ``hot'' (C+HDM) \cite{Klypin:1992sf,Dolgov:2002wy}
.
More recent models feature only \textit{warm} dark matter particles (WDM)~\cite{Bode:2000gq}, meaning that they are semi-relativistic during the earliest stages of structure formation with non-negligible free-streaming particle length
. 
WDM models feature an intermediate velocity dispersion between HDM and CDM that results in a suppression of structures at small scales due to free-streaming~\cite{Bond:1980ha}
. 
If this free streaming scale today is smaller than the size of galaxy clusters, it can provide a solution to the missing satellites problem~\cite{Joung:2008zy,Lovell_2012,Lovell:2013ola,Arg_elles_2018}
. 
However, thermally produced WDM suffers from the so called \textit{catch-22} problem when studied within N-body simulations \cite{Maccio_2012,Shao_2013}. Such WDM-only simulations either show unrealistic core-sizes for particle masses above the keV range, or they acquire the right halo sizes though for sub-keV masses, in direct conflict with phase-space constraints \cite{2009JCAP...03..005B}. It is important to remark that this may be due to shortcomings on the simulations themselves, and could be alleviated by including baryon feedback \cite{Fitts_2019}
. 
A particular, promising realization of these WDM models has been the minimal extension of the SM by intermediate-mass sterile neutrinos in the $\mathcal{O}$(keV) range known as $\nu$MSM (see, for example, \cite{Boyarsky:2018tvu} for a review)
. 

From the astrophysical point of view, fermion masses in this range and up to $\mathcal{O}$(0.1 MeV) seem to also be favored by recent elementary particle based DM halo studies \cite{Arg_elles_2018,argelles2018novel}
, where self-gravitating equilibrium systems were shown to be both in excellent agreement with rotation curve observations while thermodynamically stable (coarse-grained entropy maxima) within cosmological timescales~\cite{daky}
. 
From current cosmological data it is possible to constrain these models analizing observable properties, such as from Lyman-$\alpha$ forest and sub-structure observations in the Local Group. Comprehensive reviews of the constraints for sterile neutrinos can be found in e.g. \cite{Adhikari2016,Boyarsky_2019}
. 

Another compelling alternative to colissionless CDM, apart from WDM, is to consider interactions in CDM
. 
This consideration relaxes the assumption that CDM interacts only gravitationally after early decoupling, and includes interactions either between DM and SM particles or additional hidden particles, or among DM particles themselves. These later models are denominated as ``self-interacting'' DM models (SIDM) (see \cite{Tulin2017,amrr} for reviews)
.
Born out of N-body simulations \cite{2000ApJ...534L.143B}, SIDM halos could explain the cores of galaxies when a $2 \leftrightarrow 2$ interaction is assumed, with cross-sections constrained to be roughly of $\sigma/m \sim 0.5 - 10 \ \mathrm{cm}^2/\mathrm{g}$~\cite{Tulin2017,Zavala_2013,Elbert:2014bma}
. However, certain tensions have been raised about the upper limits in the self-interaction cross section, based on a more refined analysis of the Bullet Cluster~\cite{Randall:2007ph}. This has motivated the consideration of velocity dependent cross sections (i.e. $\sigma$ as a function of the rms velocity of DM particles) which are sensitive to the baryonic environment \cite{2018ApJ...853..109E}.

Most SIDM studies assume a cosmological evolution identical to CDM on large scales, and that the linear matter power spectrum remains unchanged. However, many models include other ingredients that can produce small scale damping~\cite{Cyr-Racine2015,Bringmann:2016ilk,Binder2016,Bringmann2006}
. 
A good example of the latter are the DM + Dark Radiation (DR) models considered by the ETHOS collaboration~\cite{Cyr-Racine2015}, who created a framework for structure formation that encompasses several microphysical interaction models via an effective theory. Interestingly, interacting scenarios combining DM+DR interactions with SIDM effects, are able to generate a truncation in the power spectrum while producing shallower inner density profiles~\cite{2016MNRAS.460.1399V}, alleviating the core-cusp and missing satellite problems altogether%
.

So far, we have mentioned both WDM and SIDM as possible solutions to the tensions between $\Lambda$CDM and observations on small scales, and discussed about their possible realizations
. 
Here, we take both approaches into consideration%
. 
Previous studies have shown that the inclusion of self-interactions among WDM particles in quasi-relaxed DM halos can alleviate some constraints, as shown in~\cite{Arguelles:2015baa,Yunis:2018eux} for the case of self-interacting right handed neutrinos%
.
Also in~\cite{Yunis:2018eux} it is discussed the possibility of novel sterile neutrino production mechanisms through heavy mediators, while further effects of including a scalar self-interaction in the $\nu$MSM active-sterile mixing production scenarios, were considered in~\cite{Johns:2019cwc}%
.


We focus here on the description and treatment of the linear theory of cosmological perturbations for self-interacting WDM (SI-WDM) scenarios, and provide explicit expressions for the Boltzmann hierarchies for different self-interacting sterile neutrino DM scenarios, with its corresponding beyond SM field mediators
. 
Efforts on calculating the evolution of these perturbations either in traditional CDM or WDM scenarios (see~\cite{MaBertschinger95} for a summary), have been outlined either via semi-analytic methods such as in~\cite{2012PhRvD..85d3516D,2012PhRvD..85d3517D}, or via numerical integration of the coupled Einstein-Boltzmann system~\cite{MaBertschinger95,Seljak:1996is,Lewis:1999bs,lesgourgues2011cosmic}
. 
For the latter, freely available numerical routines such as CAMB~\cite{Lewis:1999bs} or CLASS~\cite{lesgourgues2011cosmic} exist as general purpose tools, or more specialized ones as the (CDM-based) ETHOS code~\cite{Cyr-Racine2015} for interacting DM+DR models
. 
An earlier work~\cite{Hannestad00} pioneered the inclusion of SI-WDM on numerical Einstein-Boltzmann solvers (though under important simplifications, see also \cite{2001PhRvD..64h8301A}), finding an enhanced suppression of power in small scales when compared to WDM only evolution
. 


The objective of this work is to contribute to the findings of these early realizations of SI-WDM structure formation. To this aim we provide here a systematic and accurate treatment of collisions in WDM models extending \cite{Hannestad00}, and at the same time retaining certain independence from a particular Lagrangian self-interacting model
. 
Our procedure is motivated by the tools provided by the Boltzmann hierarchies for interactive (active) neutrinos~\cite{Oldengott2014,Oldengott2017,Kreisch:2019yzn}. They are used and generalized to perform an accurate framework for the collision term in the linearized Boltzmann equation for the SI-WDM species, and derive an explicit and analytical expression for the equations of motion
. 
Motivated by~\cite{Cyr-Racine2015}, we do not commit to a particular form of the scattering amplitude, but provide a general parametrization in terms of model dependent coefficients that naturally includes several interaction mediators such as a massive scalar (as seen in~\cite{Oldengott2017,Kreisch:2019yzn,Johns:2019cwc}) or a vector field (as proposed from first principles in~\cite{Arguelles:2015baa,Yunis:2018eux})%
. 
The general results here presented are aimed (but not limited) to further evaluate the SI-WDM effects in the matter power spectrum, CMB anisotropies, halo models and production mechanisms, and may also be useful beyond the study of DM such as the study of active neutrino physics and their anomalies~\cite{Kreisch:2019yzn}
. 


In order to set our notation and conventions, in what remains of this section we briefly introduce the cosmological perturbation theory and Einstein-Boltzmann equations.
\subsection{Cosmological Perturbation Theory}
\label{sec:Intro_CPT}

In cosmology, the evolution of perturbations to the isotropic homogeneous background, which are originated through a primordial power spectrum and will eventually collapse to form the myriad of observed structures today, is handled through the Einstein equations. There, the universal spacetime metric is split into a background Friedmann-Robertson-Walker (FRW) metric and a small perturbation to said metric. The Einstein equations govern the evolution of this perturbation with the perturbed energy-momentum tensors acting as sources. Several choices exist in order to describe these metric perturbations: a ``gauge freedom'' in the equations. Here, we will use the so called synchronous gauge, where the line element is defined as 

\begin{equation}
ds^2 = a^2(\tau) \left\{ -d\tau^2 + (\delta_{ij} + h_{ij}) dx^i dx^j \right\} \ ,
\label{eq:Intro_CPT_ds2_Syn}
\end{equation}

\noindent where the scalar mode of the perturbation $h_{ij}$ can be described in terms of two fields $h(\vec{k}, \tau)$ and $\eta(\vec{k}, \tau)$ as 

\begin{equation}
h_{ij}(\vec{x},\tau) = \int d^3 k e^{i \vec{k}.\vec{x}} \left\{ \hat{k}_i \hat{k}_j h(\vec{k}, \tau ) + \left( \hat{k}_i \hat{k}_j - \frac{1}{3}\delta_{ij} \right) 6 \eta(\vec{k}, \tau ) \right\} \ , \quad \vec{k}=k \hat{k} \ .
\label{eq:Intro_CPT_ds2_Syn_etahdef}
\end{equation}

\noindent A discussion on gauge freedom and gauge modes in the context of perturbations to the FRW metric can be found in \cite{MaBertschinger95,NotasCosmologia,Dodelson_book}. Here, we quote the final form of the Einstein equations in the synchronous gauge, in Fourier space:

\begin{equation}
\begin{split}
k^2 \eta - \frac{1}{2} \frac{\dot{a}}{a} \dot{h} &= 4 \pi G a^2 \delta T^0_{\hphantom{0} 0} \ , \\
k^2 \dot{\eta} &= 4 \pi G a^2 (\overline{\rho} + \overline{P} )  \theta \ ,\\
\ddot{h} + 2 \frac{\dot{a}}{a} \dot{h} - 2 k^2 \eta &= - 8 \pi G a^2 \delta T^i_{\hphantom{i} i} \ , \\
\ddot{h} + 6 \ddot{\eta} + 2 \frac{\dot{a}}{a} \left( \dot{h} + 6\dot{\eta} \right) - 2k^2 \eta &= -24 \pi G a^2 (\overline{\rho} + \overline{P} ) \sigma \ ,
\end{split}
\label{eq:Intro_CPT_Einstein_Syn}
\end{equation}

\noindent where

\begin{equation}
(\overline{\rho} + \overline{P} ) \theta \equiv i k^j \delta T^0_{\hphantom{0} j} \quad , \quad (\overline{\rho} + \overline{P} )\sigma \equiv - \left( \hat{k_i} \hat{k_j} - \frac{1}{3} \delta_{ij} \right) \Sigma^i_{\hphantom{i} j} \ ,
\label{eq:Intro_CPT_Einstein_Syn_Defs}
\end{equation}

\noindent with $\Sigma$ the traceless component of $T^i_j$, $\overline{\rho}$ and $\overline{P}$ the background density and pressure respectively; and the metric perturbation functions $h$, $\eta$ in Synchronous gauge are defined as in \cref{eq:Intro_CPT_ds2_Syn_etahdef}.

\subsection{The Relativistic Boltzmann Equation}
\label{sec:Intro_Boltz}

In order to close the system of equations in \eqref{eq:Intro_CPT_Einstein_Syn} without the assumption of a perfect fluid, the perturbations in a given energy component can be obtained in a more general way by making use of the Boltzmann equation, which governs the evolution of the phase space distribution function (DF). As a relativistic invariant, this function is used to describe the number of particles of a given fluid in a differential unit of volume:

\begin{equation}
d N = f( x^i , P_j , \tau ) d x^1 d x^2 d x^3 dP_1 dP_2 dP_3 \ ,
\label{eq:Boltz_dN}
\end{equation}

\noindent where $x^i$ are the spatial coordinates and $P^i$ refers to the spatial components of the conjugate momentum, defined as $P \equiv (E/a, a[\delta_{ij} + h_{ij}/2] p^j ) $ in terms of the 4-momentum $p^j$ measured by an observer comoving with the FLRW coordinates. In practice, it is  convenient to describe the perturbations to this function as a function of comoving proper momentum $q_i \equiv a(\tau) p_i $ (with $p_i$ measured in a comoving frame) as:

\begin{equation}
f( x^i , P_j , \tau ) = f_0 (q, \tau) + F ( x^i , q , n_j , \tau ) \ ,
\label{eq:Boltz_PsiDef}
\end{equation}

\noindent where $\vec{q}=q \hat{n}$ is the comoving momentum, and $n_j$ its $j$ direction component and $f_0$ is the background DF. The phase space density evolves according to the relativistic Boltzmann equation. In terms of these new variables, this is:


\begin{equation}
P^\alpha \frac{\partial f}{\partial x^\alpha} - \Gamma^\gamma_{\alpha \beta}P^\alpha P^\beta \frac{\partial f}{\partial P^\gamma} = \left( \frac{\partial f}{\partial \tau}\right)_{col} ,
\label{eq:Boltz_Boltz}
\end{equation}

\noindent where $\Gamma^\gamma_{\alpha \beta}$ is the general relativistic metric connection and $h_{ij}$ the metric perturbation in the synchronous gauge (see \cite{MaBertschinger95} for details) . The right hand side of the equation involves the terms due to collisions (referred here as the \textit{collision term}), whose form depends on the type of particle interactions involved. In the case of a general-relativistic formulation of perturbations, the derivatives with respect to the coordinates $df/dx^i$ and $df/dq$ depend explicitly on the way one chooses to express the perturbed metric: the so called ``gauge choice''. We refer the reader to \cite{NotasCosmologia} for a comprehensive explanation on perturbed FRW metrics and the different gauge choices, and \cite{MaBertschinger95} for a ``canonical'' application to most of the cosmological fluids in more than one gauge. In k-space, the equation that dictates the evolution of the perturbation to the phase space distribution $F$ can be obtained from \eqref{eq:Boltz_Boltz} and \eqref{eq:Boltz_PsiDef}, to first order in $F$ as:

\begin{equation}
\frac{\partial F}{\partial \tau} + i \frac{q k}{\epsilon} ( \hat{k} . \hat{n} ) F + \frac{d \  f_0}{d \ln q} \left[ \dot{\eta} - \frac{\dot{h} + 6 \dot{\eta}}{2} ( \hat{k} . \hat{n} )^2 \right] = \left( \frac{\partial f}{\partial \tau}\right)_{col}^{(1)} \ ,
\label{eq:Boltz_BoltzCon}
\end{equation} 

\noindent with $\epsilon = a E$ the comoving energy and $h$, $\eta$ the potential functions describing the scalar mode of $h_{ij}$ defined as in \cref{eq:Intro_CPT_ds2_Syn_etahdef}. This equation is to be solved together with the zero order Boltzmann equation \cite{Oldengott2014} 

\begin{equation}
\frac{\partial f_0}{\partial \tau} = \left( \frac{\partial f}{\partial \tau}\right)_{col}^{(0)} \ ,
\label{eq:Boltz_BoltzZero}
\end{equation} 

\noindent and the equations for the other relevant species together with Einstein equations, to give a closed system. The equations for the metric perturbations are obtained from the  Einstein equations with the perturbations of the total energy-momentum tensor (built as the sum of the contributions for all relevant species acting as a source term. 

\section{The Boltzmann Equation for SI-WDM: Interaction Terms}
\label{sec:Boltz}

Here, we focus on the right-hand-side (RHS) of equation \eqref{eq:Boltz_BoltzCon}. This term describes the interaction between the different particle species, and the eventual self-interactions between the same species. Some species can be considered as collisionless during most of their lifetime such as CDM \cite{MaBertschinger95}: for some approaches on interactions and collision terms see e.g. \cite{Ali-Haimoud2018,Bertschinger2006,Binder2016,Cyr-Racine2015}. For most other species (such as photons or baryons) the collision term plays a major role in their evolution.

There has been recent progress in dealing with the collision term in cosmological simulations from first principles. See for example \cite{Cyr-Racine2015,Binder2016} for a streamlining on the treatment of the term in CDM models, and \cite{Ali-Haimoud2018,Bertschinger2006} for an approximation of the collision term in terms of a Fokker-Planck operator. The focus of the following section is to extend the works of Oldengott et. al. \cite{Oldengott2014,Oldengott2017}, where the collision term has been uniquely calculated for ultrarelativistic species (active neutrinos) and scalar field-mediators. Our extension is to the case of SI-WDM, including a more general scattering amplitude for species that are neither ultrarelativistic nor fully non-relativistic at decoupling, with emphasis in self-interacting sterile neutrinos.

The RHS of equation \eqref{eq:Boltz_BoltzCon} counts the number of collisions a particle species $i$ undergoes in a time interval $dt$ per unit phase space. For a CP invariant two body scattering process $i + j \leftrightarrow m + n$, the full expression for the collision term is:

\begin{align}
\begin{split}
\left( \frac{\partial f_i}{\partial t} \right)_{coll}&(\vec{k},\vec{q},\tau) = \frac{g_j g_m g_n}{2E_q}\int \frac{d^3 l}{(2\pi)^3 2 E_l} \frac{d^3 q'}{(2\pi)^3 2 E_{q'}}  \frac{d^3 l'}{(2\pi)^3 2 E_{l'}} \delta_{D}^{(4)} (\mathbf{q}+\mathbf{l}-\mathbf{q'}-\mathbf{l'})  \\
&\times (2\pi)^4 |\mathcal{M}|^2_{ij \leftrightarrow mn} \big\{ f_m(\vec{k},\vec{q'},\tau) f_n(\vec{k},\vec{l'},\tau)[1 \pm f_i(\vec{k},\vec{q},\tau)] [1 \pm f_j(\vec{k},\vec{l},\tau)] \\
& \phantom{(2\pi)^4 |\mathcal{M}|^2_{ij \leftrightarrow mn} \big(}- f_i(\vec{k},\vec{q},\tau) f_j(\vec{k},\vec{l},\tau)[1 \pm f_m(\vec{k},\vec{q'},\tau)] [1 \pm f_n(\vec{k},\vec{l'},\tau)] \big\} \ ,
\end{split}
\label{eq:FO_Collsion}
\end{align}

\noindent where $g$ is the number of internal degrees of freedom of each species, $|\mathcal{M}|^2$ is the squared Feynmann amplitude for the process, and $\delta_D^{(4)}$ is the Dirac delta functional over the energy-momentum 4-vectors labeled with boldface. The collision term as measured in the time-interval $dt$ is related to the expression in \eqref{eq:Boltz_BoltzCon} as $(\partial f_i / \partial \tau)_{col} = a (\partial f_i / \partial t)_{col}$ \cite{Oldengott2014} 

The zero-order integral, which dictates the evolution of the background phase space distribution $f_0$ is simplified, under the same assumptions, as:

\begin{equation}
\begin{split}
\left( \frac{\partial f_i}{\partial t} \right)_{ii \leftrightarrow ii}^{(0)}(|\vec{q}|,\tau) = \frac{g_i^3}{2E_q (2\pi)^5} &\int \frac{d^3 l}{2 E_{l}} \frac{d^3 q'}{2 E_{q'}} \frac{d^3 l'}{2 E_{l'}} \delta_{D} (q+l-q'-l') |\mathcal{M}|^2  \\
& \times \big[ f_0(l', \tau) f_0(q', \tau) - f_0(l, \tau) f_0(q, \tau) \big]  \\
& \equiv \mathcal{D}_1[f] + \mathcal{D}_2[f] \ .
\end{split}
\label{eq:ZO_DDef}
\end{equation}

The first order collision integral, which involves the first order perturbation $F(\vec{k}, \vec{q}, \tau) $ can be simplified in the case of interactions $ii \leftrightarrow ii$ to:

\begin{equation}
\begin{split}
\left( \frac{\partial f_i}{\partial t} \right)_{ii \leftrightarrow ii}^{(1)}&(\vec{k},\vec{q},\tau) = \frac{g_i^3}{2E_q (2\pi)^5} \int \frac{d^3 l}{2 E_{l}} \frac{d^3 q'}{2 E_{q'}} \frac{d^3 l'}{2 E_{l'}} \delta_{D}^{(4)} (\mathbf{q}+\mathbf{l}-\mathbf{q'}-\mathbf{l'}) |\mathcal{M}|^2  \\
& \times \big[ 2 f_0(q', \tau) F(\vec{k}, \vec{l'}, \tau) - f_0(q, \tau) F(\vec{k}, \vec{l}, \tau) - f_0(l, \tau) F(\vec{k}, \vec{q}, \tau) \big]  \\
& \equiv \mathcal{C}_1 [f] + \mathcal{C}_2 [f] + \mathcal{C}_3 [f] \ ,
\end{split}
\label{eq:FO_Collision_SI}
\end{equation}

\noindent where we have made use of the symmetry of $|\mathcal{M}|^2$ under the exchange $q' \leftrightarrow l'$, and under the assumption that Bose enhancement and Pauli blocking are negligible as is customary done for DM candidates on such early epochs \cite{1990S&T....80S.381K}. In the case of a cosmological component that only interacts with itself, this would provide a source term in the RHS of equation \eqref{eq:Boltz_BoltzCon}, the equation of motion for the phase space perturbation $F$.

Here, we maintain a general form for $f_0$ and provide the necessary collision term to obtain its evolution via the zero-order Boltzmann equation \eqref{eq:Boltz_BoltzZero}. Concerning applications of the results, a few comments are in order. 
In most interacting DM studies it is common that either an equilibrium form (ultra relativistic, maxwellian or Juttner, see for example \cite{GondoloGelmini91,Cannoni_2014}) or a "frozen-out" form for $f_0$ can be assumed for most of the dynamical evolution of perturbations \cite{Oldengott2014,Hannestad00}. Alternatively it turns out to be enough to compute the evolution of a pseudo-temperature of the DM component as described for instance in \cite{Cyr-Racine2015,Bringmann:2016ilk}. An equilibrium distribution would implicitly assume either a thermal decoupling history of DM or a period of strong coupling in self-interactions
\footnote{For beyond SM neutrinos (assuming relativistic decoupling of Self-Interactions), a typical example is to set $f_0 \propto \exp (-q/T_{\mathrm{dec},0})$, where $q$ is the comoving momentum and $T_{\mathrm{dec},0}$ is the SI decoupling temperature today. In \cite{Oldengott2014} an extra normalization factor is included to provide a correction accounting for the effects of Fermi statistics in the number density.}

In \cite{Oldengott2014} both the first \eqref{eq:FO_Collsion} and zero order \eqref{eq:Boltz_BoltzZero} collision integrals have been considered for the case of active neutrinos with a scalar interaction. In that case, a specific interaction model has been evaluated and the particle mass of the neutrinos has been neglected, given that they remain ultrarelativistic until late times. Here, we maintain certain level of generality in the choice of interaction amplitudes, and explicitly include the mass of the particle. This generalization of the collision term can be useful in certain WDM models that include self-interactions between dark particles. In particular, for those models where the ultra relativistic to non relativistic transition takes place in the radiation dominated era, and neither the massless or very massive DM particle limits properly account for the WDM features \cite{CLASSIV}. These topics are more thoroughly discussed in the following sections.

Besides the above mentioned assumptions for the collision terms in \eqref{eq:ZO_DDef} and \eqref{eq:FO_Collision_SI}, we focus here on the case where the only relevant source of interaction  is the self-interaction among the DM particles themselves (i.e. DM-DM collisions). However, if other interactions are relevant our results can be generalized by adding the corresponding collision term to the RHS of \eqref{eq:Boltz_BoltzCon}. Moreover, the evolution of the mediator fields should in principle be studied self-consistently.  Nevertheless, in certain situations one can neglect the backreaction of those fields. For instance, in the case of very massive mediator particles this assumption is justified as the population of the mediators should be Boltzmann-suppressed at the times of interest. This is generically not true, however, in the case of a massless mediator: the contribution of the mediator population to the energy-momentum tensor may not be negligible and the interactions between these two components should be properly accounted for.
Here we do not address the dynamics of the mediator fields and restrict our analysis to the effects of the self-interactions of WDM. We focus below on the massive mediator cases, and relegate to \cref{sec:Appx_Massless} the computation for massless mediators.

\subsection{Scattering Amplitude}

Further assumptions enter the expression we will use for the spin-averaged scattering amplitude $|\mathcal{M}|$. We will assume that this amplitude can be expressed as a second degree polynomial in the Maldestam variables $s,t$ as defined in \eqref{eq:Appx_FO_C3_s2_MandDef}:

\begin{equation}
|\mathcal{M}|^2 \equiv m_{(2,0)}s^2 + m_{(1,1)} s t + m_{(0,2)}t^2 + m_{(1,0)}s + m_{(0,1)} t + m_{(0,0)}.
\label{eq:FO_C3_s3_M2Def}
\end{equation}

\noindent This parametrization leaves the $m_{(i,j)}$ coefficients free as model dependent constants and allows us to recover a few relevant cases for our study, such as the ones to be considered in \ref{sec:Solutions_Models}. This assumption, together with the approaches taken in describing both the collision terms and the Boltzmann equations, allow us to complement previous works \cite{Cyr-Racine2015} aiming to describe self-interacting species in cosmology. It is in this way that we maintain some model independence, being able to describe a wide array of (elastic) interaction cross sections either in an exact or approximate way. 

This parametrization encompasses  most tree level interactions due to massive mediators with $m_{\mathrm{med}} \gg m$, where $m$ is the DM mass and $m_{\mathrm{med}}$ is the one of the mediator. Notably, this includes both of the examples studied in \cite{Oldengott2014} as well as many more.
Particularly, in the limit $m \rightarrow 0$, this parametrises the tree level self-interactions due to a massless scalar mediator, which turns out to be a constant scattering amplitude (as detailed in \cref{sec:Solutions_Models_s0}). 
However, in a general case with $m \neq 0$ this parametrization does \emph{not} account for massless mediators (of interest for self interacting CDM models, see \cite{Tulin2017}). This is discussed further in \cref{sec:Appx_Massless}, where we consider the DM-DM collision term for a massless scalar mediator which cannot be modelled as \eqref{eq:FO_C3_s3_M2Def}.

In order to explicitly perform the collision term integrations, it is necessary to recast this expression into their respective powers of $t$, which reads (with $B_t$ and $C_t$ trivial functions of $s$),

\begin{equation}
|\mathcal{M}|^2 = A_t t^2 + B_t t + C_t,
\label{eq:FO_C3_s3_M2Coeff}
\end{equation}

\noindent as relevant in the case of the $\mathcal{C}_3$, $\mathcal{C}_2$ integrals as demonstrated in appendix \ref{sec:Appx_FO}. A similar expression works in the case of the $\mathcal{C}_1$ integrals, this time involving $u$ (having used the relation $s^2+t^2+u^2=4 m^2$, with $A_u$, $B_u$ and $C_u$ simple functions of t as shown in appendix \ref{sec:Appx_FO}):

\begin{equation}
|\mathcal{M}|^2 = A_u u^2 +B_u u +C_u.
\label{eq:FO_C1_s3_M2Coeff}
\end{equation}

\section{Solutions to the Collision Terms}
\label{sec:Solutions}

\subsection{The First Order Collision Integral}
\label{sec:Solutions_FO}

In this section we write down the final results of the first order collision term in \eqref{eq:FO_Collision_SI}, and refer to the reader to appendix \ref{sec:Appx_FO} for the detailed derivation. In terms of integrations in energies and Mandelstam variables, $\mathcal{C}_3$ can be expressed as:

\begin{equation}
\mathcal{C}_3 = - \frac{F(\vec{q},\vec{k},\tau) g_i^3}{32 (2\pi)^3 E_q q} \int dE_l ds f_0 (E_l,\tau) \chi (s)\,,
\label{eq:FO_C3_s3_intFinal}
\end{equation}

\noindent with $\chi(s)$ defined as

\begin{equation}
\chi(s) = \sqrt{1-\frac{4m^2}{s}} \frac{1}{3} A_t (s-4m^2)^2 + \frac{1}{2} B_t (s-4m^2) +C_t \, ,
\label{eq:FO_C3_s3_ChiDef}
\end{equation}

\noindent and $\{A_t, B_t, C_t\}$ given in \cref{eq:FO_C3_s3_M2Coeff}. Here and in what follows we use the convention that all integrals run over the full range of the respective variables unless it is explicitly specified.

The calculations for the $\mathcal{C}_2$ term are identical to the ones developed in \ref{sec:Appx_FO_C3} for $\mathcal{C}_3$. The only difference is that the roles of the background and perturbed DF are reversed. This can easily be seen from the definition of the term in \eqref{eq:FO_Collision_SI}. So, the final expression for the integral is

\begin{equation}
\mathcal{C}_2 = - \frac{f_0 (E_q,\tau) g_i^3}{32 (2\pi)^3 E_q q} \int dE_l ds F(\vec{l},\vec{k},\tau) \chi (s)\,,
\label{eq:FO_C2_intDef}
\end{equation}

\noindent where we have implicitly used that $\vec{l}$ is a function of only $(E_l,s)$. Given $\vec{q}$, this is straightforward to check from the definitions of $s$, $E_l$. In the case of $\mathcal{C}_1$, the calculation diverges greatly from the one of $\mathcal{C}_3$. In this case, both the background DF and the perturbation are integrated over, and to perform the integration it is necessary to know $f_0(E_q, \tau)$. This integral can be expressed in terms of time-dependent collision kernel $K(E_q, E_{q'}, t, \tau)$ as 

\begin{equation}
\mathcal{C}_1 = \frac{g_i^3}{16 (2\pi)^3 E_q q} \int dE_{q'} dt F (E_{q'},t) \, K(E_q, E_{q'}, t, \tau)\, ,
\label{eq:FO_C1_s3_KDef}
\end{equation}

\noindent where the kernel is given by:

{\footnotesize
\begin{align*}
K(E_q, E_{q'}, t, \tau) = \Bigg\{ \frac{A_u}{8 |\vec{q} - \vec{q'}|^5} \Bigg\{ & \left< f_0 \right>_2 \Bigg[ 4 t \left(3 (E_q + E_{q'})^2 t - ((E_q - E_{q'})^2 - t) (-4 m^2 + t)\right)\Bigg] \\
	+ & \left< f_0 \right>_1 \Bigg[ 4 t (4 (E_q - E_{q'})^2 (E_q + 3 E_{q'}) m^2  \\
	  & \hphantom{\left< f_0 \right>_1 \Bigg[} - 4 (E_q (E_q - E_{q'}) (E_q + 2 E_{q'}) + (E_q + 3 E_{q'}) m^2) t + (E_q + 3 E_{q'}) t^2) \Bigg]  \\
	+ & \left< f_0 \right>_0 \Bigg[ (48 (E_q - E_{q'})^4 m^4 - 16 (E_q - E_{q'})^2 m^2 (2 E_q^2 - 3 E_q E_{q'} + 6 m^2) t  \\
	  & \hphantom{\left< f_0 \right>_0 \Bigg[} + 8 (E_q^2 (E_q - E_{q'})^2 + (7 E_q^2 - 12 E_q E_{q'} + 3 E_{q'}^2) m^2 + 6 m^4) t^2  \\ 
	  & \hphantom{\left< f_0 \right>_0 \Bigg[} - 4 (2 E_q^2 - 3 E_q E_{q'} + 6 m^2) t^3 + 3 t^4)  \Bigg] \Bigg\}  \\
	  +\frac{B_u}{2 |\vec{q} - \vec{q'}|^3} \Bigg\{ & \left< f_0 \right>_1 \Big[ t (E_q+E_{q'}) \Big]+ \left< f_0 \right>_0 \Big[ 2 (E_q - E_{q'})^2 m^2 + 2 E_q (-E_q + E_{q'}) t - 4 m^2 t + t^2 \Big] \Bigg\}  \\
	  + \frac{C_u}{|\vec{q}-\vec{q'}|} \hphantom{\Bigg\{ } & \left< f_0 \right>_0 \quad \Bigg\}
\label{eq:FO_C1_s3_intFinal}
\end{align*}
\par}

\noindent put in terms of moments of the background distribution function $f_0$, that take the form

\begin{equation}
\left< f_0 \right> _n (E_q , E_{q'}, t , \tau) = \int_{R_2}^{\infty} d E_{l'} f_0 (E_{l'}, \tau) E_{l'}^{n}\ ,
\label{eq:FO_C1_s3_f0MeanDef}
\end{equation}

\noindent which are functions of $(E_q,E_{q'},t)$ only through $R_{2}$, defined as

\begin{equation*}
R_{1,2} = \frac{1}{2} \left\{ E_q - E_{q'} \pm |\vec{q}-\vec{q'}|\sqrt{1 - \frac{4m^2}{t}} \right\} \ .
\end{equation*}

This kernel is the most complex part of the collision term, mainly due to its explicit dependence  on time through the momenta of the background DF. 
However, once the scattering amplitude $|\mathcal{M}|^2$ is specified, it should be numerically feasible to evaluate the integrals.

\subsection{The Zero Order Collision Integral}
\label{sec:Solutions_ZO}

The treatment of the term $\mathcal{D}_2[f]$ mimics exactly the one for $\mathcal{C}_3[f]$ but with the simplification $F(\vec{q},\vec{k},\tau) \rightarrow f_0(E_q)$. Thus, this term can be expressed as

\begin{equation}
\mathcal{D}_2 = - \frac{f_0(E_q) g_i^3}{32 (2\pi)^3 E_q q} \int dE_l ds f_0 (E_l,\tau) \chi (s) \ .
\label{eq:ZO_D2_intFinal}
\end{equation}

The term $\mathcal{D}_1[f]$ is much more complicated. The key to solve this integral is to define a method to recast an integration in an angular variable by an integral in energy, as described in appendix \ref{sec:Appx_ZO_D1}. As shown in such appendix, this procedure leads to a  collision integral that can be expressed as

\begin{equation}
\mathcal{D}_1 = \frac{g_i^3}{16 E_q q (2\pi)^3} \sum_{i=1}^{4} \int_{\mathcal{I}_i} dE_{q'} dE_{l'} f_0(E_{q'}, \tau) f_0(E_{l'}, \tau) k_i (E_q, E_{q'}, E_{l'}, \tau)\ ,
\label{eq:ZO_D1_intKDef}
\end{equation}

\noindent in terms of 4 integrals of kernel functions: the integration limits $\mathcal{I}_i$ are defined in \eqref{eq:Appx_ZO_D1_limitsTheta_final} schematically as

{\small
\begin{equation}
\int_{m}^{E_q} dE_{q'} \left[ \int_{E_q-E_{q'}+m}^{E_q} dE_{l'} + \int_{E_q}^{\infty} dE_{l'} \right] + \int_{E_q}^{\infty} dE_{q'} \left[ \int_{m}^{E_q} dE_{l'} + \int_{E_q}^{\infty} dE_{l'} \right] \equiv \sum_{i=1}^{4} \int_{\mathcal{I}_i}\ ,
\label{eq:ZO_D1_limitsEnergy_final}
\end{equation}
\par}

\noindent and the kernels $k_i (E_q, E_{q'}, E_{l'}, \tau)$ are given by

{\small
\begin{align}
\begin{split}
k_i (E_q, E_{q'}, E_{l'}, \tau) &= \int dt \ \kappa (E_q, E_{q'}, E_{l'}, t, \tau ) \quad , \quad i=2,4 \\
k_i (E_q, E_{q'}, E_{l'}, \tau) &= \int_{t_1}^{t_2} dt \ \kappa (E_q, E_{q'}, E_{l'}, t, \tau ) \quad , \quad i=1,3
\end{split}
\label{eq:ZO_D1_intKDef_expr}
\end{align}
\par}

\noindent with 

{\footnotesize
\begin{align}
\begin{split}
\kappa (E_q, E_{q'}, E_{l'}, t, \tau ) = \Bigg\{ \frac{A_u}{8 |\vec{q} - \vec{q'}|^5} \Bigg\{ & E_{l'}^2 \Bigg[ 4 t \left(3 (E_q + E_{q'})^2 t - ((E_q - E_{q'})^2 - t) (-4 m^2 + t)\right)\Bigg]  \\
	+ & E_{l'} \Bigg[ 4 t (4 (E_q - E_{q'})^2 (E_q + 3 E_{q'}) m^2  \\
	  & \hphantom{ E_{l'} \Bigg[} - 4 (E_q (E_q - E_{q'}) (E_q + 2 E_{q'}) + (E_q + 3 E_{q'}) m^2) t + (E_q + 3 E_{q'}) t^2) \Bigg]  \\
	+ & \hphantom{E_{l'}} \Bigg[ (48 (E_q - E_{q'})^4 m^4 - 16 (E_q - E_{q'})^2 m^2 (2 E_q^2 - 3 E_q E_{q'} + 6 m^2) t  \\
	  & \hphantom{ E_{l'} \Bigg[ } + 8 (E_q^2 (E_q - E_{q'})^2 + (7 E_q^2 - 12 E_q E_{q'} + 3 E_{q'}^2) m^2 + 6 m^4) t^2  \\ 
	  & \hphantom{ E_{l'} \Bigg[ } - 4 (2 E_q^2 - 3 E_q E_{q'} + 6 m^2) t^3 + 3 t^4)  \Bigg] \Bigg\}  \\
	  + \frac{B_u}{2 |\vec{q} - \vec{q'}|^3} \Bigg\{ & E_{l'} \Big[ t (E_q+E_{q'}) \Big]+ \Big[ 2 (E_q - E_{q'})^2 m^2 + 2 E_q (-E_q + E_{q'}) t - 4 m^2 t + t^2 \Big] \Bigg\} \\
	  + \frac{C_u}{|\vec{q}-\vec{q'}|} \hphantom{\Bigg\{ } & \quad \Bigg\} \ ,
\end{split}
\label{eq:ZO_D1_intFinal_kappadef_big}
\end{align}
\par}

\noindent and $t_{1,2}$ defined as the two solutions to the following equation:

\begin{equation*}
P(t_{1,2})= |\vec{q}-\vec{q'}|\sqrt{1-\frac{4m^2}{t_{1,2}}} = 2E_{l'}-E_q+E_{q'} \ .
\end{equation*}

Thus, we have arrived at a somewhat general expression for the collision integrals 
which depends on the scattering amplitude only through the coefficients $A_{t,u}$, $B_{t,u}$ and $C_{t,u}$ defined in equations \eqref{eq:FO_C3_s3_M2Coeff} and  \eqref{eq:FO_C1_s3_M2Coeff}.
After this, in order to obtain a Boltzmann hierarchy that can be in principle solved numerically one can follow a procedure analogous to the one described in \cite{Oldengott2014}. We perform this procedure in \cref{sec:Hier}, while in next we provide some examples of the kernel functions obtained from different models of the self-interaction.

Collision terms at the level of the zero-order distribution function are common in other applications of DM such as in $\Lambda$CDM (thermal production and decoupling) \cite{MoBoschWhiteBook}. The tools developed in this section, together with an accurate treatment of inelastic collision terms could help to discern the effects of self-interactions in early DM production, though remains as an interesting avenue for future research.





\subsection{Kernel Functions for Different Mediator Models}
\label{sec:Solutions_Models}

In this section, we calculate the different kernel functions involved in the collision integrals for a small subset of self-interaction models. We need to compute the coefficients $[A_t, B_t, C_t]$ in \cref{eq:FO_C3_s3_M2Coeff} for the $\mathcal{C}_3[f]$ integral and $[A_u, B_u, C_u]$ given in \cref{eq:FO_C1_s3_M2Coeff} for the $\mathcal{C}_1[f]$ and $\mathcal{D}_1[f]$ integrals.

Motivated by the possibility that the DM constituents are sterile neutrinos, we consider the following three cases: the first two evaluated by \cite{Oldengott2014} which are interactions mediated by scalar particles; and the case of a heavy vector field proposed in \cite{amrr}.

For the first case, the interaction Lagrangian can be written as (further information about the scattering processes can be found in \cite{Oldengott2014}):

\begin{equation}
\mathcal{L}_{int} = \mathfrak{g} \overline{\nu}_R \nu_R \phi\,,
\label{eq:Models_Scalar_Lagrangian}
\end{equation}

\noindent where $\mathfrak{g}$ is the scalar coupling constant, $\phi$ is the scalar field and $\nu_R$ is the DM field modelled as a right handed neutrino. In the case studied in \cite{Oldengott2014} the massless scalar limit reduces to a constant amplitude, however, this does not happen generally. We refer to \cite{Srednicki2007} for an expression of the scattering amplitude for scalar mediators of arbitrary mass. In this study, as an example, we only consider a \textit{constant amplitude} case, reminding that only in the limit of massless DM it corresponds to a zero mass scalar mediator (see \cref{sec:Appx_Massless}). Our main focus here is a \textit{massive scalar} mediator, meaning that $m_{\mathrm{med}} = m_\phi \gg E$ (with $E$ denoting the mean energy of the colliding DM particles).

The vectorial model of \cite{amrr} also assumes DM is given by right handed neutrinos but with an interaction Lagrangian given by

\begin{equation}
\mathcal{L}_{int} = -g_V V_{\mu} \overline{\nu}_R \gamma^{\mu} \nu_R \ ,
\label{eq:Models_Vector_Lagrangian}
\end{equation}

\noindent with $g_V$ acting as a coupling constant and $V_{\mu}$, the massive vector field. In the cases considered here, and under the assumption $g_V \lesssim 1$ all mediators fall into the \textit{massive} case $m_{\mathrm{med}} = m_V \gg E$ (see \eqref{eq:Models_vm_MCoeffsT}).

The authors of ref. \cite{amrr} have proposed this effective interaction-Lagrangian to derive a self-interacting system of self-gravitating sterile-neutrinos on galaxy scales. When applied to the Milky Way, it was there shown how a $\mathcal{O}(10^1)$ keV-fermionic DM concentration at the center of the DM halo (i.e. forming a degenerate condensate), could work as an alternative to the super massive black hole (SMBH) in SgrA*. At the same time such fermionic halo model provides a plausible (and alternative) explanation to the small scale structure observables.

It is important to note that, while motivated by the study of sterile neutrinos, the framework and the interaction models here presented, remain general and can be used in other applications such as (massive) active neutrino cosmology.

\subsubsection{Constant Amplitude}
\label{sec:Solutions_Models_s0}

We start with a simple toy model: a constant scattering amplitude $|\mathcal{M}|^2$. We adopt the notation used in \cite{Oldengott2014} for the massless scalar mediator. This constant amplitude can be expressed as 

\begin{equation}
|\mathcal{M}_{\nu \nu \leftrightarrow \nu \nu}|^2_{0} = 6 \mathfrak{g}^4 \ ,
\label{eq:Models_s0_Msquare}
\end{equation}

\noindent where $\mathfrak{g}$ is the scalar coupling constant in the ultra-relativistic case. Being constant in the involved momenta, the coefficients of the expansion in Mandelstam variables are, simply:

\begin{equation}
\begin{cases}
A_t = B_t = A_u = B_u = 0 \\
C_u = C_t = C_0 \equiv 6 \mathfrak{g}^4
\end{cases} \ .
\label{eq:Models_s0_MCoeffs}
\end{equation}

For the $\chi (s)$ function appearing in the final form for $\mathcal{C}_3$ and $\mathcal{C}_2$, we obtain

\begin{equation}
\chi (s) = C_0 \sqrt{1 - \frac{4m^2}{s}} \ .
\label{eq:Models_s0_chi}
\end{equation}

For the time dependent kernel function $K(E_q, E_{q'}, t, \tau)$ in $\mathcal{C}_1$, we find 

\begin{equation}
K(E_q, E_{q'}, t, \tau ) = \frac{C_0 \left<f_0\right>_0}{\sqrt{(E_q - E_{q'})^2 -t}} \ ,
\label{eq:Models_s0_K}
\end{equation} 

\noindent with $\left<f_0\right>_0$ as defined in \cref{eq:FO_C1_s3_f0MeanDef} and

\begin{equation}
\begin{cases}
k_i (E_q, E_{q'}, E_{l'}) = 2 C_0 \left( \sqrt{(E_q-E_{q'})^2 - t_1} - \sqrt{(E_q-E_{q'})^2 - t_2}\right) \quad , \quad i=1,3 \\
k_i (E_q, E_{q'}) = 4 C_0 \min ( q , q' ) \quad , \quad i=2,4 \ ,
\end{cases}
\label{eq:Models_s0_ki}
\end{equation}

\noindent with $t_{1,2}$ defined as the roots of the equation $P(t) = 2E_{l'} - E_{q} + E_{q'}$ with $t_2>t_1$, as used in \cref{sec:Solutions_ZO} for the kernel functions to calculate the background DF in $\mathcal{D}_1$. When comparing these expressions with the ones used in \cite{Oldengott2014}, both collision integrals coincide in the limit $m \rightarrow 0$, showing explicitly that our more general expression for the collision term reduces to this known limiting case.

\subsubsection{Massive Scalar Mediator}
\label{sec:Solutions_Models_sm}

We follow here the considerations of \cite{Oldengott2014} for the case of a scalar mediator which is considerable more massive than the mean scattering energy. In this case the population of scalar particles would be Boltzmann suppressed, so there would be no need to track the evolution of their population. Moreover, in this scenario, the neutrinos would be initially in thermal equilibrium (as noted in \cite{Oldengott2014} and references therein). In this case, the interaction amplitude reduces to:

\begin{equation}
|\mathcal{M}_{\nu \nu \leftrightarrow \nu \nu}|^2_{m} = \frac{\mathfrak{g}^4}{2 m_\Phi^4}\left( s^2 + t^2 + u^2 \right) \ ,
\label{eq:Models_sm_Msquare}
\end{equation}

\noindent with $m_\phi$ denoting the scalar mediator mass. Here, by using the identity $s+t+u= 4m^2$ we can either replace $u$ or $s$ in the scattering amplitude to obtain the two sets of scattering coefficients $[A,B,C]$:

\begin{equation}
\begin{cases}
A_t= 2 \frac{\mathfrak{g}^4}{2 m_\Phi^4} \equiv 2 C_m\\
B_t= 2 C_m (s-4m^2) \\
C_t= 2 C_m (s^2 -s4m^2 +8m^4) 
\end{cases} \ ,
\label{eq:Models_sm_MCoeffsT}
\end{equation}

\begin{equation}
\begin{cases}
A_u= 2 C_m\\
B_u= 2 C_m (t-4m^2) \\
C_u= 2 C_m (t^2 -t4m^2 +8m^4) 
\end{cases} \ .
\label{eq:Models_sm_MCoeffsU}
\end{equation}

Making use of these coefficients, the kernel functions $\chi$, $k_i$ and $K$ as defined in \cref{sec:Solutions_ZO} and \ref{sec:Solutions_FO} respectively, read as follows:

\begin{equation}
\chi (s) = \frac{1}{3} C_m \sqrt{1 - \frac{4m^2}{s}} \left(256 m^4 - 128 m^2 s + 19 s^2 \right) \ ,
\label{eq:Models_sm_chi}
\end{equation}

{\footnotesize
\begin{equation}
\begin{split}
K (E_q, E_l, s, \tau) =& \frac{C_m}{4 ((E_l - E_q)^2 - s)^{5/2}} \Bigg\{ \left<f_0\right>_2 4 s \Bigg[ (4 (E_l - E_q)^2 m^2 + 2 (E_l^2 + 4 E_l E_q + E_q^2 - 2 m^2) s + s^2 \Bigg]  \\
+ & \left<f_0\right>_1 4 s(E_q -E_l) \Bigg[-4 (E_l - E_q)^2 m^2 - 2 (E_l^2 + 4 E_l E_q + E_q^2 - 2 m^2) s - s^2\Bigg]  \\
+ & \left<f_0\right>_0 \Bigg[ 48 (E_l - E_q)^4 m^4 - 16 (E_l - E_q)^2 m^2 (E_l^2 - 3 E_l E_q + E_q^2 + 6 m^2) s  \\
& \hphantom{\left<f_0\right>_0 \Bigg[} + 8 ((E_l - E_q)^2 (E_l^2 - E_l E_q + E_q^2) + (5 E_l^2 - 12 E_l E_q + 5 E_q^2) m^2 + 6 m^4) s^2 \\
& \hphantom{\left<f_0\right>_0 \Bigg[} - 4 (3 E_l^2 - 7 E_l E_q + 3 E_q^2 + 6 m^2) s^3 + 7 s^4 \Bigg] \Bigg\} \ ,
\end{split}
\label{eq:Models_sm_K}
\end{equation}
\par}

{\footnotesize
\begin{equation}
\begin{split}
k_i (E_q, E_{q'}, E_{l'}) =& \int_{\mathcal{I}_i} dt \frac{C_m}{4 ((E_{q'} - E_q)^2 - t)^{5/2}} \Bigg\{ E_{l'}^2 4 t \Bigg[ (4 (E_{q'} - E_q)^2 m^2 + 2 (E_{q'}^2 + 4 E_{q'} E_q + E_q^2 - 2 m^2) t + t^2 \Bigg]  \\
+ & E_{l'} 4 t(E_q -E_{q'}) \Bigg[-4 (E_{q'} - E_q)^2 m^2 - 2 (E_{q'}^2 + 4 E_{q'} E_q + E_q^2 - 2 m^2) t - t^2\Bigg]  \\
+ & \Bigg[ 48 (E_{q'} - E_q)^4 m^4 - 16 (E_{q'} - E_q)^2 m^2 (E_{q'}^2 - 3 E_{q'} E_q + E_q^2 + 6 m^2) t  \\
& \hphantom{\Bigg[} + 8 ((E_{q'} - E_q)^2 (E_{q'}^2 - E_{q'} E_q + E_q^2) + (5 E_{q'}^2 - 12 E_{q'} E_q + 5 E_q^2) m^2 + 6 m^4) t^2 \\
& \hphantom{\Bigg[} - 4 (3 E_{q'}^2 - 7 E_{q'} E_q + 3 E_q^2 + 6 m^2) t^3 + 7 t^4 \Bigg] \Bigg\} \ ,
\end{split}
\label{eq:Models_sm_ki}
\end{equation}
\par}

\noindent where the integration regions in the $t$ variable for these last integrals are in the range $[t_{\cos \theta = -1} , t_{\cos \theta =1} ]$ for $i=2,4$; and $[t_1,t_2]$ where $t_{1,2}$ are the solutions of $P(t)=2E_{l'} - E_{q} + E_{q'}$ for $i=1,3$ (see appendix \ref{sec:Appx_ZO_D1} for details). Here the background DF kernels $k_i$ have analytical (though complicated) expressions in $(E_q,E_{q'},E_{l'},t_{1,2})$, which are not very illuminating to write down.

\subsubsection{Massive Vector Field}
\label{sec:Solutions_Models_vm}

In \cite{amrr} the authors calculate the $4$-fermion self-scattering amplitude for the right handed sterile neutrinos with the interaction \eqref{eq:Models_Vector_Lagrangian}, and reach the following result:

\begin{equation}
|\mathcal{M}_{\nu \nu \leftrightarrow \nu \nu}|^2_{V} = \left( \frac{g_V}{m_V} \right)^4 \frac{1}{\cos^4 \theta_W'} \left[ 44 ( \mathbf{q} \mathbf{l} )^2 - ( \mathbf{q} \mathbf{q'} )^2 - ( \mathbf{q} \mathbf{l'} )^2 \right] \ ,
\label{eq:Models_vm_Msquare}
\end{equation}

\noindent where $m_V$ is the mediator mass, $\theta_W'$ the (dark sector) Weinberg angle and $\mathbf{p}\mathbf{q} \equiv p^\mu q_\mu$ on a 4-vector notation. By making use of the following Mandelstam variables properties

\begin{equation}
\begin{cases}
 \mathbf{q} \mathbf{l}  = s - 2 m^2 \\
 \mathbf{q} \mathbf{q'} = 2 m^2 - t \\
 \mathbf{q} \mathbf{l'} = 2 m^2 - u 
\end{cases} \ ,
\label{eq:Models_vm_MandProps}
\end{equation}

\noindent the equation \eqref{eq:Models_vm_Msquare} can be put in terms of $(s,t,u)$. As for massive scalars, the scattering coefficients can be calculated using $s+t+u= 4m^2$:

 \begin{equation}
\begin{cases}
A_t= - 2 \left( \frac{g_V}{m_V} \right)^4 \frac{1}{\cos^4 \theta_W'} \equiv -2 C_V \\
B_t= C_V (8 m^2 - 2 s)  \\
C_t= C_V (168 m^4 - 172 m^2 s + 43 s^2) 
\end{cases} \ ,
\label{eq:Models_vm_MCoeffsT}
\end{equation}

\begin{equation}
\begin{cases}
A_u= 43 C_V \\
B_u= C_V (88 t - 172 m^2) \\
C_u= C_V (168 m^4 - 172 m^2 t + 43 t^2) 
\end{cases} \ ,
\label{eq:Models_vm_MCoeffsU}
\end{equation}

\noindent and the expressions for the kernels are:

\begin{equation}
\chi (s) = \frac{4}{3} C_V \sqrt{1 - \frac{4 m^2}{s}} (74 m^2 - 29 s) (m^2 - s) \ ,
\label{eq:Models_vm_chi}
\end{equation}

{\footnotesize
\begin{equation}
\begin{split}
K & (E_q, E_l, s, \tau) = \frac{C_V}{8 ((E_l - E_q)^2 - s)^{5/2}} \Bigg\{172 \left<f_0\right>_2 s \Bigg[4 (E_l - E_q)^2 m^2 + 2 (E_l^2 + 4 E_l E_q + E_q^2 - 2 m^2) s + s^2 \Bigg]  \\
+& 4 \left<f_0\right>_1 s \Bigg[ 172 (E_l - E_q)^3 m^2 + 4 (-E_l + E_q) (-44 E_l^2 - 86 E_l E_q + E_q^2 + 43 m^2) s - (47 E_l + 133 E_q) s^2 \Bigg]  \\
+& \left<f_0\right>_0 \Bigg[ 656 (E_l - E_q)^4 m^4 + 16 (E_l - E_q)^2 m^2 (2 E_l^2 + 39 E_l E_q + 2 E_q^2 - 82 m^2) s  \\
& \hphantom{\left<f_0\right>_0 \Bigg[} - 8 ((E_l - E_q)^2 (-43 E_l^2 - 2 E_l E_q + 2 E_q^2) + (-35 E_l^2 + 156 E_l E_q - 35 E_q^2) m^2 - 82 m^4) s^2  \\
& \hphantom{\left<f_0\right>_0 \Bigg[} + 4 (-84 E_l^2 + 121 E_l E_q + 6 E_q^2 - 78 m^2) s^3 + 121 s^4 \Bigg]\Bigg\} \ ,
\end{split}
\label{eq:Models_vm_K}
\end{equation}
\par}

{\footnotesize
\begin{equation}
\begin{split}
k_i & (E_q, E_{q'}, E_{l'}) = \int_{\mathcal{I}_i} dt \frac{C_V}{8 ((E_{q'} - E_q)^2 - t)^{5/2}} \Bigg\{\\
\hphantom{+}& 172 E_{l'}^2 t \Bigg[4 (E_{q'} - E_q)^2 m^2 + 2 (E_{q'}^2 + 4 E_{q'} E_q + E_q^2 - 2 m^2) t + t^2 \Bigg]  \\
+& 4 E_{l'} t \Bigg[ 172 (E_{q'} - E_q)^3 m^2 + 4 (-E_{q'} + E_q) (-44 E_{q'}^2 - 86 E_{q'} E_q + E_q^2 + 43 m^2) t - (47 E_{q'} + 133 E_q) t^2 \Bigg]  \\
+& \Bigg[ 656 (E_{q'} - E_q)^4 m^4 + 16 (E_{q'} - E_q)^2 m^2 (2 E_{q'}^2 + 39 E_{q'} E_q + 2 E_q^2 - 82 m^2) t  \\
& \hphantom{\Bigg[} - 8 ((E_{q'} - E_q)^2 (-43 E_{q'}^2 - 2 E_{q'} E_q + 2 E_q^2) + (-35 E_{q'}^2 + 156 E_{q'} E_q - 35 E_q^2) m^2 - 82 m^4) t^2  \\
& \hphantom{\Bigg[} + 4 (-84 E_{q'}^2 + 121 E_{q'} E_q + 6 E_q^2 - 78 m^2) t^3 + 121 t^4 \Bigg]\Bigg\} \ ,
\end{split}
\label{eq:Models_vm_ki}
\end{equation}
\par}

\noindent where, again, the integration regions in the $t$ variable for these last integrals are in the range $[t_{\cos \theta = -1} , t_{\cos \theta =1} ]$ for $i=2,4$ and $[t_1,t_2]$, where $t_{1,2}$ are the solutions of $P(t)=2E_{l'} - E_{q} + E_{q'}$ for $i=1,3$. These last kernel functions for the background DF have analytical forms but are not illuminating, just as in the massive scalar case (see appendix \ref{sec:Appx_FO} for details).

\section{Boltzmann Hierarchy}
\label{sec:Hier}

Once having obtained the expressions for the collision integral kernels, it is a standard practice to perform a Legendre expansion in \eqref{eq:Boltz_BoltzCon} in order to construct a so called Boltzmann hierarchy of equations, which is independent of the angle between $\vec{k}$ and $\vec{q}$. 
In order to calculate the time dependent kernels above, the full solution to the background DF $f_0 (E_q, \tau)$ must be obtained. If we assume that, in the time scales of interest, the collision term is only due to self-scattering, then the evolution of $f_0$ is governed by:

\begin{equation}
\begin{split}
&\frac{\partial f_0}{\partial \tau}(E_q, \tau) = a \left(\mathcal{D}_1[f_0] + \mathcal{D}_2[f_0]\right) =\\
&= G_0 a \Bigg\{ -f_0(E_q) \int dE_l f_0(E_l) \kappa^{(0)} (E_q, E_l) + 2 \sum_{j=1}^{4} \int_{\mathcal{I}_j} dE_{q'} dE_{l'} f_0(E_{q'}) f_0(E_{l'}) \mathcal{K}^{(0)}_j (E_q, E_{q'}, E_{l'}) \Bigg\} \ ,
\end{split}
\label{eq:Hier_f0_motion}
\end{equation}

\noindent where $G_0 = 1/[4(2\pi)^3]$ and

\begin{align}
\kappa^{(0)} (E_q, E_l) =& \frac{1}{E_q q} \int ds \chi(s) \ , \label{eq:kappa0} \\
\mathcal{K}^{(0)}_j (E_q, E_{q'}, E_{l'}) =& \frac{1}{E_q q} k_j (E_q, E_{q'}, E_{l'}) \ .
\label{eq:Hier_f0_kernelDef} 
\end{align}

\noindent Here, we will assume that the initial conditions for $f_0$ are set beforehand at some early time and that its subsequent evolution is only governed by self-interactions as said before. This ansatz, notably, excludes the production mechanism that should give rise to the initial population of these particles: we follow \cite{Hannestad00} and implicitly assume that the mechanism for production is not significanly affected by the self-interaction mechanism. 
Once the evolution of this distribution is known, the various moments $\left< f_0 \right>_i$ defined in \cref{eq:Appx_FO_C1_s3_f0MeanDef} can be calculated, thus allowing to obtain the time dependent kernels $K$.

As the coefficients in the LHS of the Boltzmann \cref{eq:Boltz_BoltzCon} are only functions of $|\vec{q}|$, $|\vec{k}|$ and $\cos \epsilon \equiv \hat{k}.\hat{q}$, it is assumed that the RHS also depends only on these parameters. Thus, in order to express the relevant equations of both sides in a Legendre series we first write

\begin{equation}
\begin{split}
F(|k|, |q|, \cos \epsilon ) &= \sum _{l=0}^{\infty} (-i)^l (2l+1) F_l(|k|, |q|) P_l(\cos \epsilon )\ ,\\
F_l(|k|, |q|) &= \frac{i^l}{2} \int_{-1}^{1} d \cos \epsilon F(|k|, |q|, \cos \epsilon) P_l(\cos \epsilon ) \ ,
\end{split}
\label{eq:Hier_F_LegendreQ} 
\end{equation}

\noindent where $P_l(\cos \epsilon ) $ is the $l$-th Legendre polynomial and $F_l$ is the $l$-th multiple of the perturbed DF. In this case however, a residual dependence on the azimuthal angle between $\vec{q}$ and $\vec{k}$, $\psi$, is still present (as caused by the first order collision terms). In \cite{Oldengott2014} the authors argue that an averaging over $\psi$ in the collision terms has no effect on the expressions for the collision integrals, and they perform such average. The argument is based on the fact that the LHS of \eqref{eq:Boltz_BoltzCon} is not affected by such averaging. Also, from a phenomenological viewpoint, the only observable is the integrated effect of the perturbation, further strengthening the claim. Therefore, in what follows, we perform such average as well. The moment decomposition in the LHS of the Boltzmann equation is well known, and the reader can consult \cite{Bertschinger2006} for the expressions for massive neutrinos in the collisionless case on both typical gauge choices.

In order to calculate the moment expansions of the collision integral, we make use of the following property: given a collision term with the form

\begin{equation}
\left( \frac{\partial f}{\partial \tau}\right)_{k}^{(1)} (\vec{k}, \vec{q}, \tau) = \int d \cos \theta d|q'| \mathcal{K} (|q|, |q'|, \cos \theta, \tau) F(\vec{k}, \vec{q}, \tau) \ ,
\label{eq:Hier_F_Prop1} 
\end{equation}

\noindent the ($\psi$ averaged) $l$-th multipole can be written as:

\begin{equation}
\frac{i^l}{2} \int_0^{2\pi} \frac{d\psi}{2\pi} \int_{-1}^{1} d\cos \epsilon P_l(\cos \epsilon) \left( \frac{\partial f}{\partial \tau}\right)_{k}^{(1)} = \int d|q'| \mathcal{K}_l (|q|, |q'|,  \tau ) F_l (|k|, |q'|, \tau ) \ ,
\label{eq:Hier_F_Prop2} 
\end{equation}

\noindent with

\begin{equation}
\mathcal{K}_l (|q|, |q'|, \tau ) \equiv \int_{-1}^{1} d \cos \theta \mathcal{K} (|q|, |q'|, \cos \theta , \tau ) P_l (\cos \theta) \ .
\label{eq:Hier_F_Prop3} 
\end{equation}  

\noindent However, our expressions for the collision integrals are expressed in terms of Mandelstam variables and energies instead of angles and momenta. Our kernels are also expressed in terms of these variables. We can solve both problems by recasting the integration in Mandelstam variables in the definition of the kernel moments. So, for a collision integral of the form  

\begin{equation}
\left( \frac{\partial f}{\partial \tau}\right)_{k}^{(1)} (\vec{k}, \vec{q}, \tau) = \int ds dE_l \mathcal{K} (E_q, E_l, s, \tau) F(\vec{k}, \vec{l}, \tau)
\label{eq:Hier_F_PropEq1} 
\end{equation}

\noindent the property \eqref{eq:Hier_F_Prop2} would be modified as follows:

\begin{equation}
\frac{i^l}{2} \int_0^{2\pi} \frac{d\psi}{2\pi} \int_{-1}^{1} d\cos \epsilon P_l(\cos \epsilon) \left( \frac{\partial f}{\partial \tau}\right)_{k}^{(1)} = \int dE_l \mathcal{K}_l (E_q, E_l, \tau ) F_l (|k|, |l|, \tau) \ ,
\label{eq:Hier_F_PropEq2} 
\end{equation}

\noindent with

\begin{equation}
\mathcal{K}_l (E_q, E_l, \tau ) \equiv \int ds \mathcal{K} (E_q, E_l, s, \tau ) P_l (\cos \theta (s) ) \ .
\label{eq:Hier_F_PropEq3} 
\end{equation}  

\noindent Then, putting together the results of sections \ref{sec:Appx_FO_C3}, \ref{sec:Appx_FO_C2} and \ref{sec:Appx_FO_C1} we arrive at the following moment expansion, in Synchronous gauge:

\begin{equation}
\mbox{\small $
\begin{split}
\dot{F_0} (k, E_q, \tau) =& - \frac{q k}{E_q} F_1 (k, E_q, \tau ) + \frac{\dot{h}}{6} \frac{\partial f_0}{\partial \ln q} \\
& - G_0 a F_0 (k, E_q, \tau ) \Gamma (E_q, \tau) + G_0 a \int dE_l F_0 (k, E_l, \tau ) \mathcal{K}^{(1)}_0 (E_q, E_l, \tau) \\
\dot{F_1} (k, E_q, \tau) =& \frac{q k}{3 E_q} F_0 (k, E_q, \tau ) - \frac{2 q k}{3 E_q} F_2 (k, E_q, \tau ) \\
& - G_0 a F_1 (k, E_q, \tau ) \Gamma (E_q, \tau) + G_0 a \int dE_l F_1 (k, E_l, \tau ) \mathcal{K}^{(1)}_1 (E_q, E_l, \tau) \\
\dot{F_2} (k, E_q, \tau) =& \frac{q k}{5 E_q} \Big[ 2 F_1 (k, E_q, \tau ) - 3 F_3 (k, E_q, \tau ) \Big] - \frac{\partial f_0}{\partial \ln q} \Bigg[ \frac{1}{15} \dot{h} + \frac{2}{5} \dot{\eta} \Bigg]\\
& - G_0 a F_2 (k, E_q, \tau ) \Gamma (E_q, \tau) + G_0 a \int dE_l F_2 (k, E_l, \tau ) \mathcal{K}^{(1)}_2 (E_q, E_l, \tau) \\
\dot{F_l} (k, E_q, \tau) =& \frac{q k}{(2l+1) E_q} \Big[ l F_{(l-1)} (k, E_q, \tau ) - (l+1) F_{(l+1)} (k, E_q, \tau ) \Big] \\
& - G_0 a F_l (k, E_q, \tau ) \Gamma (E_q, \tau) + G_0 a \int dE_l F_l (k, E_l, \tau ) \mathcal{K}^{(1)}_l (E_q, E_l, \tau) \quad , \quad l \geq 3\\
\end{split}
$}
\label{eq:Hier_F_motion} 
\end{equation}

\noindent with the various kernel moments defined as:

\begin{equation}
\Gamma (E_q, \tau) = \int dE_l f_0 (E_l) \kappa^{(0)} (E_q, E_l ) \ ,
\label{eq:Hier_F_kernelDef_gamma} 
\end{equation}

\begin{equation}
\mathcal{K}^{(1)}_l (E_q, E_l, \tau ) = -\chi_l (E_q, E_l) f_0 (E_q)  + 2 \frac{1}{E_q q} K_l (E_q, E_l, \tau )  \ ,
\label{eq:Hier_F_kernelDef_K} 
\end{equation}

\noindent with $\kappa^{(0)}$ defined as \eqref{eq:kappa0}, and $K_l$, $\chi_l$ the Legendre transforms of the $K$, $\chi$ kernel functions defined as in \eqref{eq:Hier_F_PropEq3}, and where the $l$-th moment of the perturbed DF $F_l$ is defined as in \eqref{eq:Hier_F_LegendreQ}, and we have chosen to express the momentum dependence in terms of energy for consistency. In order to solve this hierarchy, the kernel functions for the interaction model must be specified.

\section{Relaxation Time Approximation}
\label{sec:RTA}

Even if the evolution of the background DF $f_0$ may seem complicated due to the collisions in play, its effect may be accounted for in a much simpler way depending on the particularities of the interaction. Concretely, if the rate of particle interactions is much higher than the rate of expansion of the universe, measured roughly by $H$, the Hubble rate, we may assume that the shape of the distribution function is one in equilibrium. That is to say, a DF that obeys

\begin{equation}
\left( \frac{\partial f_0^{eq} (E_q, t) }{\partial t} \right)_{coll} = 0 \ ,
\label{eq:RTA_Equil_Cond}
\end{equation}  

\noindent such as Maxwell-Boltzmann, Fermi-Dirac or Bose-Einstein distributions, depending on the particle model used. It is possible to construct a substitute collision operator for $f_0$ that reconstructs the expected behavior for small departures from thermal equilibrium:

\begin{equation}
\left( \frac{\partial f_0}{\partial t} \right)_{coll} \approx \frac{f_0(E_q, t) - f_0^{eq} (E_q, t) }{\tau (E_q) } \ .
\label{eq:RTA_RTA_CollTerm}
\end{equation}

\begin{figure}
    \centering
    \begin{subfigure}{0.32\textwidth}
        \centering
        \includegraphics[width=\textwidth]{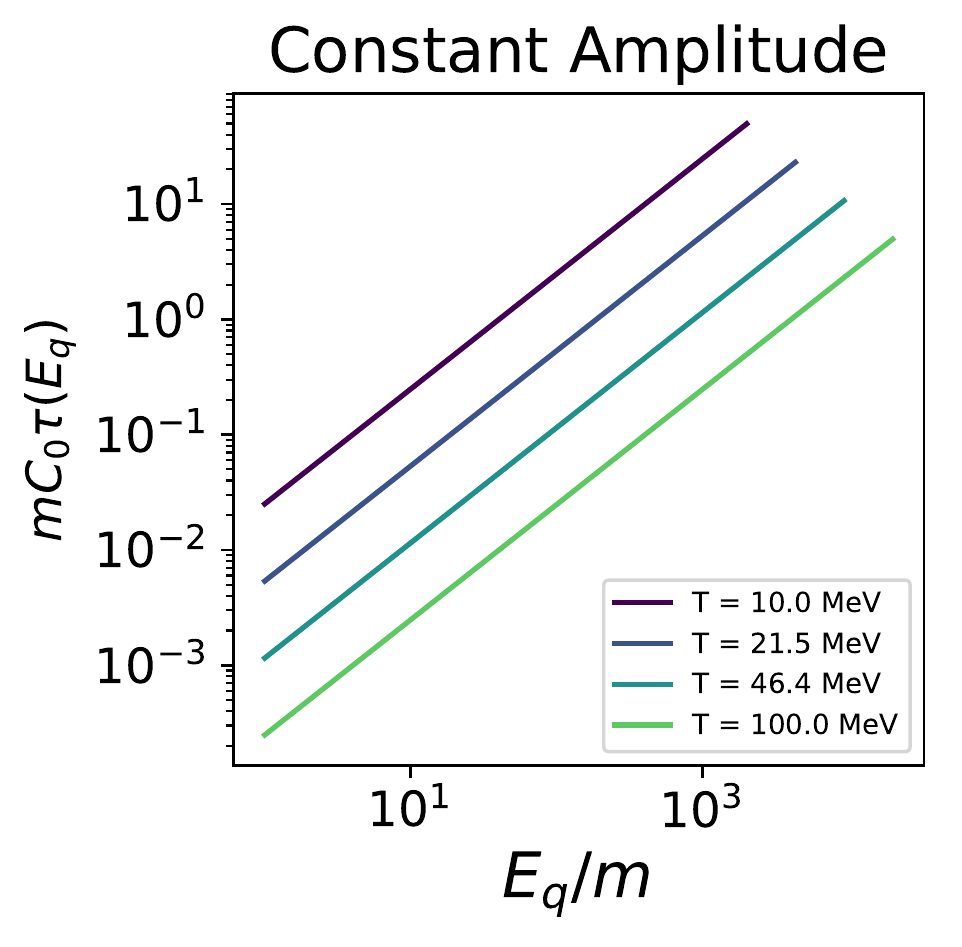}
    \end{subfigure}
    \begin{subfigure}{0.32\textwidth}
        \centering
        \includegraphics[width=\textwidth]{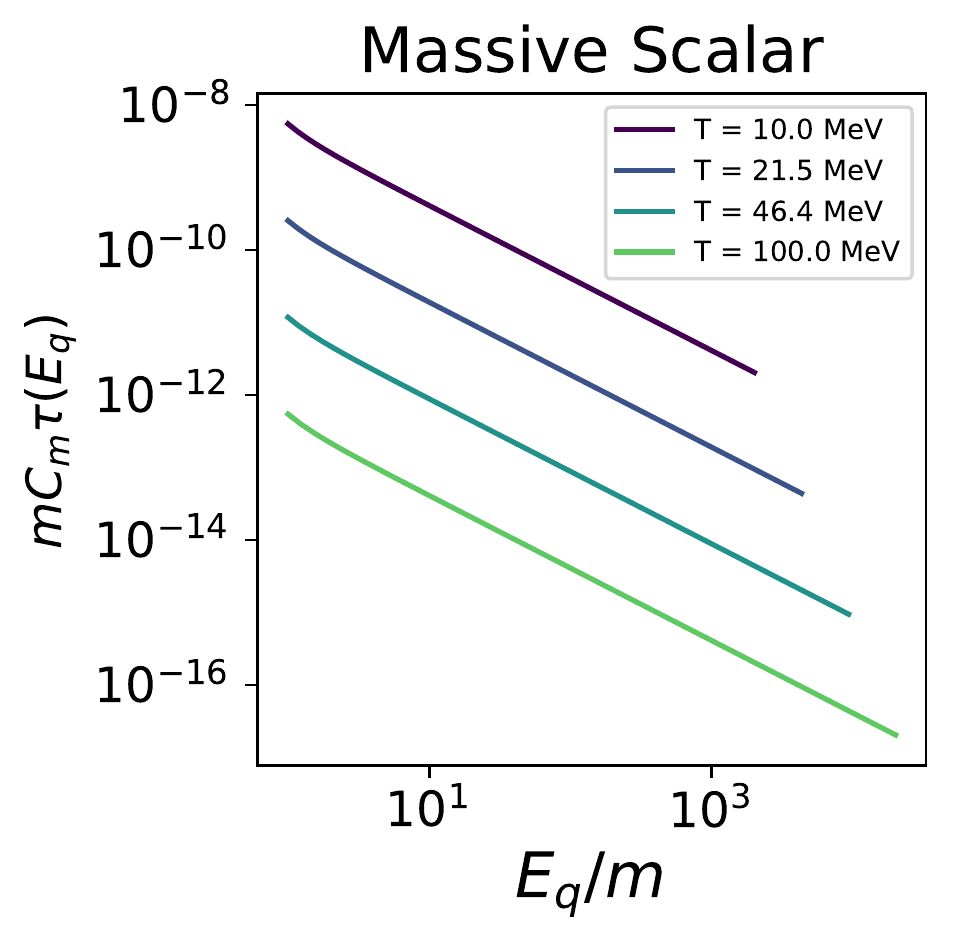}
    \end{subfigure}
    \begin{subfigure}{0.32\textwidth}
        \centering
        \includegraphics[width=\textwidth]{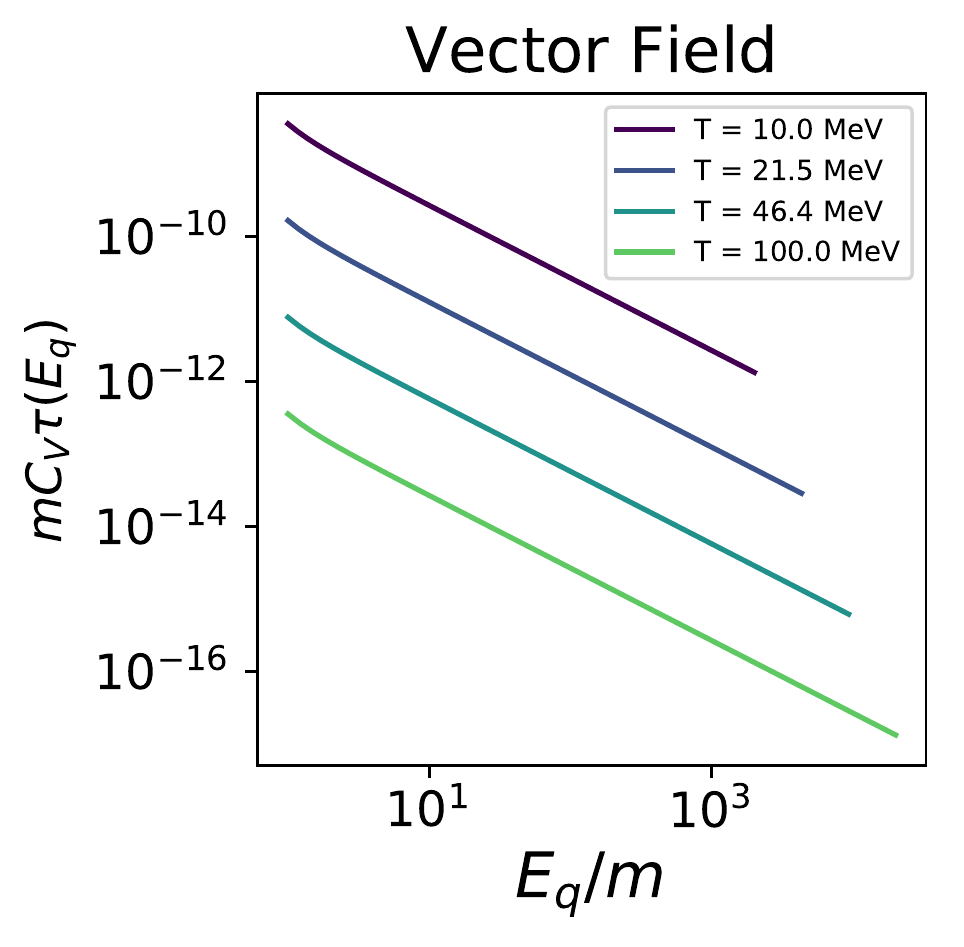}
    \end{subfigure}
\caption{\small Relaxation times for three different interaction models: Constant Amplitude (left), Massive Scalar (center) and Vector Field (right), calculated for a Maxwell-Boltzmann background DF for different temperatures. 
The definitions on the interaction constants for the models considered (with interaction Lagrangians defined in \ref{sec:Solutions_Models}) can be found in \eqref{eq:Models_s0_MCoeffs} for the constant amplitude model, in \eqref{eq:Models_vm_MCoeffsT} for the massive scalar model and in \eqref{eq:Models_vm_MCoeffsT} for the massive vector field.}
\label{fig:RTA_Tau_Eq}
\end{figure}

\noindent This is know as the \textit{relaxation time approximation} of the collision operator. The relaxation time $\tau$ is the timescale in which the system is expected to relax to equilibrium. This parameter can in principle have a $q$ dependence and is commonly defined as \cite{VereshchaginAksenov17}\footnote{In the first equality we approximate the \textit{relaxation time} (the timescale for the system to relax towards equilibrium) by the \textit{collision time} (the mean time between collisions). While in the cases considered in \cref{sec:Solutions_Models} it can be considered as a good approximation, cases where many collisions produce small changes in momenta (such as long range interactions) require additional care (see \cite{Hofmann_2001} for a discussion).}:

\begin{equation}
\tau (E_q) \approx \left< \sigma v \right>^{-1} = - \frac{f_0(E_q, t)}{\mathcal{D}_2[f_0]} \approx - \frac{f_0^{eq}(E_q, t)}{\mathcal{D}_2[f_0^{eq}]} \ ,
\label{eq:RTA_D1_Equiv}
\end{equation}

\noindent which involves the integral of the kernel $\chi$, evaluated in the thermal equilibrium background DF. It is straighforward to evaluate these integrals, as they only involve known functions. We have performed these numerically for the interaction models posed in \ref{sec:Solutions_Models}, and the results can be seen in figure \ref{fig:RTA_Tau_Eq} (we refer to appendix \ref{sec:Appx_RTA} for more details).

For all of these models, we follow \cite{Hannestad00} and assume that the abundance of WDM and its primordial distribution function are already set deep into the radiation dominated epoch and the effects of self-interactions in these initial conditions can be effectively decoupled from the evolution of perturbations.

\subsection{Application to Self-Interaction Decoupling}

In order to evaluate whether or not a thermal $f_0$ can be assumed, for a given particle physics model for the interaction, one may look at the ensemble averages of the interaction rate $\Gamma (E_q) = \tau^{-1} (E_q)$. This value is to be compared to $H$ at this point: if $\Gamma \gg H$, the system is effectively in thermal equilibrium and adopts an equilibrium background distribution function $f_0$. 

This is equivalent to the traditional approach used to determine if a species has decoupled from the rest of the cosmic plasma in the standard sector (see for example \cite{Oldengott2017}). In other words, a given interaction is considered to cease being relevant if the interaction rate per particle $\Gamma \sim \left< \tau^{-1} \right>_{th} $ (see \cref{sec:Appx_RTA}) is overtaken by the Hubble expansion rate, which in the radiation dominated era is $H \sim T^2/m_{pl}$ where $m_{pl}$ is the Planck mass. So, in summary, a thermal background DF can be assumed if at some point during the evolution of the perturbations, the self-interactions were a dominant phenomenon in the sense $\Gamma > H$.

\begin{figure}
    \centering
    \begin{subfigure}{0.32\textwidth}
        \centering
        \includegraphics[width=\textwidth]{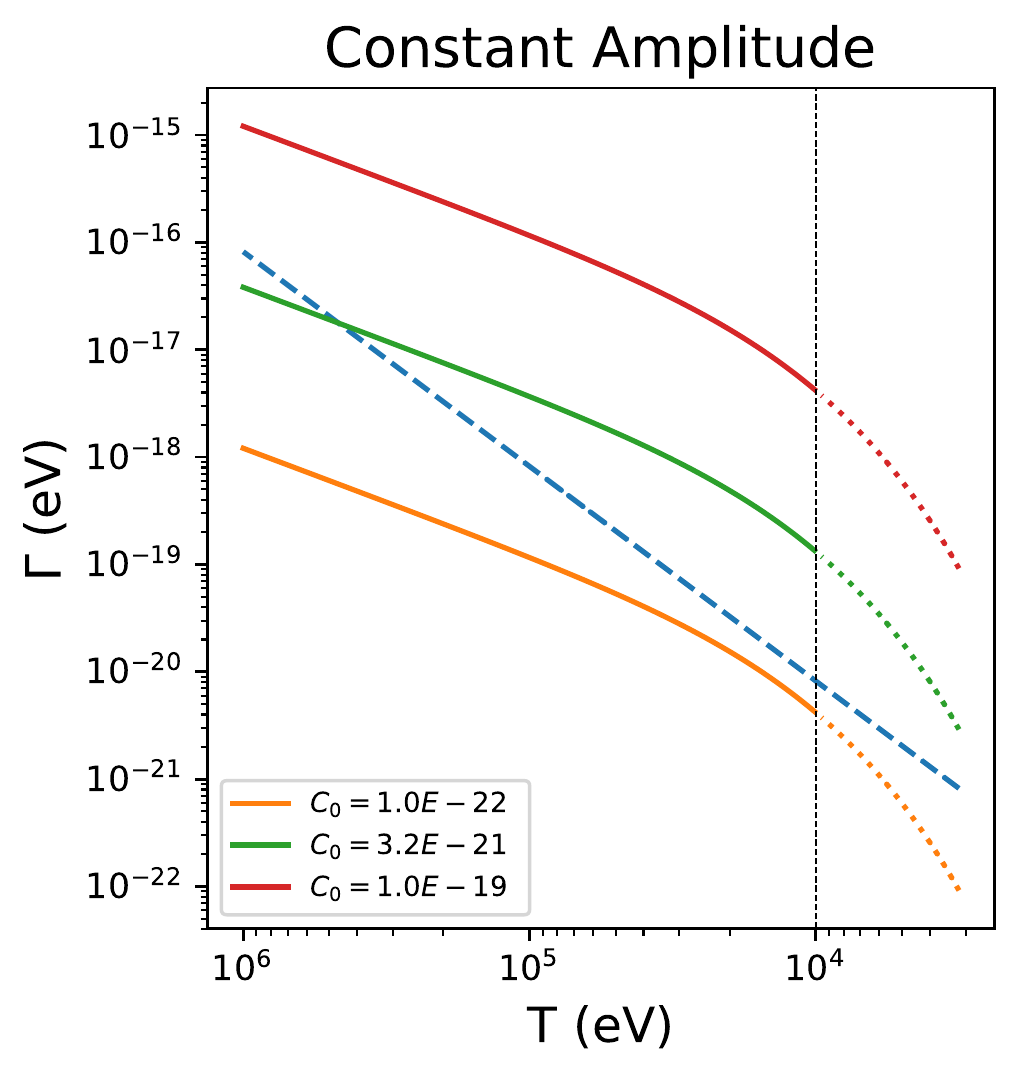}
    \end{subfigure}
    \begin{subfigure}{0.32\textwidth}
        \centering
        \includegraphics[width=\textwidth]{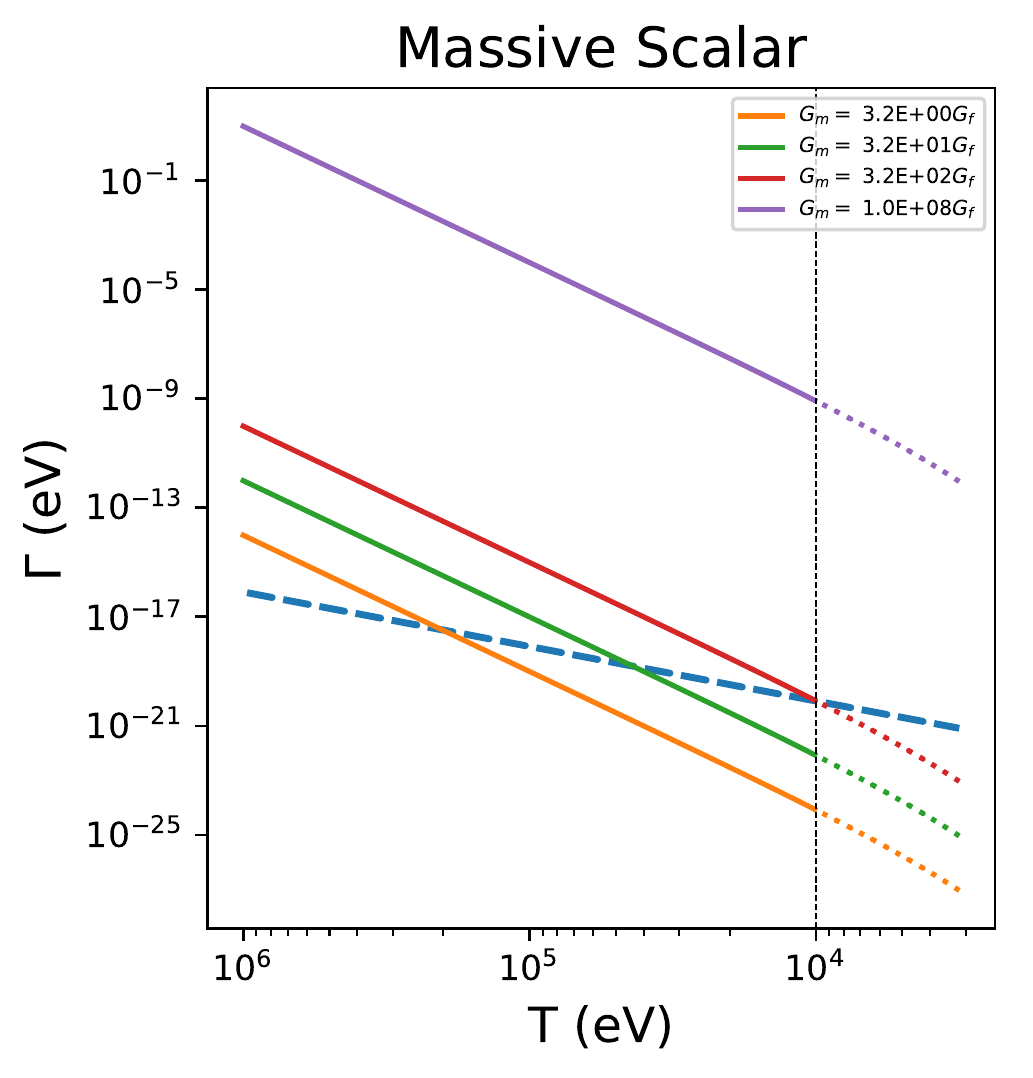}
    \end{subfigure}
    \begin{subfigure}{0.32\textwidth}
        \centering
        \includegraphics[width=\textwidth]{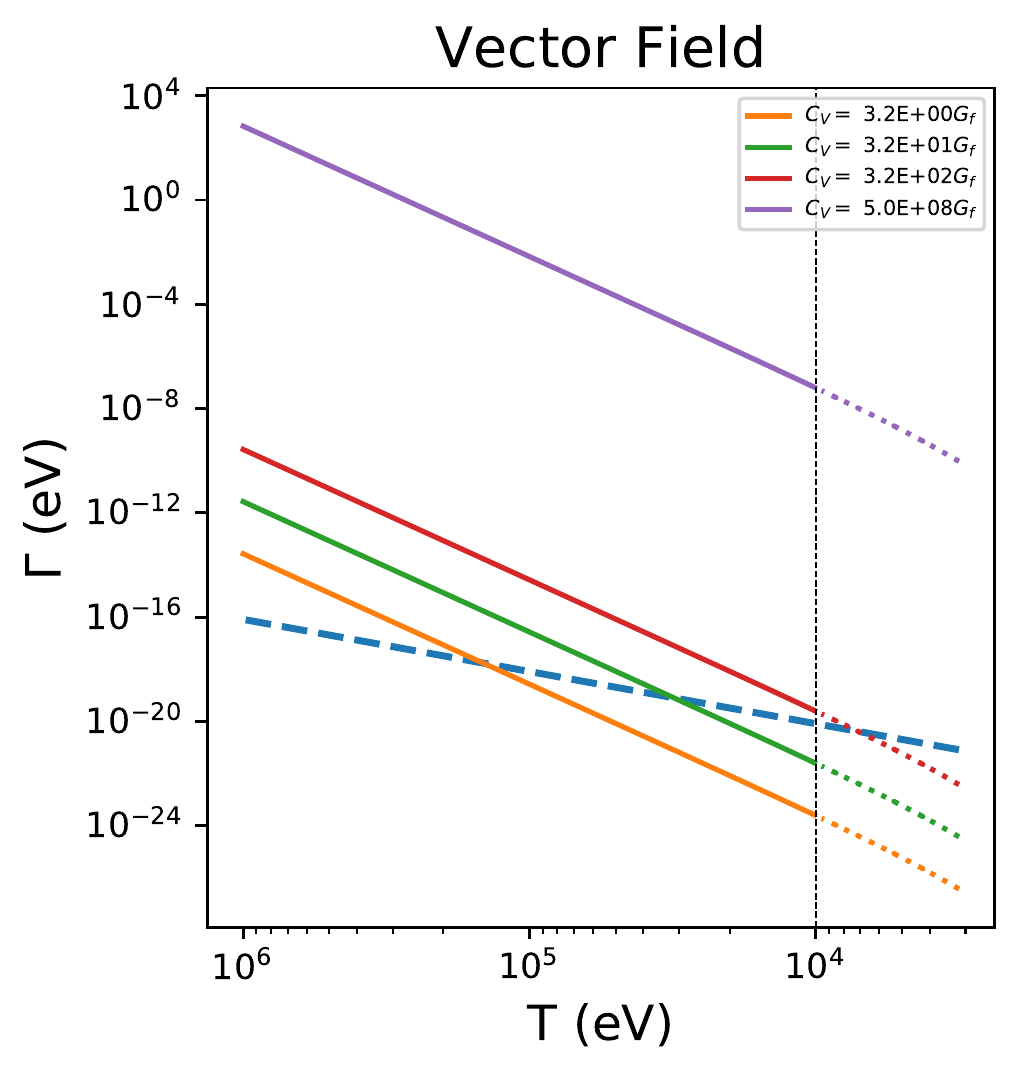}
    \end{subfigure}
\caption{\small Evolution of interaction rate per particle and Hubble expansion rate (in dashed line) for three different interaction models: 
Constant Amplitude, Massive Scalar and (massive) Vector Field, 
with interaction Lagrangians defined in \cref{sec:Solutions_Models} and interaction constants defined in \eqref{eq:Models_s0_MCoeffs}, \eqref{eq:Models_sm_MCoeffsT} and \eqref{eq:Models_vm_MCoeffsT} respectively, 
and calculated for a Maxwell-Boltzmann background DF. 
Several interaction strengths are evaluated for each model: both the ones relevant for a relativistic decoupling/recoupling as well as interaction strengths satisfying Bullet Cluster constraints (see \cite{amrr}). 
The calculations are performed for a DM particle mass of $10$ keV, and the vertical line marks the relativistic-nonrelativistic transition temperature.}
\label{fig:RTA_Decoupling}
\end{figure}

We can see how this interaction rate evolves along with the temperature of the plasma in \cref{fig:RTA_Decoupling}, using the models of \cref{sec:Solutions_Models}. We work here on the assumption that the self-interaction decouples (this is to say, $\Gamma>H$) while the particle itself is still relativistic. 
At the moment the self-interaction decouples its distribution function remains ``frozen-out'': the function itself remains unchanged and the evolution is just given by the redshift in physical momenta $p \propto a$. If decoupled while relativistic but well after the initial production of these particles, the distribution is frozen with a form $f_0 \propto e^{-p_{\mathrm{dec}}/T_{\mathrm{dec}}}$ and the redshift in momenta can be reinterpreted as a temperature evolution of the form $T \propto a^{-1}$. 

Depending on the coupling strength, it is possible that the self-interaction decouples while the particle is non relativistic. This opens the possibility of a species that undergoes chemical and kinetic decoupling from the plasma while still relativistic, but remains in equilibrium (with itself) until a later stage. After the decoupling of the self-interactions, the background distribution would be frozen out on a Maxwell form $f_0 \propto e^{-p_{\mathrm{dec}}^2/(2 m T_{\mathrm{dec}})}$ and the temperature is interpreted to evolve as $T \propto a^{-2}$, while preserving the number density at chemical decoupling.
.

We can see in \cref{fig:RTA_Decoupling} that the assumption of relativistic decoupling of the self-interaction does not necessarily hold for some interaction constants (e.g. $C_V \sim 10^8 G_f$). In that case, the self-interactions should decouple while non-relativistic, if at all, and they alter the distribution function which in turn renders the method we used to obtain \cref{fig:RTA_Decoupling} (described in \eqref{eq:Appx_RTA_Sigma_InvTau}) inapplicable, as it assumes $f_0 \propto e^{-E/T}$. In particular, couplings $C_V \sim 10^8 G_f$ in the vector field case were shown in \cite{amrr} to correspond to cross sections in the range $\sigma/m \sim 0.1 - 1\ \mathrm{cm}^2/\mathrm{g}$, which are usually considered to alleviate various problems in N-body simulations on self-interacting CDM, and are strongly constrained by observations\footnote{Interestingly, those cross sections may be large enough to spoil the assumption  that the DF corresponds to particles that decouple being relativistic, as used in previous applications of SI-WDM such as \cite{Hannestad00}. An interaction constant that large may cause the particle to remain in thermal equilibrium well into a non-relativistic regime.}  \cite{Tulin2017}.

\subsection{An Approximate Form for the Collision Integrals}

The full form of the Boltzmann hierarchy for these species \eqref{eq:Hier_F_motion} can be reduced by making use of the relaxation time approximation. The most straightforward way to do this is by simply replacing expression \eqref{eq:RTA_RTA_CollTerm} into the collision term and calculating the new hierarchies, through expression $D_2[f_0]$ \eqref{eq:ZO_D2_intFinal}. In \cite{Hannestad00} such an approach is taken in a simplified way: instead of the full (energy dependent) relaxation time $\tau_{rel}$, its thermal average is used (see \cref{sec:Appx_RTA}):

\begin{equation}
C[F] \approx - a \frac{F(\vec{q}, \vec{k}, \tau)}{\left< \tau \right>_{th}} \ .
\end{equation}

\noindent This simple approach however leads to an important conceptual error: this approximation (and to a certain extent \eqref{eq:RTA_RTA_CollTerm} as well) qualitatively simply ``erase'' the perturbations to the DF $F(\vec{q}, \vec{k}, \tau)$ \cite{Krapivsky10}. This violates conservation of particle, momentum and energy densities, resulting in a poor approximation to the full collision term in the case of perturbations. In \cite{Hannestad00} it is noted that this can be avoided by setting the $C_{l=0,1}[f]=0$, and these conservation laws are recovered, thus arriving to a Boltzmann hierarchy of the form:

{\small
\begin{equation}
\begin{split}
\dot{F_0} (k, E_q, \tau) \simeq& - \frac{q k}{E_q} F_1 (k, E_q, \tau ) + \frac{\dot{h}}{6} \frac{\partial f_0}{\partial \ln q}  \\
\dot{F_1} (k, E_q, \tau) \simeq& \frac{q k}{3 E_q} F_0 (k, E_q, \tau ) - \frac{2 q k}{3 E_q} F_2 (k, E_q, \tau ) \\
\dot{F_2} (k, E_q, \tau) \simeq& \frac{q k}{5 E_q} \Big[ 2 F_1 (k, E_q, \tau ) - 3 F_3 (k, E_q, \tau ) \Big] - \frac{\partial f_0}{\partial \ln q} \Bigg[ \frac{1}{15} \dot{h} + \frac{2}{5} \dot{\eta} \Bigg] - a \frac{F_2(k, E_q, \tau)}{\left< \tau \right>_{th}} \\
\dot{F_l} (k, E_q, \tau) \simeq& \frac{q k}{(2l+1) E_q} \Big[ l F_{(l-1)} (k, E_q, \tau ) - (l+1) F_{(l+1)} (k, E_q, \tau ) \Big] - a \frac{F_l(k, E_q, \tau)}{\left< \tau \right>_{th}} \quad , \quad l \geq 3 \ .
\end{split}
\label{eq:RTA_ApproxHier_Motion_Hann}
\end{equation}
\par}

\noindent This \textit{relaxation time approximation} \cite{Oldengott2017,Hannestad00,BGK} has the advantage of localizing the equations in momenta, which results in more efficient numerical integration by eliminating all coupling between different momentum bins and allowing for sparse evaluation. 

\begin{figure}
\centering
\includegraphics[width=\textwidth]{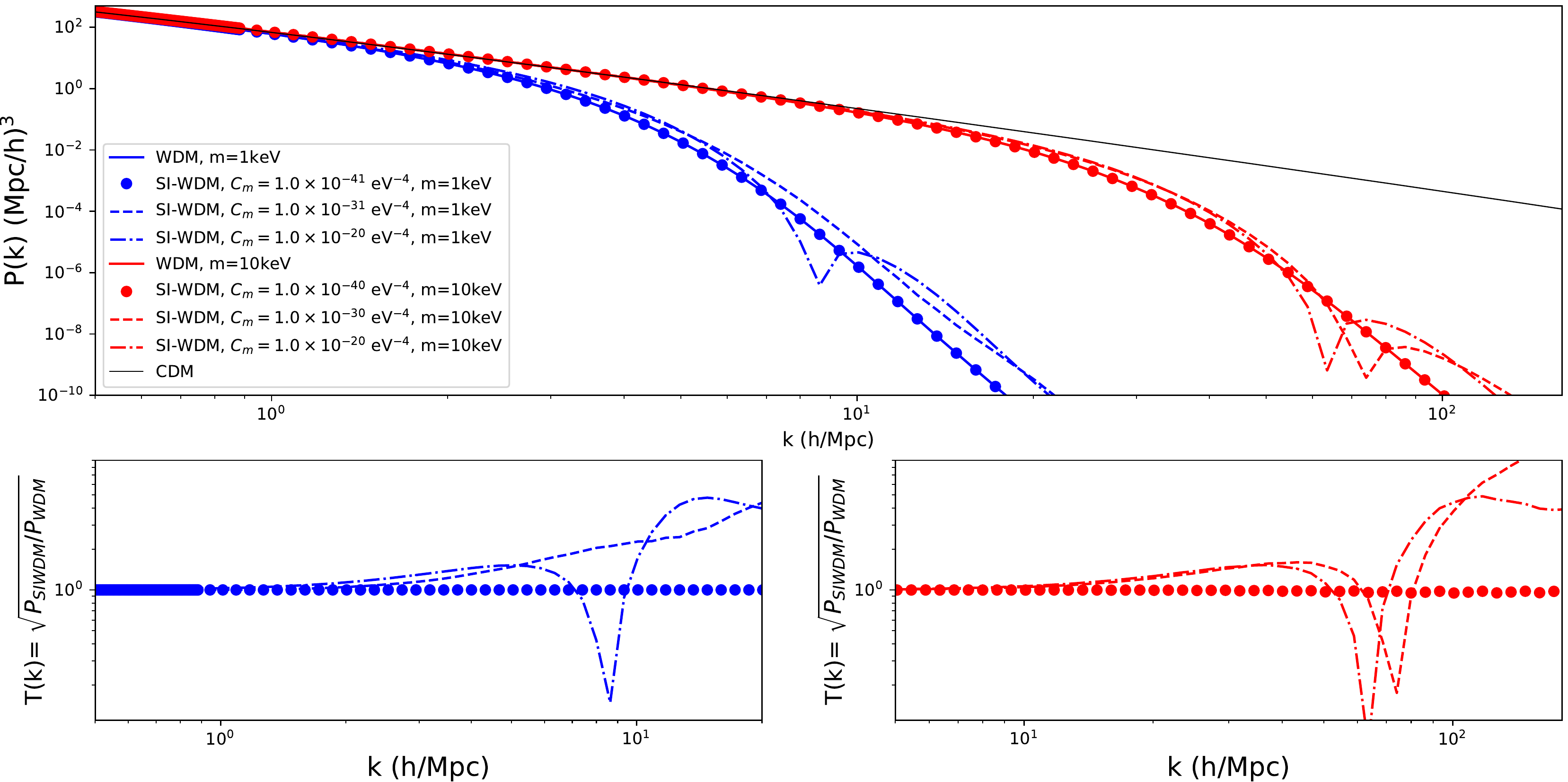}
\caption{\small 
Power Spectrum (\textit{top panel}) and Transfer Functions with respect to standard WDM (\textit{bottom panels}) for a massive scalar SI-WDM model under the relaxation time approximation \eqref{eq:RTA_ApproxHier_Motion_Hann} for two values of the DM particle mass: $1$ and $10$ keV. Also plotted are the power spectra of CDM and of a $1$ and $10\ {\rm keV}$ WDM model. Notice all the calculations assume relativistic interaction decoupling, however dashed and dashed-dotted lines refer to coupling strengths which do not fulfill this hypothesis and should undergo non relativistic self-interaction decoupling.
These results have been taken from \cite{Yunis_TBP} with permission from the authors.}
\label{fig:RTA_PS}
\end{figure}

As an illustration of the effects of self-interactions in the matter power spectrum, we provide in \cref{fig:RTA_PS}  specific examples for the case of a massive scalar field-mediator \eqref{eq:Models_sm_Msquare}, under the relaxation time approximation \eqref{eq:RTA_ApproxHier_Motion_Hann}. We have used an extended version of CLASS 2.7.2 \cite{lesgourgues2011cosmic,CLASSIV} where we include our results for SI-WDM models with particle masses in the $\sim$ keV range.  

In order to compare the results for standard CDM, WDM and the SI-WDM model, both WDM and SI-WDM components were assumed to have a relativistic Fermi-Dirac equilibrium distribution function $f_0$ with a given temperature $T_{dec}$ (i.e. a DF that corresponds to relativistically decoupled thermal relics), and  their abundances were adjusted to match that of CDM in the best fit data from Planck 2018 \cite{Planck2018}. It is important to notice that the assumption that the background DF is given at all times by the
relativistic Fermi-Dirac distribution may not apply if the self-interaction is sufficiently large. For instance,  for some interaction strengths that are  favored by SIDM N-body simulations for CDM \cite{Tulin2017}, it is necessary to consider  non-relativistic self-interaction decoupling.

In \cref{fig:RTA_PS} we recover the results of \cite{Hannestad00} for the case of $m=1$ keV, but for different coupling strengths. This is because of a missing scale factor in their calculation of the relaxation time, see \cite{Oldengott2017}. As shown here for the first time,  for the masses considered, those couplings actually correspond to non-relativistic self-interaction decoupling. In \cite{2001PhRvD..64h8301A} similar results are obtained, but using a fluid approximation  leading to spurious oscillations at high enough $k$ (see \cite{Oldengott2017,CLASSIV} for a discussion). 


In the power spectra shown in \cref{fig:RTA_PS} it can be seen that for the smaller interactions, in which the assumption of relativistic self-interaction decoupling is fulfilled, the results are practically indistinguishable from standard WDM. It is only for the cases of higher coupling constants (where the relativistic decoupling assumption is no longer valid) that certain features appear: both the modification in the transfer functions observed in \cite{Hannestad00} and acoustic oscillations at higher $k$ reminiscent of fluid approximations (see \cite{CLASSIV,Cyr-Racine2015}). Such results are explicitly shown in dashed or dot-dashed lines in \cref{fig:RTA_PS}, for given self-interaction strengths either for the case of $m=1$ keV and $m=10$ keV. \footnote{In the case of non relativistic self-interaction decoupling, the power spectrum damping is expected to shift to higher $k$ as the distribution function becomes ``colder''. A detailed analysis of this effect in terms of realistic cosmological effects on small-scales are the subject of future analysis \cite{Yunis_TBP}.}. 

For the case of $m\sim$ few keV as of typical WDM models under the thermal decoupling assumption \cite{Lovell:2013ola}, it can be seen that our SI-WDM power spectrums do not exhibit the steep trend at large $k$ typical of those standard WDM scenarios. This less abrupt suppression of power at typical (comoving) wave numbers of $k\sim 10$ h/Mpc (i.e. short scales relevant for sub-halo structures), can be  better visualized in the transfer function of \cref{fig:RTA_PS} (bottom pannel). Such an effect should point to a better agreement with small-scale structure constraints for the lower end of the (thermal relic) keV particle-mass range. All in all, a more general behaviour of the suppression in the power spectrum (relevant for sub-structure number counts), together with the self-interacting nature of the $\sim$ keV DM candidates (relevant to the inner shape of DM halos), could bring the SI-WDM paradigm into an appealing alternative to the CDM paradigm.  

Before closing, we point out that the simplified approach to the hierarchies given here, can be generalized slightly by using the \textit{separable ansatz} instead. Namely, assuming that the  ``temperature perturbation'' $\mathcal{F}_l(E_q,k,\tau)$ is independent of momentum
, that is

\begin{equation}
F_l (k, E_q, \tau) \approx -\frac{1}{4} \frac{d \ln f_0 }{d \ln q} f_0(E_q, \tau) \mathcal{F}_l (k, \tau) \ .
\label{eq:RTA_ApproxHier_SepAnsatz}
\end{equation}

\noindent Then, the $l$-th collision term can be reduced to:

\begin{equation}
C_l[f] = - a F_l (k, E_q, \tau) (\Gamma_{\mathrm{rel}} (E_q, \tau) - \Gamma_{\mathrm{exch}, l} (E_q, \tau)) \ ,
\label{eq:RTA_ApproxHier_CollisionTerm_SepAnsatz}
\end{equation}

\noindent with:

\begin{equation}
\begin{split}
\Gamma_{\mathrm{rel}} (E_q, \tau) &= - D_2[f_0] / f_0 = \tau^{-1} (E_q, \tau) \\
\Gamma_{\mathrm{exch}, l} (E_q, \tau) &= G_0 \int dE_l \left( \chi_l (E_q, E_l) f_0 (E_q) - 2 \frac{1}{E_q q} K_l (E_q, E_l, \tau ) \right) \frac{d \, f_0 (l) / d \ln l}{d \, f_0 (q) / d \ln q} \ .
\end{split}
\label{eq:RTA_ApproxHier_CollisionTerm_SepAnsatz_GammaDef}
\end{equation}

\noindent While this is a significant simplification to the collision term, as proposed in \cite{Oldengott2017}, further simplifications can be done by performing a momentum average

\begin{equation}
\left< \Gamma_{\mathrm{rel}} - \Gamma_{\mathrm{exch}, l} \right>_{\mathrm{avg}} \equiv \frac{\int dq q^3 f_0(q) (\Gamma_{\mathrm{rel}} - \Gamma_{\mathrm{exch}, l} )}{\int dq q^3 f_0(q)} \equiv \alpha_l \left< \tau^{-1} \right>_{avg} \ .
\label{eq:RTA_ApproxHier_CollisionTerm_SepAnsatz_OldengottAvg}
\end{equation}

Under this approximation the Boltzmann hierarchy reduces to\footnote{This form holds under the assumption that conservation laws for number density, momentum and energy are fulfilled. It has been explicitly checked in \cite{Oldengott2017} for the case of massless particles under massive scalar mediators, but it remains to be checked in the more general cases.}:

{\small
\begin{equation}
\begin{split}
\dot{F_0} (k, E_q, \tau) \simeq& - \frac{q k}{E_q} F_1 (k, E_q, \tau ) + \frac{\dot{h}}{6} \frac{\partial f_0}{\partial \ln q}  \\
\dot{F_1} (k, E_q, \tau) \simeq& \frac{q k}{3 E_q} F_0 (k, E_q, \tau ) - \frac{2 q k}{3 E_q} F_2 (k, E_q, \tau ) \\
\dot{F_2} (k, E_q, \tau) \simeq& \frac{q k}{5 E_q} \Big[ 2 F_1 (k, E_q, \tau ) - 3 F_3 (k, E_q, \tau ) \Big] - \frac{\partial f_0}{\partial \ln q} \Bigg[ \frac{1}{15} \dot{h} + \frac{2}{5} \dot{\eta} \Bigg] - a \frac{\alpha_2 F_2(k, E_q, \tau)}{\left< \tau \right>_{avg}} \\
\dot{F_l} (k, E_q, \tau) \simeq& \frac{q k}{(2l+1) E_q} \Big[ l F_{(l-1)} (k, E_q, \tau ) - (l+1) F_{(l+1)} (k, E_q, \tau ) \Big] - a \frac{\alpha_l F_l(k, E_q, \tau)}{\left< \tau \right>_{avg}} \quad , \quad l \geq 3 \ ,
\end{split}
\label{eq:RTA_ApproxHier_Motion_SepAnsatz}
\end{equation}
\par}

\noindent provided $C_{l=0,1=0}$. This approximation 
can be further reduced to the form \eqref{eq:RTA_ApproxHier_Motion_Hann} by assuming $\alpha_{l \geq 2}=1$.
If it results on a better overall approximation than \eqref{eq:RTA_ApproxHier_Motion_Hann}, remains to be explored in future works.

\section{Summary and Outlook}

Throughout this work, we aimed to fill a gap in the description and treatment of linear theory of scalar perturbations in cosmology by including the case of a self-interacting warm dark matter component. 
Motivated by the possible impact of these self-interactions in large and small structure formation scales, we provide an accurate treatment of collisions in the early universe, extending previous works on the subject while maintaining a phenomenological approach that allows us to retain certain model independence on the particular interaction Lagrangian.
By extending the treatment in \cite{Oldengott2014,Oldengott2017,Kreisch:2019yzn}  for active neutrinos, we calculated the first and zero order collision terms and provided a general framework in order to include these collision terms in the coupled Einstein-Boltzmann system. This was done with the objective of accurately evaluating the effect of WDM self-interactions on the linear power spectrum and the CMB anisotropies.

In \cref{sec:Boltz} we provide a short summary about the assumptions used in this framework, as well as the general form \eqref{eq:FO_C3_s3_M2Def} for the interaction amplitude that was used. It is shown there that this form can \added[id=dln]{accurately} describe several models of massive mediator interactions between sterile neutrinos, though not limited to those cases, and including for Majoron-like scalar mediators between right handed neutrinos.
The main calculations are given in \cref{sec:Solutions_FO} and \ref{sec:Solutions_ZO} where we show the results for the first and zero order collision terms respectively in the SI-WDM scenario. Also, we provide in \cref{sec:Solutions_Models} some examples for a handful of specific interaction models, along with the corresponding coefficients for the collision terms.

A detailed treatment on how to include these collision terms in a Boltzmann hierarchy is shown in \cref{sec:Hier}, and several possible simplifications in order to treat the evolution of the background and perturbed distributions based on the relaxation time approximation are shown in \cref{sec:RTA}. In this last section we discuss the effects of self-interactions in the matter power spectrum for $\sim$ keV DM thermal relics, by providing an specific example for the case of massive scalar field-mediators. Besides acoustic oscillations arising at large $k$, it is shown a less abrupt suppression of power (relative to standard WDM) for typical comoving wave numbers of $k\sim 10$ h/Mpc, relevant for small-scale structure constraints.

While developed with the intent of being used in the calculation of cosmological perturbations in the case of SI-WDM, the forms of the collision terms themselves are quite general and they can be used also in several other applications, for example in the case of (massive) active neutrino cosmology, as noticed in \cref{sec:Solutions_Models}.
The implementation of this formalism in a CMB Boltzmann solver developed to explore the effects of the SI-WDM framework in cosmology, has been partially used in \cref{sec:RTA} through an explicit example (as explained above). A detailed exploration for other field-mediators, interaction strengths, as well as other quantitative small-scale structure effects are left for a future work \cite{Yunis_TBP}.  
In this direction, further exploration of possible approximation schemes or efficient computation methods is key in order to successfully implement the hierarchies \eqref{eq:Hier_F_motion} in a practical way.
Moreover, the phenomenological model in \eqref{eq:FO_C3_s3_M2Def} for the interaction amplitude can be generalized by including other types of interaction Lagrangians, such as light mediators or more complex models accurately. 
For light mediators, in \cref{sec:Appx_Massless}, we extend the calculation of the DM-DM collision term. This calculation is a necessary step to further generalize the equations to include models where other collision terms involving light mediators are relevant. In order to explore  this kind of WDM interactions, further modeling it is required, since in general the population of light mediators cannot be neglected, and a consistent generalization should also model the evolution of their distribution function and collision terms.
Also, it should be possible to extend this formalism to include the effects of Bose enhancement or Pauli blocking by generalizing the collision kernels, as was argued in \cite{Oldengott2014}.

\acknowledgments
We thank G. V. Vereshchagin for a critical reading of this work, and N. E. Mavromatos for useful comments on an earlier version of the manuscript. We also thank C. Sc\'occola for discussions.  CRA has been supported by CONICET, Secretary of Science and Technology of FCAG and UNLP. DNL has been supported by CONICET, ANPCyT and UBA.

\clearpage
\bibliographystyle{jhep}
\bibliography{SIDM,SIDM-INTRO,SIDM-DIANA,SIDM-REV1}

\clearpage
\appendix

\section{First Order Collision Integral Terms}
\label{sec:Appx_FO}

\subsection{Calculation of $\mathcal{C}_3[f]$}
\label{sec:Appx_FO_C3}

The term $\mathcal{C}_3$ can be expressed as:

\begin{equation}
\mathcal{C}_3 [f] = - \frac{g_i^3}{2 E_q (2\pi)^5} \int \frac{d^3 l}{2 E_{l}} \frac{d^3 q'}{2 E_{q'}} \frac{d^3 l'}{2 E_{l'}} |\mathcal{M}|^2 \delta_D^{(4)}(\mathbf{q}+\mathbf{l}-\mathbf{q'}-\mathbf{l'}) f_0 (l) F(\vec{q}) \ ,
\label{eq:Appx_FO_C3_Def}
\end{equation}

\noindent where $g_i$ is the number of degrees of freedom of the particle and we have omitted the $ \vec{k} $, $\tau$ dependencies on the distribution functions for compactness.

\tocless\subsubsection{Solving for $\vec{l'}$ and $\beta$}

The energy conservation Dirac delta $\delta_D^{(4)}(\mathbf{q}+\mathbf{l}-\mathbf{q'}-\mathbf{l'})$ can be used to solve directly the $l'$ integral using the definition of the invariant integration measure:

\begin{equation}
\begin{split}
\int \frac{d^3 l'}{2 E_{l'}}  \delta_D^{(4)} (\mathbf{q}+\mathbf{l}-\mathbf{q'}-\mathbf{l'}) & = \int d^3 l' \theta (E_{l'}) \Theta (E_q + E_l - E_{q'} ) \times \\
& \qquad \times \delta_D^{(3)} (\vec{q}+\vec{l}-\vec{q'}-\vec{l'})\delta_D (E_{l'}^2 - (E_q+E_l-E_{q'})^2 )  \\
& = \Theta( E_q + E_l -E_{q'} ) \delta_D (E^2_{\vec{q}+\vec{l}-\vec{q'}} - (E_q+E_l-E_{q'})^2)\\
& \equiv \Theta( E_q + E_l -E_{q'} ) \delta ( g (\vec{q},\vec{l},\vec{l'} ) )
\end{split}
\label{eq:Appx_FO_C3_s1_deltaLp}
\end{equation}

\noindent where

\begin{equation}
g(\vec{q},\vec{l},\vec{l'}) \equiv - E^2_{\vec{q}+\vec{l}-\vec{q'}} + (E_q+E_l-E_{q'})^2 \ ,
\label{eq:Appx_FO_C3_s1_gDef}
\end{equation}

\noindent and $\Theta$ is the Heaviside theta function. The following parametrization is used for the momentum 3-vectors:

\begin{equation}
\begin{cases}
\vec{q}= q (0, 0, 1)\\
\vec{l}= l (0, \sin \alpha, \cos \alpha ) \\
\vec{q'}= q' (\sin \beta \sin \theta, \cos \beta \sin \theta, \cos \theta) 
\end{cases} \ ,
\label{eq:Appx_FO_C3_s1_param}
\end{equation}

\noindent So the argument of the Dirac delta in \eqref{eq:Appx_FO_C3_s1_gDef} can be expressed as:

\begin{equation}
\begin{split}
g(\vec{q},\vec{l},\vec{l'}) =& 2 m^2 + 2 E_q E_l - 2 E_q E_{q'} - 2 E_l E_{q'} + 2 l q' (\cos \alpha \cos \theta + \sin \alpha \cos \beta \sin \theta ) \\
& + 2 q q' \cos \theta - 2 q l \cos \alpha \ .
\end{split}
\label{eq:Appx_FO_C3_s1_gParam}
\end{equation}

\noindent In this parametrization for the momentum 3-vectors, the integrals in the collision term can be expressed as 

\begin{equation}
\begin{split}
\mathcal{C}_3 [f] = &-\frac{F(\vec{q}) g_i^3}{8 (2\pi)^4 E_q} \int dq' dl d (\cos \theta) d (\cos \alpha) \frac{l^2}{E_l} \frac{{q'}^2}{E_{q'}} f_0(l) \Theta (E_q+E_l-E_{q'}) |\mathcal{M}|^2 \\
& \times \int_0^{2\pi} d\beta \delta_D(g(\vec{q},\vec{l},\vec{q'})) \ ,
\end{split}
\label{eq:Appx_FO_C3_s1_intDeltaG}
\end{equation}

\noindent where we have used $d^3 l= 2\pi l^2 dl d(\cos \alpha)$ and $d^3 q'= d \beta {q'}^2 dq' d(\cos \theta)$, we have omitted the relevant integration bounds except on the $\beta$ integral and assumed that the scattering amplitude $|\mathcal{M}|$ does not depend on the azimuthal angle $\beta$. For this integral, we use the following identity of the Dirac delta:

\begin{equation}
\int_0^{2\pi} d\beta \delta_D(g) = \int_0^{2\pi} d\beta \sum_i \delta_D(\beta - \beta_i) \left| \frac{\partial g}{\partial \beta} \right|_{\beta_i}^{-1} \ ,
\label{eq:Appx_FO_C3_s1_dbetaG}
\end{equation}

\noindent that allows calculations of compositions between delta functionals and functions, where $\beta_i$ are real roots of the real function $g(...,\beta,...)$. For this integral, we have

\begin{align}
\frac{\partial g}{\partial \beta} &= -2 l q' \sin \alpha \sin \theta \sin \beta \\
\cos \beta_i &= \left( l q' \sin \alpha \sin \theta \right)^{-1} \big[ -m^2 +(E_q E_{q'} - q q' \cos \theta ) + (E_l E_{q'} - l q' \cos \alpha \cos \theta ) \nonumber \\
& + (q l \cos \alpha - E_q E_l ) \big] \ .
\label{eq:Appx_FO_C3_s1_dgdbeta/cosbetai} 
\end{align}

\noindent From this, we can infer that two solutions $\beta_i$ exist: one in the interval $[0,\pi]$ and one in $[\pi,2\pi]$. As the absolute value of the derivative of $g$ is the same in both solutions, we can express the integral in \eqref{eq:Appx_FO_C3_s1_intDeltaG} as

\begin{equation}
\int_0^{2\pi} d\beta \delta_D (g) = 2 \int_0^{\pi} \left| \frac{\partial g}{\partial \beta} \right|_{\cos \beta_i}^{-1} \delta_D (\beta-\beta_i) = 2 \left| \frac{\partial g}{\partial \beta} \right|_{\cos \beta_i}^{-1} \ .
\label{eq:Appx_FO_C3_s1_dbetaGSol}
\end{equation} 

To ensure the physical condition that $|\cos \beta_i| < 1$, we add a Heaviside step function in $\cos^2 \beta_i$. The following property follows from \eqref{eq:Appx_FO_C3_s1_dgdbeta/cosbetai} :

\begin{equation}
\Theta (1-\cos^2 \beta_i) = \Theta \left( \left| \frac{\partial g}{\partial \beta} \right|_{\cos \beta_i}^{2} \right) \ .
\label{eq:Appx_FO_C3_s1_ThetaProp}
\end{equation}

\noindent So, the $\mathcal{C}_3$ integral in \eqref{eq:Appx_FO_C3_s1_intDeltaG} can be expressed as

\begin{equation}
\begin{split}
\mathcal{C}_3 [f] &= -\frac{g_i^3 F(\vec{q})}{4 (2\pi)^4 E_q} \int dl d (\cos \alpha) \frac{l^2}{E_l} f_0(l) \int d q' d(\cos \theta) \frac{{q'}^2}{E_{q'}} \Theta (E_q+E_l-E_{q'}) |\mathcal{M}|^2  \\
& \times \Theta \left( \left| \frac{\partial g}{\partial \beta} \right|_{\cos \beta_i}^{2} \right) \left| \frac{\partial g}{\partial \beta} \right|_{\cos \beta_i}^{-1} \ .
\end{split}
\label{eq:Appx_FO_C3_s1_int2Theta}
\end{equation}

Here, an important point in the calculation is reached. The remaining angular integrals are in $\cos \theta$ and $\cos \alpha$, the angles between $\vec{q}$ and $\vec{q'}$, and $\vec{q}$ and $\vec{l}$, respectively. Now, without any knowledge of the background function $f_0$, up to three of the remaining four integrals could be solved. However to continue solving from here on we need to know about the scattering amplitude $\mathcal{M}$. We will go as far as possible without specifying this, and then we will assume an ansatz for a general form of $\mathcal{M}$. To continue we express the term $|\partial g / \partial \beta|_{\cos \beta_i}$ in terms of the variables $\{q,l,\theta,\alpha\}$ as follows:

\begin{equation}
\left| \frac{\partial g}{\partial \beta} \right|_{\cos \beta_i}^{2} = a_3^{(\theta)} \cos^2 \theta + b_3^{(\theta)} \cos \theta + c_3^{(\theta)} \ ,
\label{eq:Appx_FO_C3_s1_coeffDgDbeta_def}
\end{equation}

\noindent with coefficients:

\begin{equation}
\begin{cases}
a_3^{(\theta)} = -4 {q'}^2 |\vec{l} + \vec{q}|^2 \leq 0\\
b_3^{(\theta)} = 8 q' [q+l\cos \alpha] [E_q E_{q'} + E_l E_{q'} - E_q E_l +ql\cos \alpha -m^2 ] \\
c_3^{(\theta)} = 4 \left\{ l^2 {q'}^2 \sin^2 \alpha - [E_q E_{q'} + E_l E_{q'} - E_q E_l +ql\cos \alpha -m^2 ]^2 \right\} 
\end{cases} \ .
\label{eq:Appx_FO_C3_s1_coeffDgDbeta_expr}
\end{equation}

Now, let us consider in greater detail the integration of the Heaviside theta function in equation \eqref{eq:Appx_FO_C3_s1_int2Theta}. The argument of the function is a quadratic function in $\cos \theta$ with a negative leading coefficient. Thus, the function will only be non zero if two real roots of the polynomial $|\partial g / \partial \beta|^2_{\cos \beta_i}$ exist and it will be unity between them. So, the $\Theta$ function can be translated into a border condition for the $\cos \theta$ integral and an existence condition for the roots:

\begin{equation}
\begin{split}
\int d (\cos \theta) & \frac{|\mathcal{M}|^2}{\sqrt{a_3^{(\theta)} \cos^2 \theta + b_3^{(\theta)} \cos \theta + c_3^{(\theta)}}} \Theta (a_3^{(\theta)} \cos^2 \theta + b_3^{(\theta)} \cos \theta + c_3^{(\theta)}) \\
&= \Theta \big[ ( b_3^{(\theta)} )^2 - 4 a_3^{(\theta)} c_3^{(\theta)} \big] \int_{x_1}^{x_2} d (\cos \theta) \frac{|\mathcal{M}|^2}{\sqrt{a_3^{(\theta)} \cos^2 \theta + b_3^{(\theta)} \cos \theta + c_3^{(\theta)}}} \ ,
\end{split}
\label{eq:Appx_FO_C3_s1_thetaConsProp}
\end{equation}

\noindent with $x_{1,2}$ the roots of the polynomial $|\partial g / \partial \beta|^2_{\cos \beta_i}$,

\begin{equation}
x_{1,2} = \frac{b_3^{(\theta)}}{2 |a_3^{(\theta)}|} \pm \sqrt{\left( \frac{b_3^{(\theta)}}{2 |a_3^{(\theta)}|} \right)^2 + \frac{c_3^{(\theta)}}{|a_3^{(\theta)}|}} \ .
\label{eq:Appx_FO_C3_s1_dgDbetaRoots}
\end{equation}

The argument of the Heaviside step function can be expressed as

\begin{equation}
\begin{split}
( b_3^{(\theta)} )^2 - & 4 a_3^{(\theta)} c_3^{(\theta)} = 64 {q'}^2 l^2 \sin ^2 \alpha \big[ E_q E_l (1 - \cos y) \big]  \\
& \times \Bigg\{ -2E_{q'}^2 + 2(E_q + E_l) E_{q'} - \left[ E_q E_l ( 1 - \cos y ) + \frac{m^2 |\vec{l} + \vec{q}|^2 }{E_q E_l (1 - \cos y)} \right] \Bigg\} \ ,
\end{split}
\label{eq:Appx_FO_C3_s1_discrExpr}
\end{equation}

\noindent where

\begin{equation}
1 - \cos y = 1 + \frac{m^2}{E_q E_l} - \frac{ql}{E_q E_l} \cos \alpha \ .
\label{eq:Appx_FO_C3_s1_cosyDef}
\end{equation}

\noindent So this argument has 4 real roots in $E_{q'}$: $\{ -m, +m, R_1, R_2\}$, with $R_{1,2}$ given by

\begin{equation}
R_{1,2} = \frac{1}{2} \left\{ E_q + E_l \pm |\vec{l} + \vec{q}| \sqrt{1 - \frac{2 m^2}{E_q E_l (1-\cos y ) }} \right\} \ .
\label{eq:Appx_FO_C3_s1_discrRoots}
\end{equation}

\noindent The $-m$ root can already be discarded as non physical. In order to to obtain the ordering for the rest of the roots, thus the non-zero intervals for the Heaviside step functions, we will next develop some alternative notation for the angular variables $\theta$, $\alpha$, making use of Mandelstam variables.

\tocless\subsubsection{Mandelstam variables for $\mathcal{C}_3$}

Mandesltam variables are Lorentz invariant quantities constructed with the relevant information on a two on two scattering process and are defined as:

\begin{equation}
\begin{cases}
s \equiv (\mathbf{q} + \mathbf{l})^2 = (E_q + E_l)^2 - |\vec{q} + \vec{l}|^2 > 0 \\
t \equiv (\mathbf{q} - \mathbf{q'})^2 = (E_q - E_{q'})^2 - |\vec{q} - \vec{q'}|^2 < 0 \ . \\
u \equiv (\mathbf{q} - \mathbf{l'})^2 = (E_q - E_{l'})^2 - |\vec{q} - \vec{l'}|^2 < 0 
\end{cases}
\label{eq:Appx_FO_C3_s2_MandDef}
\end{equation}

\noindent We can make use of these quantities to advance in the remaining integrals for the collision integral $\mathcal{C}_3$. First of all it is important to note that at this stage the derivations here and in \cite{Oldengott2014} start to diverge. Some of the interesting properties of Mandelstam variables, which allows them to be of use in these calculations, are related to the center of momentum (CoM) frame that cannot be properly defined in the case of collisions between massless particles. Indeed, for the case of identical particles the Mandelstam variables can be calculated in the CoM frame making use of their Lorentz invariance:

\begin{equation}
\begin{cases}
s = 4 \left( E^{CoM}_m \right)^2 \\
t = -2  \left( p^{CoM}_m \right)^2 \left( 1 - \cos \theta^{CoM} \right) \ , \\
u = -2  \left( p^{CoM}_m \right)^2 \left( 1 + \cos \theta^{CoM} \right) 
\end{cases}
\label{eq:Appx_FO_C3_s2_MandCom}
\end{equation}

\noindent with $E^{CoM}_m$, $p^{CoM}_m$ the individual particle's energy and momentum magnitude measured in the CoM frame and $\theta^{CoM}$ the scattering angle measured in the same frame. We can also express these variables in terms of the quantities we have used throughout the calculation of $\mathcal{C}_3$ as:

\begin{equation}
\begin{cases}
s = 2 E_q E_l \left( 1 + \frac{m^2}{E_q E_l} - \frac{ql}{E_q E_l}\cos \alpha \right) = 2 E_q E_l (1 - \cos y)\\
t = -2 E_q E_{q'} \left( 1 - \frac{m^2}{E_q E_{q'}} - \frac{qq'}{E_q E_{q'}}\cos \theta  \right) \ ,
\end{cases}
\label{eq:Appx_FO_C3_s2_MandParam}
\end{equation}

\noindent where here, the quantities are measured in the ``lab'' frame, that is to say, the fixed frame in which we have measured $\vec{q}$. It would not be possible to change the whole integral to CoM quantities making use of the Lorentz invariant measure $d^3 p / 2 E_p$ because $\vec{q}$ is fixed by the LHS of the Bolztmann equation. As a first use for these quantities, let us consider the roots $R_{1,2}$ in the polynomial above. The roots can be immediately recast as a function of $s$:

\begin{equation}
R_{1,2} = \frac{1}{2} \left\{ E_q + E_l \pm |\vec{l} + \vec{q}| \sqrt{1 - \frac{4m^2}{s}} \right\} \ .
\label{eq:Appx_FO_C3_s2_discrRoots}
\end{equation}

From expression \eqref{eq:Appx_FO_C3_s2_MandCom} we can see that $s \geq 4m^2$, thus these roots are always real. As for the ordering of the roots, it should be quite obvious that $R_2>{m, R_1}$. It is also possible to prove that $R_1>m$: 

\begin{itemize}

\item Let us first define the total momentum 4 vector: $p^\mu = ( E_q + E_l, \vec{q} + \vec{l} ) $. If we recall the form of a Lorentz transformation over a 4-vector,

\begin{equation}
\begin{cases}
{p^0} ' = \gamma ( p^{0} - \vec{\beta} . \vec{p} ) \\
{\vec{p}} ' = \gamma ( - |\vec{\beta}| p^{0} + \vec{p} ) 
\end{cases}
\end{equation}

\noindent we can infer that the form of $R_1$ is very similar to the 0 component of this 4-vector, measured in a different coordinate system (Lorentz boosted). The magnitude of this boost from the ''laboratory'' system ( where we measure $\vec{q},\ \vec{l}$ ) to this new system would be $|\vec{\beta}| = \sqrt{1-4m^2/s}$ in the direction of the total 3-momentum $\vec{q} + \vec{l}$ \footnote{This is a valid boost rapidity $\beta$, as using the properties of the Mandelstam variables one can find $|\vec{\beta}| = p_m^{CoM} / E_m^{CoM}$, with these quantities being the individual particle's energies and momenta measured in the center of momentum system. }. The total energy $p^0$ measured in this new frame of reference is

\begin{equation}
E_{p^\mu} ' = 2 \gamma R_1
\end{equation} 

\item On the other hand this energy, measured in the boosted system, can be also written as

\begin{equation}
E_{p^\mu} ' = m_{p^\mu} \gamma ' \ ,
\end{equation}

\noindent where, this time, $m_{p^\mu} = \sqrt{|p_\mu p^\mu|}$ and $\gamma '$ is the Lorentz factor corresponding to the boost from the center of momentum system to this new system of reference. 

\item Equating these two expressions, one finds:

\begin{equation}
R_1 = \frac{\sqrt{|p_\mu p^\mu|}}{2} \frac{\gamma '}{\gamma} = \frac{\sqrt{s}}{2} \frac{2m \gamma'}{\sqrt{s}} = m \gamma' \geq m  \ ,
\end{equation}

\noindent because $\gamma' \geq 1$. \hfill $\blacksquare$

\end{itemize}

So, going back to the expression for the last Heaviside theta \eqref{eq:Appx_FO_C3_s1_discrExpr} we can conclude that it is both non zero and physical only between $R_1$ and $R_2$. We can then express the whole $\mathcal{C}_3$ integral as

\begin{equation}
\mathcal{C}_3 = - \frac{g_i^3 F(q)}{4 (2\pi)^4 E_q} \int dl d (\cos \alpha) \frac{l^2}{E_l} f_0 (l)\int_{q'(R_1)}^{q'(R_2)} dq' \frac{{q'}^2}{E_{q'}} \int_{x_1}^{x_2} d (\cos \theta) \frac{|\mathcal{M}|^2 (s,t) }{\sqrt{\left| \frac{\partial g}{\partial \beta} \right|^2_{\cos \beta_i}}}\ ,
\label{eq:Appx_FO_C3_s2_intMand}
\end{equation}

\noindent where $q'(R) = \sqrt{R^2-m^2}$ and we have made use of the fact that $ R_2 \geq E_q + E_l $ to eliminate $\Theta(E_q + E_l - E_{q'})$. The observant reader may have noticed that not only we omitted the dependency on the dynamical variables of $|\mathcal{M}|^2$, but we have also simply carried it outside of both integrations, on $\vec{l'}$ and $\beta$. As the Mandelstam variables encompass all of the relevant invariant quantities involved on the kinematics of the process itself, it is reasonable to expect the scattering amplitude $|\mathcal{M}|^2$ to only depend on $(s,t,u)$. Now, we have solved first a three momentum integral in $\vec{l'}$ imposing the momentum conservation Dirac delta. When calculating scattering amplitudes momentum conservation is explicitly imposed, so we can consider  $|\mathcal{M}|$ to already be evaluated at $\vec{l'} = -\vec{q} - \vec{l} + \vec{q'}$. Now for the $\beta$ integral we have solved a Dirac delta in the function $g$ defined in \eqref{eq:Appx_FO_C3_s1_gDef}. If we express this function in Mandelstam variables we can see that it is simply $\delta_D (s+t+u-4m^2)$. Thus, we can impose the condition $s+t+u = 4 m^2$ in $|\mathcal{M}|$ and bring it out of the $\beta$ integral. Note that this condition is trivially fulfilled given the definition of $s$, $t$ and $u$. 

\tocless\subsubsection{Change of variables to $\{s,t,E_q,E_{q'}\}$}

At this point, as mentioned above, further integration is not possible without specific knowledge of the scattering amplitude $|\mathcal{M}|$. However in terms of the variables we have used so far, namely $\{\vec{q},\vec{q'},\vec{l},\vec{l'}\}$, the scattering amplitude may indeed have a very complicated and ultimately redundant expression. The Mandelstam variables contain all of the Lorentz invariant quantities that are involved in the expression of $|\mathcal{M}|$ so that its only dependencies would be on $(s,t)$, after having applied the identity $s+t+u = 4 m^2$. In order to accommodate to a more general expression for the scattering amplitude, it is convenient to change variables in the integrals of $\mathcal{C}_3$ to obtain an expression involving $(s,t)$. Indeed, we can make the following variable change:

\begin{equation}
\begin{cases}
q' \rightarrow E_{q'} = \sqrt{{q'}^2+m^2} \\
l \rightarrow E_l = \sqrt{l^2+m^2} \\
\cos \alpha \rightarrow s = 2 E_q E_l \left( 1 + \frac{m^2}{E_q E_l} - \frac{ql}{E_q E_l}\cos \alpha \right) \\
\cos \theta \rightarrow t = -2 E_q E_{q'} \left( 1 - \frac{m^2}{E_q E_{q'}} - \frac{qq'}{E_q E_{q'}}\cos \theta  \right) \\
\end{cases} \ .
\label{eq:Appx_FO_C3_s3_COV}
\end{equation}

\noindent The integration measures change accordingly as:

\begin{equation}
dq' d \cos \theta = \frac{E_{q'}}{2 q {q'}^2} dE_{q'} dt \quad , \quad dl d \cos \alpha = \frac{E_{l}}{2 q {l}^2} dE_{l} ds \ .
\label{eq:Appx_FO_C3_s3_measure}
\end{equation}

\noindent So that the whole integral can be expressed as

\begin{equation}
\mathcal{C}_3 = - \frac{g_i^3 F(q)}{16 (2\pi)^4 E_q q^2} \int dE_l ds f_0 (E_l) \int_{R_1}^{R_2} dE_{q'} \int_{t(x_1) \equiv t_1}^{t(x_2) \equiv t_2} dt \frac{|\mathcal{M}|^2 (s,t) }{\sqrt{\left| \frac{\partial g}{\partial \beta} \right|^2_{\cos \beta_i}}}
\label{eq:Appx_FO_C3_s3_intDMand}
\end{equation}

Now, the expression \eqref{eq:Appx_FO_C3_s1_coeffDgDbeta_def} that states $|\partial g/ \partial \beta|^2_{\cos \beta_i} $ as a second degree polynomial in $\cos \theta$ can be recast into a polynomial in $t$ using \eqref{eq:Appx_FO_C3_s2_MandParam}:

\begin{equation}
\left| \frac{\partial g}{\partial \beta} \right|^2_{\cos \beta_i} = a_3^{(t)} t^2 + b_3^{(t)} t + c_3^{(t)} \ ,
\label{eq:Appx_FO_C3_s3_coeffDgDbeta_def}
\end{equation}

\noindent with

\begin{equation}
\begin{cases}
a_3^{(t)} = -\frac{1}{q^2} |\vec{l}+\vec{q}|^2 \\
b_3^{(t)} = - \frac{2}{q^2} \left\{ s(E_q-E_l) (E_q-E_{q'}) + 2 q^2 l^2 \sin^2 \alpha \right\} \\
c_3^{(t)} = \frac{s}{q^2} (4m^2 - s) (E_q-E_{q'})^2
\end{cases} \ .
\label{eq:Appx_FO_C3_s3_coeffDgDbeta_expr}
\end{equation} 

The ingredients are all set to perform the $t$ integral given an expression for $|\mathcal{M}|$. We then assume that this amplitude can be expressed as a second degree polynomial in $(s,t)$, as in \eqref{eq:FO_C3_s3_M2Def}:

\begin{equation}
|\mathcal{M}|^2 \equiv m_{(2,0)}s^2 + m_{(1,1)} s t + m_{(0,2)}t^2 + m_{(1,0)}s + m_{(0,1)} t + m_{(0,0)} \ .
\label{eq:Appx_FO_C3_s3_M2Def}
\end{equation}

In order to explicitly perform the $t$ integral, we group the coefficients in \eqref{eq:Appx_FO_C3_s3_M2Def} in their respective powers of $t$:

\begin{equation}
|\mathcal{M}|^2 = A_t t^2 + B_t t + C_t \ .
\label{eq:Appx_FO_C3_s3_M2Coeff}
\end{equation}

So the $t$ integral becomes

\begin{equation}
\begin{split}
\int_{t_1}^{t_2} dt \frac{|\mathcal{M}|^2}{\sqrt{a_3^{(t)} t^2 + b_3^{(t)} t + c_3^{(t)}}} &= \frac{1}{\sqrt{- a_3^{(t)} }} \Bigg\{ A_t \int_{t_1}^{t_2} dt \frac{t^2}{\sqrt{(t-t_1)(t_2-t)}} + \\
& + B_t \int_{t_1}^{t_2} dt \frac{t}{\sqrt{(t-t_1)(t_2-t)}} + C_t \int_{t_1}^{t_2} dt \frac{1}{\sqrt{(t-t_1)(t_2-t)}} \Bigg\} \ .
\end{split}
\label{eq:Appx_FO_C3_s3_dtFormalSol}
\end{equation}

Then, we use the following identity:

\begin{equation}
\int_{x_-}^{x_+} dx \frac{Ax^2 + Bx + C}{\sqrt{(x_+-x)(x-x_-)}} = \pi \left\{ A \left( \frac{3b^2 -4 a c}{8a^2} \right) - B \left( \frac{b}{2a} \right) + C \right\} \ ,
\label{eq:Appx_FO_C3_s3_dtIdent}
\end{equation}

\noindent where $x_{(+,-)}$ are the solutions of $ax^2+bx+c=0$ to arrive at

\begin{equation}
\begin{split}
\mathcal{C}_3 = - \frac{g_i^3 F(q)}{16 (2\pi)^4 E_q q} & \int dE_l ds f_0 (E_l) \int_{R_1}^{R_2} dE_{q'} \frac{\pi}{|\vec{q}+\vec{l}|}  \\
& \times \left[ A_t \left( \frac{3 \left( b_3^{(t)} \right)^2 - 4 a_3^{(t)} c_3^{(t)}}{8 \left( a_3^{(t)} \right)^2 } \right) - B_t \left( \frac{b_3^{(t)}}{2 a_3^{(t)}} \right) + C_t \right] \ .
\end{split}
\label{eq:Appx_FO_C3_s3_intIdent}
\end{equation}

\noindent The expressions in brackets have fairly complicated forms, but mostly polynomial in nature in $(E_{q'},E_q,E_l,s,m)$. They can be factored into powers of $E_{q'}$ in order to integrate them as:

\begin{equation}
\mbox{\small $
\begin{split}
\frac{3 \left( b_3^{(t)} \right)^2 - 4 a_3^{(t)} c_3^{(t)}}{8 \left( a_3^{(t)} \right)^2 } = \frac{1}{8 |\vec{l} + \vec{q}|^4} \Bigg\{ & E_{q'}^2 \Bigg[ 4 s \left(2 s \left(E_l^2-4 E_l E_q+E_q^2-2 m^2\right)+4 m^2 (E_l+E_q)^2+s^2\right)\Bigg]  \\
	+ & E_{q'} \Bigg[ 4s \big(4 m^2 s (3 E_l-E_q)-4 m^2 (3 E_l-E_q) (E_l+E_q)^2 \\
	& \hphantom{E_{q'}^2 \Bigg[} + s^2 (E_q-3 E_l) -4 E_q s (E_q-2 E_l) (E_l+E_q)\big) \Bigg]  \\
	+ & \hphantom{E_{q'}} \Bigg[ 8 s^2 \left(m^2 \left(3 E_l^2+12 E_l E_q+7 E_q^2\right)+E_q^2 (E_l+E_q)^2+6 m^4\right) \\
	& \hphantom{E_{q'}^2 \Bigg[} - 4 s^3 \left(3 E_l E_q+2 E_q^2+6 m^2\right)-16 m^2 s (E_l+E_q)^2  \\
	& \hphantom{E_{q'}^2 \Bigg[} \times \left(3 E_l E_q+2 E_q^2+6 m^2\right) + 48 m^4 (E_l+E_q)^4+3 s^4  \Bigg] \Bigg\} \ ,
\end{split}
$}
\label{eq:Appx_FO_C3_s3_atTerm}
\end{equation}

\begin{equation}
\mbox{\small $
\begin{split}
\frac{b_3^{(t)}}{2 a_3^{(t)}} = - \frac{1}{|\vec{l}+\vec{q}|^2} \Bigg\{ E_{q'} &\Big[ s (E_q-E_l) \Big] \\
+ &\Big[ - s \left(E_q (E_l+E_q)+2 m^2\right)+ 2 m^2 (E_l+E_q)^2+s^2/2 \Big] \Bigg\} \ .
\end{split}
$}
\label{eq:Appx_FO_C3_s3_btTerm}
\end{equation}

Once these are replaced into \eqref{eq:Appx_FO_C3_s3_dtIdent}, we can perform the integral in $E_{q'}$ using

\begin{equation}
\begin{cases}
\int_{R_1}^{R_2} dE_{q'} = |\vec{q}+\vec{l}| \sqrt{1 - 4m^2/s} \\
\int_{R_1}^{R_2} dE_{q'} E_{q'} = |\vec{q}+\vec{l}| \sqrt{1 - 4m^2/s} (E_q+E_l) / 2\\
\int_{R_1}^{R_2} dE_{q'} E_{q'}^2 =  |\vec{q}+\vec{l}| \sqrt{1 - 4m^2/s} \big[ (E_q+E_l)^2/4 + (1-4m^2/s)|\vec{q}+\vec{l}|^2/12 \big]
\end{cases} \ .
\label{eq:Appx_FO_C3_s3_EqpIntProp}
\end{equation}

\noindent In this way we arrive at the following expression for the integral $\mathcal{C}_3$:

\begin{equation}
\mathcal{C}_3 = - \frac{g_i^3 F(q)}{32 (2\pi)^3 E_q q} \int dE_l ds f_0 (E_l) \left\{ \sqrt{1-\frac{4m^2}{s}} \left[ \frac{1}{3} A_t (s-4m^2)^2 + \frac{1}{2} B_t (s-4m^2) +C_t \right] \right\} \ ,
\label{eq:Appx_FO_C3_s3_intFinal}
\end{equation}

\noindent which can be recast into a more compact notation:

\begin{equation}
\mathcal{C}_3 = - \frac{g_i^3 F(\vec{q},\vec{k},\tau)}{32 (2\pi)^3 E_q q} \int dE_l ds f_0 (E_l,\tau) \chi (s)\ ,
\label{eq:Appx_FO_C3_s3_ChiDef}
\end{equation}

\noindent where the function $\chi$ is defined in \eqref{eq:FO_C3_s3_ChiDef}, and we explicitly wrote the arguments of each function. Now, in this particular integral we could in fact formally integrate in s to obtain only a last integral over $E_l$ of the background DF times an arbitrary function. We leave this as it is, because this form will become practical in calculating the next collision integral $\mathcal{C}_2$. It becomes quite remarkable that the $\chi$ function only depends on $s$ (which in turn is only proportional to the CoM energy). This might suggest that a more straightforward method  of obtaining these integrals may be available.

\medskip

\subsection{Calculation of $\mathcal{C}_2[f]$}
\label{sec:Appx_FO_C2}

The calculations for the $\mathcal{C}_2$ term are identical to the ones developed in the previous section. The only difference here is that the roles of the background and perturbed DF are reversed. This can easily be seen from the definition of the term in \eqref{eq:FO_Collision_SI}. So, the final expression for the integral is

\begin{equation}
\mathcal{C}_2 = - \frac{g_i^3 f_0 (E_q,\tau) }{32 (2\pi)^3 E_q q} \int dE_l ds F(\vec{l},\vec{k},\tau) \chi (s) \ ,
\label{eq:Appx_FO_C2_intDef}
\end{equation}

\noindent where we have implicitly used $\vec{l} \equiv \vec{l}(E_l,s)$. 
This is true given the parametrization \eqref{eq:Appx_FO_C3_s1_param} for this vector, as its magnitude can be uniquely determined by $E_l$ and its angle with the vector $\vec{q}$ by the Mandelstam variable $s$.
Therefore, unlike the previous case, the integral on $s$ cannot be performed, since $F$ also depends on it.

\subsection{Calculation of $\mathcal{C}_1[f]$}
\label{sec:Appx_FO_C1}

The procedure for calculating $\mathcal{C}_1$ are very similar to the ones used to calculate $\mathcal{C}_3$, but with some changes in the parametrization of the vectors involved. Following from the expression \eqref{eq:FO_Collision_SI}, the full integral can be expressed as

\begin{equation}
\mathcal{C}_1 [f] = \frac{2 g_i^3}{E_q (2\pi)^5} \int \frac{d^3 l}{2 E_{l}} \frac{d^3 q'}{2 E_{q'}} \frac{d^3 l'}{2 E_{l'}} |\mathcal{M}|^2 \delta_D^{(4)}(\mathbf{q}+\mathbf{l}-\mathbf{q'}-\mathbf{l'}) f_0 (q') F(\vec{l'}) \ .
\label{eq:Appx_FO_C1_intDef}
\end{equation}

\tocless\subsubsection{Solving for $\vec{l}$ and $\beta$}

Since the integrand does not depend on $\vec{l}$, we perform the first integral using the momentum conservation Dirac delta along this variable (instead of $\vec{l'}$ as it was used before). After doing this, we use a new parametrization that reflects our choice of remaining variables:

\begin{equation}
\begin{cases}
\vec{q} = q (0,0,1) \\
\vec{q'} = q' (0, \sin \theta, \cos \theta) \\
\vec{l'} = l' (\sin \beta \sin \alpha , \cos \beta \sin \alpha , \cos \alpha )
\end{cases} \ .
\label{eq:Appx_FO_C1_s1_param}
\end{equation}

Following the steps on section \ref{sec:Appx_FO_C3}, we reach the following expression:

\begin{equation}
\begin{split}
\mathcal{C}_1 [f] = & \frac{g_i^3}{ 4 E_q (2\pi)^4} \int d(\cos \theta) d(\cos \alpha) dl' dq' \frac{{q'}^2}{E_{q'}} \frac{{l'}^2}{E_{l'}} f_0 (l') F(\vec{q'}) \Theta (E_{q'} + E_{l'} - E_q) |\mathcal{M}|^2 \ \\
\times & \int_0^{2\pi} d\beta \delta_D(g(\vec{q'},\vec{l'},\vec{q})) \ .
\end{split}
\label{eq:Appx_FO_C1_s1_intDeltaG}
\end{equation}

\noindent where this time the argument of the remaining Dirac delta $g$ can be expressed as

\begin{equation}
\begin{split}
g \equiv & \  2m^2 + (2E_{q'}E_{l'} - 2E_{l'}E_q - 2E_qE_{q'} ) - 2q'l'(\sin \alpha \cos \beta \sin \theta + \cos \alpha \cos \theta)\\
& + 2ql' \cos \alpha + 2qq' \cos \theta \ .
\end{split}
\label{eq:Appx_FO_C1_s1_gParam}
\end{equation}

Again, as the $\beta$ integral only involves the Dirac delta we can rewrite this integral as in equations \eqref{eq:Appx_FO_C3_s1_dbetaG}, \eqref{eq:Appx_FO_C3_s1_dgdbeta/cosbetai} where in this case the functions involved are:

\begin{align}
\frac{\partial g}{\partial \beta} = 2q'l' \sin \alpha \sin \beta \sin \theta \ , \\
\begin{split}
\cos \beta_i = (2q'l' \sin \theta \sin \alpha)^{-1} \Big[ &  2m^2 + (2E_{q'}E_{l'} - 2E_{l'}E_q - 2E_qE_{q'} ) - 2q'l' \cos \alpha \cos \theta  \\
& + 2ql' \cos \alpha + 2qq' \cos \theta \Big] \ .
\end{split}
\label{eq:Appx_FO_C1_s1_dgdbeta/cosbetai}
\end{align}

\noindent We also add here a Heaviside theta function in $\cos^2 \beta_i$ to ensure the condition $|\cos \beta_i| <1$, but this time with the argument

\begin{equation}
\left| \frac{\partial g}{\partial \beta} \right|^2_{\cos \beta_i} = a_1^{(\alpha)} \cos^2 \alpha +b_1^{(\alpha)} \cos \alpha + c_1^{(\alpha)} \ ,
\label{eq:Appx_FO_C1_s1_coeffDgDbeta_def} 
\end{equation}

\noindent with coefficients:

\begin{equation}
\begin{cases}
a_1^{(\alpha)} = -4 {l'}^2 |\vec{q} - \vec{q'}|^2 \\
b_1^{(\alpha)} = 8 l' (q'\cos \theta - q) \big[ m^2 +E_{q'} E_{l'} -E_q E_{l'} - E_{q'} E_q +q q' \cos \theta  \big] \\
c_1^{(\alpha)} = 4  \Big\{ {q'}^2 {l'}^2 \sin^2 \theta - \big[ m^2 +E_{q'} E_{l'} -E_q E_{l'} - E_{q'} E_q +q q' \cos \theta \big]^2 \Big\}
\end{cases} \ .
\label{eq:Appx_FO_C1_s1_coeffDgDbeta_expr}
\end{equation}

\noindent Then the $\mathcal{C}_1$ integral results:

\begin{equation}
\begin{split}
\mathcal{C}_1 [f] = &\frac{g_i^3}{4 E_q (2\pi)^4}  \int d(\cos \theta) dq'  \frac{{q'}^2}{E_{q'}} F(\vec{q'}) \int dl' \frac{{l'}^2}{E_{l'}} f_0 (l')  \Theta (E_{q'} + E_{l'} - E_q)  \\
\times & \int d(\cos \alpha) \frac{|\mathcal{M}|^2}{\sqrt{a_1^{(\alpha)} \cos^2 \alpha +b_1^{(\alpha)} \cos \alpha + c_1^{(\alpha)} }} \Theta(a_1^{(\alpha)} \cos^2 \alpha +b_1^{(\alpha)} \cos \alpha + c_1^{(\alpha)} ) \ .
\end{split}
\label{eq:Appx_FO_C1_s1_int2Theta}
\end{equation}

\noindent The argument for the last theta function is a second degree polynomial, this time in $\cos \alpha$, with negative leading coefficient. As before, we solve this condition by imposing integration limits in the $\cos \alpha$ integral and adding a new Heaviside theta function to ensure the existence of real roots for the polynomial \eqref{eq:Appx_FO_C1_s1_coeffDgDbeta_def}. The discriminant of this polynomial becomes:

\begin{equation}
\begin{split}
\left(b_1^{(\alpha)}\right)^2 -4 a_1^{(\alpha)} c_1^{(\alpha)} = \  64 {l'}^2 {q'}^2 \sin^2 \theta & E_{q'} E_q (1 - \cos z) \Bigg\{ 2E_{l'}^2 + 2(E_{q'}-E_q) E_{l'} \\
& - \Bigg[ E_q E_{q'} (1-\cos z) + \frac{m^2 |\vec{q}-\vec{q'}|^2}{E_q E_{q'} (1-\cos z)} \Bigg] \Bigg\} \ ,
\end{split}
\label{eq:Appx_FO_C1_s1_discExpr}
\end{equation}

\noindent where

\begin{equation}
1-\cos z = 1 - \frac{m^2}{E_q E_{q'}} - \frac{q q'}{E_q E_{q'}} \cos \theta \ .
\label{eq:Appx_FO_C1_s1_coszDef}
\end{equation}

\noindent The argument of the function again has four real roots in $E_{l'}$ : $\{-m,+m,R_1,R_2\}$ with $R_{1,2}$ given by

\begin{equation}
R_{1,2} = \frac{1}{2} \left\{ E_q - E_{q'} \pm |\vec{q}-\vec{q'}|\sqrt{1 + \frac{2m^2}{E_q E_{q'} (1-\cos z)}} \right\} \ .
\label{eq:Appx_FO_C1_s1_discRoots}
\end{equation}

\noindent As before, we can discard any roots smaller than $m$ as non physical. This time however, the ordering of these roots is different, as we will see below once we express these roots in terms of Mandelstam variables.

\tocless\subsubsection{Mandelstam variables for $\mathcal{C}_1$}

So again, in order to assess the ordering of the roots $R_{1,2}$, as well as facilitating the integration of the collision kernels, we make use of the Mandelstam variables. The relevant variables for the integrals here are $t$ and $u$. Their expressions are given in \eqref{eq:Appx_FO_C3_s2_MandDef}, however as we have changed the parametrization of the vectors themselves for this collision term, we need their expressions in terms of the momentum variables:

\begin{equation}
\begin{cases}
t = (\mathbf{q}-\mathbf{q'})^2 = -2 E_q E_{q'} + 2 m^2 + 2 q q' \cos \theta = -2 E_q E_{q'} (1-\cos z) \\
u = (\mathbf{q}-\mathbf{l'})^2 = -2 E_q E_{l'} + 2 m^2 + 2 q l' \cos \alpha 
\end{cases} \ .
\label{eq:Appx_FO_C1_s2_MandParam}
\end{equation}

Now, the roots of $\left( b_1^{(\alpha)} \right) ^2 -4 a_1^{(\alpha)} c_1^{(\alpha)}$ in terms of these new variables is

\begin{equation}
R_{1,2} = \frac{1}{2} \left\{ E_q - E_{q'} \pm |\vec{q}-\vec{q'}|\sqrt{1 - \frac{4m^2}{t}} \right\} \ .
\label{eq:Appx_FO_C1_s2_discrRoots}
\end{equation}

\noindent We can see that the argument in the square root is always positive, because $t$ is always negative. To find out the order of the roots, we start by noticing the following three conditions:

\begin{enumerate}
\item Thanks to the fact that $t$ is spacelike, we can infer that $|E_q-E_{q'}| < |\vec{q}-\vec{q'}|$. 
\item As $t$ is negative, $\sqrt{1-4m^2/t}>1$.
\item Thus $R_1\leq 0$ for any ordering of $E_q$, $E_{q'}$. \hfill $\blacksquare$
\end{enumerate}

\noindent Then, as $R_1$ is negative, it is not a physical value for the energy integral. The proof that $R_2 \geq m^2$, hence $R_2$ is physical, is slightly more complicated but possible, and is left to the reader. It is important to note in this case the ordering of these roots in order to understand the allowed ranges for $E_{l'}$. If we focus on the discriminant \eqref{eq:Appx_FO_C1_s1_discExpr} we see that, as before, it has four roots but in this case the value is positive in the limit $E_{l'} \rightarrow \infty$. So, as the ordering of all four simple roots is in this case $\left[ \left(-m \leq R_1\right) \mathrm{ or } \left( R_1 \leq -m \right) \right] \leq m \leq R_2$, we can conclude that the only range of energies where both $E_{l'} \geq m$ and the discriminant is positive is $E_{l'} \geq R_2$. So, for $\mathcal{C}_1$ we have boundary conditions for $E_{l'}$ that are completely different from the ones for $\mathcal{C}_2$ and $\mathcal{C}_3$. We can then use these limits in energy for the full integral but first we need to change variables to Mandelstam variables to perform them, these time using $(t,u)$.

\tocless\subsubsection{Change of variables to $\{t,u,E_{q'},E_{l'}\}$}

So, we now make the appropriate change of variables proceeding in a very similar way as in the case of $\mathcal{C}_3$. In fact, the expression for the variables and differentials in $E_{q'}$ and $t$ are exactly the same as \eqref{eq:Appx_FO_C3_s3_COV}, \eqref{eq:Appx_FO_C3_s3_measure}. It is therefore immediate to see that the change of variables to $E_{l'}$ and $u$ is expressed as:

\begin{equation}
\begin{cases}
q' \rightarrow E_{q'} = \sqrt{{q'}^2+m^2} \\
l' \rightarrow E_{l'} = \sqrt{{l'}^2+m^2} \\
\cos \alpha \rightarrow u = -2 E_q E_{l'} \left( 1 - \frac{m^2}{E_q E_{l'}} - \frac{ql'}{E_q E_{l'}}\cos \alpha \right) \\
\cos \theta \rightarrow t = -2 E_q E_{q'} \left( 1 - \frac{m^2}{E_q E_{q'}} - \frac{qq'}{E_q E_{q'}}\cos \theta  \right) \\
\end{cases} \ ,
\label{eq:Appx_FO_C1_s3_COV}
\end{equation} 

\noindent with the corresponding integration measures given by

\begin{equation}
dq' d \cos \theta = \frac{E_{q'}}{2 q {q'}^2} dE_{q'} du \quad , \quad d{l'} d \cos \alpha = \frac{E_{l'}}{2 q {l'}^2} dE_{l'} du \ .
\label{eq:Appx_FO_C1_s3_measure}
\end{equation}

Now, we had expressed $|\mathcal{M}|^2$ in terms of only $(s,t)$ in \eqref{eq:Appx_FO_C3_s3_M2Def}. However, this time $s$ is not a relevant variable of integration. We can use the relation $s+t+u=4 m^2$ to obtain an analogous expression in variables $(t,u)$:

\begin{equation}
|\mathcal{M}|^2 = A_u u^2 +B_u u +C_u \ ,
\label{eq:Appx_FO_C1_s3_M2Coeff}
\end{equation}

\noindent with the $\{A_u,B_u,C_u\}$ coefficients only depending on $t$. So, the $\mathcal{C}_1$ collision integral can be expressed as:

\begin{equation}
\mathcal{C}_1 = \frac{g_i^3}{8 (2\pi)^4 E_q q^2} \int dE_{q'} dt F (E_{q'},t) \int_{R_2}^{\infty} dE_{l'} f(E_{l'}) \int_{u(y_1) \equiv u_1}^{u(y_2) \equiv u_2} du \frac{|\mathcal{M}|^2 (u,t) }{\sqrt{\left| \frac{\partial g}{\partial \beta} \right|^2_{\cos \beta_i}}} \ ,
\label{eq:Appx_FO_C1_s3_intDuDt}
\end{equation}

\noindent with $y_{1,2}$ the roots of $|\partial g /\partial \beta|^2_{\cos \beta_i}$ in $\cos \alpha$. This expression depends on $|\partial g /\partial \beta|^2_{\cos \beta_i}$, this time given by \eqref{eq:Appx_FO_C1_s1_coeffDgDbeta_def}, which can be recast into a polynomial in $u$ as:

\begin{equation}
\left| \frac{\partial g}{\partial \beta} \right|^2_{\cos \beta_i} = a_1^{(u)} u^2 + b_1^{(u)} u + c_1^{(u)} \ ,
\label{eq:Appx_FO_C1_s3_coeffDgDbeta_def}
\end{equation}

\noindent with coefficients:

\begin{equation}
\begin{cases}
a_1^{(u)} = - |\vec{q}-\vec{q'}|^2/q^2 \\
b_1^{(u)} = -2 [ t (E_q + E_{q'}) (E_q - E_{l'}) + 2 q^2 {q'}^2 \sin^2 \theta ] / q^2  \\
c_1^{(u)} = t (E_q - E_{l'})^2 (4m^2 - t) / q^2
\end{cases} \ .
\label{eq:Appx_FO_C1_s3_coeffDgDbeta_expr}
\end{equation}

The integral in $u$ can then be computed using the property \eqref{eq:Appx_FO_C3_s3_EqpIntProp}, so that the full collision integral becomes

\begin{equation}
\begin{split}
 \mathcal{C}_1 = \frac{g_i^3}{8 (2\pi)^4 E_q q} & \int dE_{q'} dt F (E_{q'},t) \int_{R_2}^{\infty} dE_{l'} f_0(E_{l'}) \frac{\pi}{|\vec{q}-\vec{q'}|}  \\
& \times \left[ A_u \left( \frac{3 \left( b_1^{(u)} \right)^2 - 4 a_1^{(u)} c_1^{(u)}}{8 \left( a_1^{(u)} \right)^2 } \right) - B_u \left( \frac{b_1^{(u)}}{2 a_1^{(u)}} \right) + C_u \right] \ .
\end{split}
\label{eq:Appx_FO_C1_s3_intFormalSol}
\end{equation}

The expressions for the coefficients accompanying $\{ A_u, B_u \}$ can be expressed in terms of powers of $E_{l'}$, in order to facilitate integration:

\begin{equation}
\mbox{\small $
\begin{split}
\frac{3 \left( b_1^{(u)} \right)^2 - 4 a_1^{(u)} c_1^{(u)}}{8 \left( a_1^{(u)} \right)^2 } = \frac{1}{8 |\vec{q} - \vec{q'}|^4} \Bigg\{ & E_{l'}^2 \Bigg[ 4 t \left(3 (E_q + E_{q'})^2 t - ((E_q - E_{q'})^2 - t) (-4 m^2 + t)\right)\Bigg]  \\
	+ & E_{l'} \Bigg[ 4 t (4 (E_q - E_{q'})^2 (E_q + 3 E_{q'}) m^2  \\
	  & \hphantom{E_{l'} \Bigg[} - 4 (E_q (E_q - E_{q'}) (E_q + 2 E_{q'}) + (E_q + 3 E_{q'}) m^2) t + (E_q + 3 E_{q'}) t^2) \Bigg]  \\
	+ & \hphantom{E_{l'}} \Bigg[ (48 (E_q - E_{q'})^4 m^4 - 16 (E_q - E_{q'})^2 m^2 (2 E_q^2 - 3 E_q E_{q'} + 6 m^2) t  \\
	  & \hphantom{E_{l'} \Bigg[} + 8 (E_q^2 (E_q - E_{q'})^2 + (7 E_q^2 - 12 E_q E_{q'} + 3 E_{q'}^2) m^2 + 6 m^4) t^2  \\ 
	  & \hphantom{E_{l'} \Bigg[} - 4 (2 E_q^2 - 3 E_q E_{q'} + 6 m^2) t^3 + 3 t^4)  \Bigg] \Bigg\} \ ,
\end{split}
$}
\label{eq:Appx_FO_C1_s3_auTerm}
\end{equation}

\begin{equation}
\mbox{\small $
\begin{split}
\frac{b_1^{(u)}}{2 a_1^{(u)}} = \frac{-1}{|\vec{q}-\vec{q'}|^2} \Bigg\{ E_{l'} &\Big[ t (E_q+E_{q'}) \Big] \\
+ &\Big[ 2 (E_q - E_{q'})^2 m^2 + 2 E_q (-E_q + E_{q'}) t - 4 m^2 t + t^2 \Big] \Bigg\} \ .
\end{split}
$}
\label{eq:Appx_FO_C1_s3_buTerm}
\end{equation}

\noindent Here the approach diverges greatly from the one we took for $\mathcal{C}_2$ and $\mathcal{C}_3$. In the previous cases we managed to also perform the integration in energy, but here this cannot be done without knowing the background DF $f_0$. Since given $f_0$, the integration in $E_{l'}$ can be performed, it is convenient to define

\begin{equation}
\left< f_0 \right> _n (E_q , E_{q'}, t , \tau) = \int_{R_2}^{\infty} d E_{l'} f_0 (E_{l'}, \tau) E_{l'}^{n} \ ,
\label{eq:Appx_FO_C1_s3_f0MeanDef}
\end{equation}

\noindent which is a function of $(E_q,E_{q'},t)$ only through $R_{2}$, and express the full collision integral in terms of these moments of $f_0$:

\begin{equation}
\mbox{\small $
\begin{split}
\mathcal{C}_1 = \frac{\pi}{(2\pi)^4 E_q q} & \int dE_{q'} dt F (E_{q'},t)  \\
\times \Bigg\{ \frac{A_u}{8 |\vec{q} - \vec{q'}|^5} \Bigg\{ & \left< f_0 \right>_2 \Bigg[ 4 t \left(3 (E_q + E_{q'})^2 t - ((E_q - E_{q'})^2 - t) (-4 m^2 + t)\right)\Bigg]  \\
	+ & \left< f_0 \right>_1 \Bigg[ 4 t (4 (E_q - E_{q'})^2 (E_q + 3 E_{q'}) m^2  \\
	  & \hphantom{\left< f_0 \right>_1 \Bigg[} - 4 (E_q (E_q - E_{q'}) (E_q + 2 E_{q'}) + (E_q + 3 E_{q'}) m^2) t + (E_q + 3 E_{q'}) t^2) \Bigg]  \\
	+ & \left< f_0 \right>_0 \Bigg[ (48 (E_q - E_{q'})^4 m^4 - 16 (E_q - E_{q'})^2 m^2 (2 E_q^2 - 3 E_q E_{q'} + 6 m^2) t  \\
	  & \hphantom{\left< f_0 \right>_0 \Bigg[} + 8 (E_q^2 (E_q - E_{q'})^2 + (7 E_q^2 - 12 E_q E_{q'} + 3 E_{q'}^2) m^2 + 6 m^4) t^2  \\ 
	  & \hphantom{\left< f_0 \right>_0 \Bigg[} - 4 (2 E_q^2 - 3 E_q E_{q'} + 6 m^2) t^3 + 3 t^4)  \Bigg] \Bigg\}  \\
	  +\frac{B_u}{2 |\vec{q} - \vec{q'}|^3} \Bigg\{ & \left< f_0 \right>_1 \Big[ t (E_q+E_{q'}) \Big]+ \left< f_0 \right>_0 \Big[ 2 (E_q - E_{q'})^2 m^2 + 2 E_q (-E_q + E_{q'}) t - 4 m^2 t + t^2 \Big] \Bigg\}  \\
	  + \frac{C_u}{|\vec{q}-\vec{q'}|} \hphantom{\Bigg\{ } & \left< f_0 \right>_0 \quad \Bigg\} \ .
\end{split}
$}
\label{eq:Appx_FO_C1_s3_intFinal}
\end{equation}

We can recast this expression in terms of a (fairly complex but mostly polynomial) integration kernel $K(E_q, E_{q'}, t, \tau)$.

\begin{equation}
\mathcal{C}_1 = \frac{g_i^3}{16 (2\pi)^3 E_q q} \int dE_{q'} dt F (E_{q'},t) \times K(E_q, E_{q'}, t, \tau)\,.
\label{eq:Appx_FO_C1_s3_KDef}
\end{equation}

This integration kernel is the most complex part of the collision term, mainly due to the explicit dependence of the integration kernel on time through the momenta of the background DF.

\section{The Zero Order Collision Term} 
\label{sec:Appx_ZO}

In this appendix we provide details on the calculation of the zero-order collision integral. As shown in \eqref{eq:ZO_DDef} this can be split into two parts: $\mathcal{D}_1[f]$ and $\mathcal{D}_2[f]$.
The treatment of the term $\mathcal{D}_2[f]$ mimics exactly the one for $\mathcal{C}_3[f]$ but with the simplification $F(\vec{q},\vec{k},\tau) \rightarrow f_0(E_q)$. Thus, it is immediate to see that this term can be expressed as in \eqref{eq:ZO_D2_intFinal}.

\subsection{Calculation of $\mathcal{D}_1[f]$}
\label{sec:Appx_ZO_D1}

The term $\mathcal{D}_1[f]$ holds some similarity to the term $\mathcal{C}_1$, as it involves the   integration of both of the vectors $\vec{q'}$, $\vec{l'}$ on which the distribution functions are evaluated. The derivation for this term closely follows the one for the perturbed DF until equation \eqref{eq:Appx_FO_C1_s3_intFormalSol}. In this case, the result can be written as

\begin{equation}
\mathcal{D}_1[f] = \frac{g_i^3}{16 E_q q (2\pi)^3} \int dt dE_{q'} f_0(E_{q'}) \int_{R_2}^{\infty} dE_{l'} \frac{f_0(E_{l'})}{|\vec{q}-\vec{q'}|}\left\{ A_u [...] + B_u [...] + C_u [...]  \right\} \ ,
\label{eq:Appx_ZO_D1_intFormalSol}
\end{equation}

\noindent where the terms accompanying the factors $\{A_u,B_u,c_u\}$ are specified in equations \eqref{eq:Appx_FO_C1_s3_auTerm}, \eqref{eq:Appx_FO_C1_s3_buTerm}. Continuing the procedure as we did for $\mathcal{C}_1$ will require the moments for the background distribution function, which is exactly the quantity we are trying to obtain. Here instead of doing that, we will express this integral in such a way that it depends on the integral of a function times the background DF itself: we want to obtain a convenient expression for the integration kernel to be used in the equation for $f_0$. To do that, we will integrate in $t$, given that the background DF does not depend explicitly on this variable. A complication in this approach comes from the fact that a simple modification of the order of integrals will not work, as the boundary for the  integral in $E_{l'}$, $R_2$, depends explicitly on $t$. In order to circumvent this, let us first define the variable $P$

\begin{equation}
P = |\vec{q}-\vec{q'}|\sqrt{1-\frac{4m^2}{t}} \ .
\label{eq:Appx_ZO_D1_PDef}
\end{equation}

\noindent We change variables from $t$ to this new variable $P$ in the integral:

\begin{equation}
\int_{t(\cos \theta = -1)}^{t(\cos \theta = 1)} dt (...) = 2 \int_{|E_q-E_{q'}|+2m}^{E_q+E_{q'}} dP \frac{2Pt^2}{|4m^2(E_q-E_{q'})^2-t^2|} (...) \ .
\label{eq:Appx_ZO_D1_PMeasure}
\end{equation}

\noindent Now, we rewrite the integration limits as a series of Heaviside theta functions

\begin{equation}
\int_{|E_q-E_{q'}|+2m}^{E_q+E_{q'}} dP \int_{R_2}^{\infty} dE_{l'} = \int dP dE_{l'} \Theta (E_{l'}-R_2)\Theta( P - ( |E_q-E_{q'}|+2m ) ) \Theta (E_q+E_{q'} - P ) \ .
\label{eq:Appx_ZO_D1_limitsTheta_expr} 
\end{equation}

\noindent Then, using the relations

\begin{align}
\begin{split}
\Theta (E_q+&E_{q'} - P ) \Theta(2E_{l'} -E_q +E_{q'} - P ) = \\
&\Theta (E_{l'} - E_q ) \Theta ( E_q+E_{q'} - P ) + \Theta (E_q - E_{l'}) \Theta (2E_{l'} -E_q +E_{q'} - P) \ ,
\end{split} \\
\begin{split}
\Theta(2E_{l'} &-E_q +E_{q'} - P ) \Theta( P - |E_q-E_{q'}| - 2m ) = \\
&\Theta (E_q - E_{q'}) \Theta(2E_{l'} -E_q +E_{q'} - P ) \Theta( P - E_q + E_{q'} - 2m ) \\
& + \Theta (E_{q'} - E_q) \Theta(2E_{l'} -E_q +E_{q'} - P ) \Theta( P - E_{q'} + E_{q'} - 2m ) \ ,
\end{split}
\label{eq:Appx_ZO_D1_limitsTheta_props}
\end{align}

\noindent it can be found out that the integration splits into four parts:

\begin{equation}
\begin{split}
&\int_{m}^{E_q} dE_{q'} \left[ \int_{E_q-E_{q'}+m}^{E_q} dE_{l'} \int_{E_q-E_{q'}+2m}^{2E_{l'}-E_q+E_{q'}} dP + \int_{E_q}^{\infty} dE_{l'} \int_{E_q-E_{q'}+2m}^{E_q+E_{q'}} dP \right] \\
& + \int_{E_q}^{\infty} dE_{q'} \left[ \int_{m}^{E_q} dE_{l'} \int_{E_{q'}-E_q+2m}^{2E_{l'}-E_q+E_{q'}} dP + \int_{E_q}^{\infty} dE_{l'} \int_{E_{q'}-E_q+2m}^{E_q+E_{q'}} dP \right] \ .
\end{split} 
\label{eq:Appx_ZO_D1_limitsTheta_final}
\end{equation}

\noindent Here, each of the four parts is first integrated in an angular variable and then in the energies. Using this procedure for the integral \eqref{eq:Appx_ZO_D1_intFormalSol} the collision integral can be expressed as:

\begin{equation}
\mathcal{D}_1[f] = \frac{g_i^3}{16 E_q q (2\pi)^3} \sum_{i=1}^{4} \int_{\mathcal{I}_i} dE_{q'} dE_{l'} f_0(E_{q'}, \tau) f_0(E_{l'}, \tau) k_i (E_q, E_{q'}, E_{l'}, \tau) \ ,
\label{eq:Appx_ZO_D1_intKDef}
\end{equation}

\noindent where $\mathcal{I}_i$ refers to the integration method in energies specified in \eqref{eq:Appx_ZO_D1_limitsTheta_final} and defined in \eqref{eq:ZO_D1_limitsEnergy_final} and the kernel functions $k_i$ are expressed as in \eqref{eq:ZO_D1_intKDef_expr}.

There, we have labeled the kernels with $i=1,...,4$ according to the integration regions $\mathcal{I}_i$ in the order in which they appear in equation \eqref{eq:Appx_ZO_D1_limitsTheta_final}. We have chosen to return the integrals to their original angular variable $t$ instead of $P$, as the integrands themselves are functions of $t$ and can't be expressed neatly in the new variable and the integration measure is also a function of both $P$ and $t$. The integrals labeled $i=2,4$ possess the original limits on $t$, as the integration scheme has not modified the bounds in $P$ appearing in \eqref{eq:Appx_ZO_D1_PMeasure}. The ones labeled $i=1,3$ are the (two) solutions of the equation $P(t_{1,2})=2E_{l'}-E_q+E_{q'}$. The integrands themselves are expressed in equation \eqref{eq:ZO_D1_intFinal_kappadef_big} on a similar way as in equation \eqref{eq:Appx_FO_C1_s3_intFinal}.

\section{The DM-DM collision term for a Massless Mediator}
\label{sec:Appx_Massless}

Here, we will give the form of the DM-DM collision term for the case of an actually massless mediator, out of the models considered in \cref{sec:Solutions_Models} and the ansatz \eqref{eq:FO_C3_s3_M2Def}. In this case, we should start by noting that, as the mediator population is not Boltzmann suppressed, this is one of the three relevant collision terms in the full evolution of the system: not only the population and DF of the DM need to be tracked but also of the mediators (hereby called DR, Dark radiation) and the other two collision terms between the DR-DR and DR-DM. 
In this appendix we calculate only the DM-DM collision term.

We consider DM scattering of light or massless mediator particles under the Lagrangian \eqref{eq:Models_Scalar_Lagrangian}. We start by considering the scattering amplitude for such a model in the massive DM particle case \cite{Srednicki2007}

\begin{equation}
|\mathcal{M}|^2 = 6\mathfrak{g}^4 + 16m^4\mathfrak{g}^4 \left( \frac{1}{s^2}+ \frac{1}{t^2} + \frac{1}{u^2} \right) - \frac{4 m^2 \mathfrak{g}^4 (s+t+u)^2 }{stu} \ .
\label{eq:Appx_DR_Msquared}
\end{equation}

We can see from this expression clearly that in the limit $m \rightarrow 0$ the amplitude reduces to \eqref{eq:Models_s0_Msquare}. As $p/m \lesssim 1$ (with $p$ the typical momentum of the particles in the CoM frame) the other terms in the scattering amplitude become more relevant, and the massless DM approximation becomes invalid. As the particle becomes highly nonrelativistic $p/m \ll 1$ the assumption of tree level diagrams being dominant breaks down as ladder diagrams become more relevant and other approaches are more well suited for the analysis, such as considering Sommerfeld Enhancement \cite{Sommerfeld_1931,Feng_2010}.

As in \cref{sec:Appx_FO} we will start by calculating $\mathcal{C}_3$ first, then calculate $\mathcal{C}_1$ on a similar fashion and rely on the derivations in \cref{sec:Appx_ZO} to relate these results to the ones for $D_1$ and $D_2$.

\addvspace{0.5cm}

\tocless{\subsection{Calculation of $\mathcal{C}_3[f]$}}

For the $\mathcal{C}_3[f]$ term, the derivations in \cref{sec:Appx_FO_C3} remain the same until the specific form of the ansatz had been used in \eqref{eq:Appx_FO_C3_s3_intIdent}. For the sake of readability, we will split the scattering amplitude \eqref{eq:Appx_DR_Msquared} into three parts: one containing only terms on $s$ (constant on the $t$ integration), and two others containing mixed $\{s,t\}$ terms and $\{s,u\}$ terms respectively

\begin{equation}
\begin{split}
|\mathcal{M}|^2 &= \mathfrak{g}^4 \left\{ \left[ 6 + \frac{16m^4}{s^2} \right] + \left[  \frac{16m^4}{t^2} + \frac{64m^6}{s t (s-4m^2)} \right] + \left[ \frac{16m^4}{u^2} + \frac{64m^6}{s u (s-4m^2)} \right] \right\} \\
&= |\mathcal{M}|^2_{\{s\}} + |\mathcal{M}|^2_{\{s,t\}} + |\mathcal{M}|^2_{\{s,u\}} \ .
\end{split}
\label{eq:Appx_DR_Msquared_C3split}
\end{equation}

The $s$ term can be integrated directly using the result \eqref{eq:Appx_FO_C3_s3_intFinal}, while the other terms require special care. We start by the mixed $\{s,t\}$ term: the integral can be performed directly

\begin{equation}
{\small
\begin{split}
&\int_{t_1}^{t_2} dt \frac{\left| \mathcal{M} \right|_{\{s,t\}}^2}{\sqrt{a_3^{(t)} t^2 + b_3^{(t)} t + c_3^{(t)}}} = \\
&\frac{8 \pi  m^4 \mathfrak{g}^4 \left(-4 m^2 \left(E_l^2+2 E_l E_q+3 E_q^2-4 E_q E_{q'}+2 E_{q'}^2\right)+2 s
   \left(E_l (E_q+E_{q'})+E_q^2-E_q E_{q'}+2 m^2\right)-s^2\right)}{s \left(4 m^2-s\right) \left|
   E_q-E_{q'}\right| ^3 \sqrt{\frac{s \left(s-4 m^2\right)}{E_q^2-m^2}}} \ .
\end{split}
}
\label{eq:Appx_DR_C3_tIntegratedtermT}
\end{equation}

In the case of the $\{s,u\}$ term integral, it is better to change variables in integral \eqref{eq:Appx_FO_C3_s3_intDMand} from $t$ to $u$. By making use of definition \eqref{eq:Appx_FO_C3_s3_coeffDgDbeta_def} and that $s+t+u=4m^2$ we can also perform the integral directly 

\begin{equation}
{\small
\begin{split}
&\int_{t_1}^{t_2} dt \frac{\left| \mathcal{M} \right|_{\{s,u\}}^2}{\sqrt{a_3^{(t)} t^2 + b_3^{(t)} t + c_3^{(t)}}} = \\
&\frac{8 \pi  m^4 \mathfrak{g}^4 \left(-4 m^2 \left(3 E_l^2 + 2 E_l ( E_q - 2 E_{q'}) + E_q^2 + 2 E_{q'}^2\right)+2 s
   \left(E_l^2+E_l (E_q-E_{q'})+E_q E_{q'}+2 m^2\right)-s^2\right)}{s \left(4 m^2-s\right) \left|
   E_l-E_{q'}\right| ^3 \sqrt{\frac{s \left(s-4 m^2\right)}{E_q^2-m^2}}} \ .
\end{split}
}
\label{eq:Appx_DR_C3_tIntegratedtermU}
\end{equation}

After the $t$ integral is done, as in \eqref{eq:Appx_FO_C3_s3_EqpIntProp} it is time to continue solving for the $E_{q'}$ integral. In the case of the $\{s,t\}$ term, it is convenient to change variables to $\overline{E}_1 = E_{q'}-E_q$. The $E_{q'}$ integral for this term then becomes

\begin{equation}
{\small
\begin{split}
&\int_{R_1}^{R_2} dE_{q'}\int_{t_1}^{t_2} dt \frac{\left| \mathcal{M} \right|_{\{s,t\}}^2}{\sqrt{a_3^{(t)} t^2 + b_3^{(t)} t + c_3^{(t)}}} = \\
& - \int_{\overline{R}_1}^{\overline{R}_2} d\overline{E}_1 \frac{8 \pi  m^4 q \mathfrak{g}^4 }{\left(\overline{E}_1^2 s \left(s-4 m^2\right)\right)^{3/2}} \left\{ -4 m^2 (E_l+E_q)^2+4 E_l E_q s+4 m^2 s-s^2 +\overline{E}_1 (2 E_l s-2 E_q s)-8 \overline{E}_1^2 m^2 \right\} \ ,
\end{split}
}
\label{eq:Appx_DR_C3_tIntegratedtermT_Eprime}
\end{equation} 

\noindent where the integration limits are

\begin{equation}
\overline{R}_{1,2} = \frac{1}{2} \left( E_l - E_q \pm \left| \vec{q} + \vec{l} \right| \sqrt{1 - \frac{4m^2}{s}} \right) \ ,
\label{eq:Appx_DR_C3_tIntegratedtermT_Eprime_Limits}
\end{equation}

\noindent and it can be demonstrated that the lower limit $\overline{R}_1$ is always negative and the upper limit $\overline{R}_2$ always positive. Remarkably, the $\{s,u\}$ term is identical upon the variable change $\overline{E}_2 = E_{q'} - E_l$.

The  above integrals in the new variable $\overline{E}$ exhibit  divergences. A proper treatment of the divergences requires the understanding of the infrared physics of  the specific model with the mediator fields. Here, we will write the divergent contributions as the following integral terms

\begin{equation}
\mathcal{I}_n = \int_{\overline{R}_1}^{\overline{R}_2} dE \frac{E^{3-n}}{\left| E \right|^3} \ .
\label{eq:Appx_DR_C3_integralEprimeResultFinal_Idef}
\end{equation}
A possible regularization is to consider that the mediator have a small mass and, following  \cite{Oldengott2014}, to introduce a (physical) mass $m_\phi$ for the mediator in the calculation of the vanishing denominator on the integrals. For a truly massless mediator case some effect, analogous to Debye shielding in plasma \cite{Buen_abad_2015,Cyr-Racine2015}, would be needed to limit the interaction range and provide an effective length scale that would act as a regulator. In view of the limited purposes of the present paper, here we leave the results in terms of the divergent integrals \eqref{eq:Appx_DR_C3_integralEprimeResultFinal_Idef}. In terms of these, the final $\mathcal{C}_3[f]$ collision term becomes

\begin{equation}
{\small
\begin{split}
&\mathcal{C}_3 = \frac{g_i^3 \mathfrak{g}^4 F(q)}{32 (2\pi)^3 E_q q} \int dE_l ds f_0 (E_l) \Bigg\{ (6+\frac{16m^4}{s^2})\sqrt{1-\frac{4m^2}{s}}  \\
& +\frac{16 \pi m^4 q}{\left[s \left(s-4m^2\right)\right]^{3/2}} \left[ \left(-4 s \left(E_l E_q+m^2\right)+4 m^2 (E_l+E_q)^2+s^2\right) \mathcal{I}_3  + 2 s (E_q-E_l) \mathcal{I}_2 - 8 m^2 \mathcal{I}_1 \right] \Bigg\} \\
& \equiv \frac{g_i^3 \mathfrak{g}^4 F(q)}{32 (2\pi)^3 E_q q} \int dE_l ds f_0 (E_l) \chi_{e} \ .
\end{split}
}
\label{eq:Appx_DR_C3_C3Final}
\end{equation}

We remark here that any Lorentz invariant regularization mechanism must fulfill the property $\chi_e \equiv \chi_e(s)$, as was outlined in \cref{sec:Appx_sigma} and \cite{GondoloGelmini91}.

\addvspace{0.5cm}

\tocless{\subsection{Calculation of $\mathcal{C}_1[f]$}}

For the $\mathcal{C}_1$ term we would have to perform a similar split in the scattering amplitude, however in this case the terms similar to \eqref{eq:Appx_FO_C1_s3_intFinal} would be the ones with only $t$ dependence, and we would have to calculate the $\{s,t\}$ and the $\{u,t\}$ terms. This split of \eqref{eq:Appx_DR_Msquared} can be expressed as

\begin{equation}
\begin{split}
|\mathcal{M}|^2 &= \mathfrak{g}^4 \left\{ \left[ 6 + \frac{16m^4}{t^2} \right] + \left[  \frac{16m^4}{s^2} - \frac{64m^6}{s t (4m^2-t)} \right] + \left[ \frac{16m^4}{u^2} - \frac{64m^6}{t u (4m^2-t)} \right] \right\} \\
&= |\mathcal{M}|^2_{\{t\}} + |\mathcal{M}|^2_{\{t,s\}} + |\mathcal{M}|^2_{\{t,u\}} \ .
\end{split}
\label{eq:Appx_DR_Msquared_C1split}
\end{equation}

Again we integrate directly the $|\mathcal{M}|^2_{\{t,u\}}$ term, this time on $u$, to find the expression

\begin{equation}
{\small
\begin{split}
&\int_{u_1}^{u_2} du \frac{\left| \mathcal{M} \right|_{\{t,u\}}^2}{\sqrt{a_1^{(u)} u^2 + b_1^{(u)} u + c_1^{(u)}}} = \\
& \frac{8 \pi  m^4 q \mathfrak{g}^4}{
   \left(\overline{E}_1^2 t \left(t-4 m^2\right)\right)^{3/2}}
   \left\{  \left(4 m^2 (E_q-E_{q'})^2+4 E_q E_{q'} t-4 m^2 t+t^2\right)
   - 2 \overline{E}_1 t \left(E_q+E_{q'}\right)
   + 8 \overline{E}_1^2 m^2 \right\} \ ,
\end{split}
}
\label{eq:Appx_DR_C1_uIntegratedtermU}
\end{equation}

\noindent where $\overline{E}_1 = E_{l'}-E_q$. For the term $|\mathcal{M}|^2_{\{t,s\}}$ we change integration variables from $u$ to $s$ using \eqref{eq:Appx_FO_C1_s3_coeffDgDbeta_def} and $s+t+u=4m^2$ to obtain

\begin{equation}
{\small
\begin{split}
&\int_{u_1}^{u_2} du \frac{\left| \mathcal{M} \right|_{\{s,u\}}^2}{\sqrt{a_1^{(u)} u^2 + b_1^{(u)} u + c_1^{(u)}}} = \\
& \frac{8 \pi  m^4 q \mathfrak{g}^4}{
   \left(\overline{E}_2^2 t \left(t-4 m^2\right)\right)^{3/2}}
   \left\{  \left(4 m^2 (E_q-E_{q'})^2+4 E_q E_{q'} t-4 m^2 t+t^2\right)
   - 2 \overline{E}_2 t \left(E_q+E_{q'}\right)
   + 8 \overline{E}_2^2 m^2 \right\} \ ,
\end{split}
}
\label{eq:Appx_DR_C1_uIntegratedtermS}
\end{equation}

\noindent where $\overline{E}_2= E_{l'}+E_{q'}$. In this case we can arrive to similar expressions as in \eqref{eq:Appx_DR_C3_C3Final} when we account for the divergent $\overline{E}_{1}$ integral in \eqref{eq:Appx_DR_C1_uIntegratedtermU}. For the case of the term \eqref{eq:Appx_DR_C1_uIntegratedtermS} the integrals themselves are very similar, however as $\overline{E}_2=E_{q'}+E_{l'}>2m$ there is no singularity in the energy integral and regularization is not needed. The final results for these integrals can be expressed as

\begin{equation}
{\small
\begin{split}
&\int_{R_2}^{\infty} dE_{l'} f(E_{l'}) \int_{u_1}^{u_2} du \frac{\left| \mathcal{M} \right|_{\{t,u\}}^2 + \left| \mathcal{M} \right|_{\{t,s\}}^2}{\sqrt{a_1^{(u)} u^2 + b_1^{(u)} u + c_1^{(u)}}} = \\
&\frac{8 \pi  m^4 q \mathfrak{g}^4}{t \sqrt{t \left(t-4 m^2\right)^3}} \Bigg\{
\left(-4 m^2 (E_q-E_{q'})^2-4 E_q E_{q'} t+4 m^2 t-t^2\right) \mathcal{J}_3
+2 (E_q+E_{q'}) t \mathcal{J}_2
-8 m^2 \mathcal{J}_1 \Bigg\} \ ,
\end{split}
}
\label{eq:Appx_DR_C1_integralEprimeResultFinal}
\end{equation}

\noindent where the results are expressed in terms of the integrals

\begin{equation}
\mathcal{J}_n (E_q,E_{q'},t,\tau) = \int_{-\overline{R}_3}^{\infty} dE \frac{f_0(E_q+E) E^{3-n}}{\left|E\right|^3} + \int_{\overline{R}_4}^{\infty} dE \frac{f_0(E-E_{q'})}{E^n} \ ,
\label{eq:Appx_DR_C1_integralEprimeResultFinal_Jdef}
\end{equation}

\noindent and the integration limits are defined as

\begin{equation}
\overline{R}_{3,4} = \frac{1}{2} \left( E_q + E_{q'} \pm \left| \vec{q} - \vec{q'} \right| \sqrt{1 - \frac{4m^2}{t}} \right)
\label{eq:Appx_DR_C1_integralEprimeResultFinal_intLimits}
\end{equation}

\noindent for the sum of all these integrals. Afterwards, we can express the various scattering functions as:

\begin{equation}
{\small
\begin{split}
& \mathcal{C}_1 = \frac{g_i^3 \mathfrak{g}^4}{8 (2\pi)^4 E_q q} \int dE_{q'} dt F (E_{q'},t) \\
& \Bigg\{ \frac{8 \pi  m^4 q }{t \sqrt{t \left(t-4 m^2\right)^3}} \Bigg[
\left(-4 m^2 (E_q-E_{q'})^2-4 E_q E_{q'} t+4 m^2 t-t^2\right) \mathcal{J}_3
+2 (E_q+E_{q'}) t \mathcal{J}_2
-8 m^2 \mathcal{J}_1 \Bigg] \\
& + \frac{\pi}{\left|\vec{q}-\vec{q'}\right|} \left<f_0\right> \left( 6 + \frac{16m^4}{t^2} \right) \Bigg\} \ .
\end{split}
}
\label{eq:Appx_DR_C1_C1Final}
\end{equation}

\addvspace{0.5cm}

\section{Numerical Integration of $\mathcal{D}_2[f_0^{eq}]$}
\label{sec:Appx_RTA}

In this section, we will provide the numerical integration scheme used to calculate $\tau$. As was justified in \ref{sec:RTA}, the expression for this quantity can be set in terms of the kernels introduced in \ref{sec:Solutions_Models}:

\begin{equation*}
\tau^{-1} (E_q) =  \frac{g_i^3}{32 (2\pi)^3 E_q |\vec{q}|} \int dE_l ds \chi (s) f_0 (E_l)
\label{eq:Appx_RTA_InvTau_Def}
\end{equation*}

As we are not considering Bose enhancement or Pauli blocking, we set the equilibrium DF as a relativistic Maxwell-Boltzmann distribution

\begin{equation}
f_0^{eq}(E_q, t) \equiv f^{MB} (E_q, T(t)) = e^{-E_q/T} \ ,
\label{eq:Appx_RTA_MB_Def}
\end{equation}

\noindent and the function $\chi (s)$ depends of the particular interaction model used. Now, in order to numerically integrate these equations, it is necessary to adimensionalize this expression. We define:

\begin{equation}
\tilde{\tau} \equiv m \tau \quad , \quad \epsilon_q \equiv E_q / m \quad , \quad \tilde{s} \equiv s / m^2 \quad , \quad \tilde{T} = T / m \ .
\label{eq:Appx_RTA_Adim_Def}
\end{equation}

\noindent Then, the expression for $\tilde{\tau}^{-1}$ is

\begin{equation}
\tilde{\tau}^{-1} (\epsilon_q) = \frac{g_i^3}{32 (2 \pi)^3 \epsilon_q \sqrt{\epsilon_q^2-1}}  \int_1^{\infty} d\epsilon_l f^{MB} (\epsilon_l) \int_{2\epsilon_q \epsilon_l + 2 - 2\sqrt{\epsilon_q^2-1} \sqrt{\epsilon_l^2-1}}^{2\epsilon_q \epsilon_l + 2 + 2\sqrt{\epsilon_q^2-1} \sqrt{\epsilon_l^2-1}}  d \tilde{s} \chi (\tilde{s}) \ .
\label{eq:Appx_RTA_InvTau_Adim}
\end{equation}

For any given choice of the kernel function $\chi(s)$, this double integral can be readily calculated. We will now summarize the kernels for the models considered in \ref{sec:Solutions_Models}:

\begin{align}
&\mathrm{Constant\ Amplitude:} \quad \chi_0 (s) = C_0 \sqrt{1 - \frac{4}{s}} \quad , \quad C_0 = 6 \mathfrak{g}^4  \ .\\
&\mathrm{Massive\ Scalar:} \quad \chi_m (s) = C_m \sqrt{1 - \frac{4}{s} } \left( 256 - 128 s + 19 s^2 \right) \quad , \quad C_m = \frac{g^4}{6} \left( \frac{m}{m_\Phi} \right)^4 \ . \\
&\mathrm{Vector\ Field:} \quad \chi_{V} (s) = C_V \sqrt{ 1 - \frac{4}{s} } (74 - 29s) (1 - s^2) \quad , \quad C_V = \frac{4}{3} \left( \frac{g_V m}{ m_V} \right)^4 \frac{1}{\cos ^4 \theta_W'} \ .
\label{eq:Appx_RTA_InvTau_Chi_Models}
\end{align}

For these models, the results of numerical integration are summarized in \cref{fig:RTA_Tau_Eq}. In order to obtain a relevant value of the relaxation time and compare it to a cosmological timescale, it is usual to obtain a thermal average of $\tau^{-1}$. For any quantity $g$, a thermal average is defined as

\begin{equation}
\left< g \right>_{th} = \frac{\int d^3 q \  g(\vec{q}) f^{eq} (E_q) }{\int d^3 q \ f^{eq} (E_q) } \ ,
\label{eq:Appx_RTA_ThAvg_Def}
\end{equation}

\noindent where $f^{eq}$ is an equilibrium distribution function. In particular, for the inverse relaxation time, this thermal averaging can be reduced to the following expression, accounting for adimensionalization:

\begin{equation}
\left< \tilde{\tau}^{-1} \right>_{th} = \frac{\int_1^{\infty} d\epsilon_q \epsilon_q \sqrt{\epsilon_q^2 - 1 } f^{MB}(\epsilon_q) \tilde{\tau}^{-1} (\epsilon_q)}{\int_1^{\infty} d\epsilon_q \epsilon_q \sqrt{\epsilon_q^2 - 1 } f^{MB}(\epsilon_q) } \ ,
\label{eq:Appx_RTA_InvTau_ThAvg}
\end{equation}

\noindent where we have assumed the background DF is given by \eqref{eq:Appx_RTA_MB_Def}.

\subsection{Parallelism with the Calculation of Abundances of Stable Species}
\label{sec:Appx_sigma}

Several similarities exist between the calculation of the thermal average of the relaxation time and the procedures in \cite{GondoloGelmini91}. In that paper, the thermal average of the quantity $\sigma v_{mol}$, i.e. the cross section times the relative M\o ller velocity for elastic $4$ fermion interaction, is calculated. This quantity differs from the relaxation time only in the normalization, and a parallelism between the original quantity $\sigma v_{mol} (\Theta, s)$ and $\chi(s)$ can be obtained.

The definition of $\sigma v_{mol}$ from \cite{DeGroot1980} can be given through the full Boltzmann equation:

\begin{equation}
(\partial_t  + \mathbf{u}.\mathbf{\nabla}) f = \frac{1}{2} \int d^3 l d\Omega (f_{q'} f_{l'} - f_q f_l) \sigma v_{mol} \ ,
\label{eq:Appx_RTA_Sigma_Def}
\end{equation}

\noindent with $f_{i} \equiv f(\vec{i})$. If we eliminate the spatial dependency of the DF we can identify this equation as the Boltzmann equation for the background DF, we can further identify the terms $\mathcal{D}_1$ and $\mathcal{D}_2$. By comparing it with the expression we had for $\mathcal{D}_2$,  it can be found that


\begin{equation}
\frac{\chi(s)}{4} = \frac{2 E_q E_l (2\pi)^3}{g_i^3} \int d\Omega \sigma v_{mol} = \frac{1}{4(2\pi)} \int d\cos \Theta_{cm} \left| \mathcal{M}\right|^2 \sqrt{1-\frac{4m^2}{s}} \ ,
\label{eq:Appx_RTA_Sigma_ChiEquiv}
\end{equation}

\noindent with $\Theta_{cm}$ the scattering angle measured in the CoM system. This property, apart from relating $\chi$ and $\sigma$, reproduces the feature that $\sigma v_{mol} E_{q} E_{l}$ is a function of $s$ only (see \cite{GondoloGelmini91}). We can use the procedures outlined in the calculation of thermal averaged annihilation cross section to obtain an alternative expression for the thermal average of the relaxation time:


\begin{equation}
\left< \tau^{-1} \right>_{th} = \frac{g_i^3}{32 (2\pi)^2} N \int ds \chi(s) \sqrt{1 - \frac{4m^2}{s}} \sqrt{s}T K_1 (\sqrt{s}/T) \ ,
\label{eq:Appx_RTA_Sigma_InvTau}
\end{equation}

\noindent where $T$ is the temperature of the background Maxwell-Boltzmann DF, $K_n$ is the modified Bessel function of the second kind of order $n$ and $N$ is a normalization constant defined as

\begin{equation}
N^{-1} = \int dE_q q E_q f_0^{MB}(E_q) \ .
\label{eq:Appx_RTA_Sigma_InvTau_Norm}
\end{equation}

\noindent This expression yields the same results as the one in \eqref{eq:Appx_RTA_InvTau_ThAvg}, but it is much easier and faster to implement numerically.

\end{document}